\documentclass[10pt]{article}

\usepackage{amssymb}
\usepackage{amsmath}
\usepackage{siunitx}
\usepackage{geometry}
\usepackage{graphicx}
\usepackage{subcaption}
\usepackage{longtable}
\usepackage{url}
\usepackage{multirow}
\usepackage{hhline}
\usepackage[font=footnotesize]{caption}
\usepackage{natbib}
\setcitestyle{round,semicolon}

\title{A Lipid-Structured Model of Atherosclerotic Plaque Macrophages with Lipid-Dependent Kinetics}
\author{Michael G. Watson, Keith L. Chambers, Mary R. Myerscough}

\begin{document}

\maketitle
\abovedisplayskip=12pt
\belowdisplayskip=12pt
\setlength{\jot}{2ex}

\begin{abstract}
Atherosclerotic plaques are fatty growths in artery walls that cause heart attacks and strokes. Plaque formation is orchestrated by macrophages that are recruited to the artery wall to consume and remove blood-derived lipids, such as low-density lipoprotein (LDL). Ineffective lipid removal, due to macrophage death and other factors, leads to the accumulation of lipid-loaded macrophages and formation of a necrotic core. Experimental observations suggest that macrophage functionality varies with the extent of lipid loading. However, little is known about the resultant influence on plaque fate. Extending work by \citet{Ford19a} and \citet{Cham22}, we develop a plaque model in which macrophages are classified by their ingested lipid content and behave in a lipid-dependent manner. The model, a system of partial-integro differential equations, considers several macrophage behaviours. These include: recruitment to the artery wall; proliferation and apotosis; ingestion of LDL, apoptotic cells and necrotic lipid; emigration from the artery wall; and necrosis of apoptotic cells. Here, we consider apoptosis, emigration and proliferation to be lipid-dependent. We model lipid-dependence in these behaviours with experimentally-informed functions of the internalised lipid load. Our results demonstrate that lipid-dependent macrophage behaviour can substantially alter plaque fate by changing both the total quantity of lipid in the plaque and the distribution of lipid between the live cells, dead cells and necrotic core. For lipid-dependent apoptosis and lipid-dependent emigration simulations, we find significant differences in outcomes for cases that ultimately converge on the same net rate of apoptosis or emigration.
\end{abstract}

\section{Introduction} \label{sIntro}
Atherosclerotic plaques are localised accumulations of cells, lipids and associated debris that form in major arteries \citep{Hans06}. Plaques are initiated when blood-borne low-density lipoprotein (LDL) penetrates the endothelium and deposits in the artery wall \citep{Taba07}. LDL accumulation elicits an immune response that attracts circulating monocytes to the artery wall. These monocytes rapidly differentiate into macrophages, which ingest (phagocytose) the LDL and stimulate further macrophage recruitment through inflammatory signalling \citep{Moor13, Tall15}. Death of lipid-loaded macrophages (known as foam cells) creates fatty deposits that may accumulate over time to form a large necrotic core \citep{Lusi00}. Rupture of a vulnerable plaque can release this necrosis into the bloodstream and trigger a clotting cascade that causes stroke or myocardial infarction \citep{Lusi00, Hans06}.

Not all plaques progress to become clinically dangerous. Many simply resolve naturally or evolve towards a benign, non-resolving state \citep{Baec19}. What largely determines plaque fate is the interaction between macrophages and lipids in the artery wall and, in particular, the relative rates at which these constituents enter and leave the tissue \citep{Moor13}. In addition to recruitment by inflammatory signalling, the plaque macrophage population can be increased by local proliferation \citep{Robb13, Lhot16}. On the other hand, plaque macrophage numbers can be reduced by death (apoptosis) or by emigration out of the wall \citep{Taba10, Llod04}. The primary source of plaque lipid is LDL infiltration from the bloodstream. Since infiltrating LDL becomes bound to the artery wall extracellular matrix \citep{Taba07}, the removal of lipid from the system requires the intervention of macrophages. Lipid internalised by macrophages can be ferried out of the plaque during macrophage emigration or by macrophage offloading to high-density lipoprotein (HDL) in a process known as reverse cholesterol transport \citep{Yvan10}. A further important mechanism, which recycles the cell and lipid content of the plaque, is macrophage efferocytosis \citep{Yin21}. Here, a macrophage phagocytoses an entire apoptotic cell and thus acquires the dying cell's ingested lipid \citep{Ford19b}. Apoptotic cells that are not efficiently cleared by efferocytosis are a source of necrotic material \citep{Koji17}. 

Many experimental results indicate that plaque macrophage behaviour can change with the extent of lipid loading \citep{Taba16}. Some of these lipid-dependent behaviours are well-recognised. For example, lipid accumulation in plaque macrophages is known to upregulate the production of the pro-inflammatory signals required for monocyte recruitment \citep{Tall15}. In addition, it has long been known that the cytotoxic effects of excessive lipid ingestion can lead to macrophage apoptosis \citep{Taba02, Feng03}. More recent evidence has implicated lipid accumulation in the modulation of several other macrophage behaviours relevant to plaque formation. Based on a detailed transcriptomic study of murine plaque macrophages, \citet{Kim18} identified that macrophage proliferation likely decreases with increasing lipid load. Results from both \emph{in vitro} and \emph{in vivo} studies further suggest that macrophages with larger lipid loads are less likely to emigrate from plaques \citep{Chen19, Gils12, Wans13}. This is due to either reduced migration capacity \citep{Chen19} or through increased expression of so-called retention factors \citep{Gils12, Wans13}. Defective macrophage efferocytosis, an often cited mechanism of atherosclerotic plaque progression \citep{Thor09, Lint16}, may also arise through lipid-dependent effects. First because lipid loading can reduce the efficiency of efferocytosis by disrupting relevant signalling pathways \citep{Yin20}, and second because excessive lipid acquisition through efferocytosis can lead to cytotoxic macrophage death \citep{Yin21}. Although these various lipid-dependent effects have been experimentally identified, little is known about how they contribute either individually or collectively to plaque formation. In this paper, we develop a mathematical model to study the impact of lipid-dependent macrophage behaviour on the dynamics and fate of atherosclerotic plaque progression.

Interest in mathematical modelling of atherosclerotic plaque formation has grown in recent years \citep{Part16, Avge19, McAu21}. Most work to date has focussed on modelling the inflammatory response of macrophages in the early plaque. Published approaches include spatially-averaged ODE models \citep{Bule12, Cohe14, Isla16, Thon18, Lui21}, spatially-resolved PDE models \citep{ElKh07, Calv09, Litt09, Yang16, Chal17, Thon19, Silv20} and agent-based models \citep{Bhui17}. To account for macrophage lipid ingestion it is common to assume that the modelled macrophages have two sub-populations: those with little or no internalised lipid (usually termed macrophages) and those with lots of internalised lipid (usually termed foam cells) \citep{Calv09, Bule12, Cill14, Hao14, Isla16, Yang16, Chal17, Silv20}. Lipid-dependent macrophage behaviour can be implicitly incorporated in this model framework by assuming that these sub-populations have, for example, different rates of lipid consumption \citep{Calv09, Bule12, Cill14, Hao14, Isla16, Silv20}, migration \citep{Calv09, Cill14, Yang16, Chal17, Silv20} or apoptosis \citep{Hao14, Isla16, Silv20}. An alternative approach is to model macrophages as a single population and track the population's total ingested lipid content \citep{Litt09, Cohe14, Thon18, Thon19, Lui21}. Here, lipid-dependent effects can be included at the population-level by assuming that macrophage behaviour depends on the average ingested lipid load \citep{Thon18, Thon19}. A more natural means to model lipid accumulation in plaque macrophage populations, including with lipid-dependent effects, is to use a structured population model in which macrophages are classified by their internalised lipid content. Lipid-structured models of plaque macrophages have been developed by \citet{Ford19a}, \citet{Cham22} and \citet{Meun19}.

The model by \citet{Ford19a} uses a system of partial integro-differential equations to study how internalised lipid loads are distributed in live and apoptotic plaque macrophage populations, and how this influences necrotic core formation. Possible behaviours of live macrophages in the model include: (i) apoptosis; (ii) emigration from the plaque; (iii) LDL and necrotic lipid phagocytosis; (iv) lipid offloading to HDL; and (v) efferocytosis of apoptotic cells. Simulations and analysis of this model demonstrate the important roles of emigration and efferocytosis in the prevention of necrotic core growth. Efferocytosis is identified as a double-edged sword, however, as it can drive ingested lipid loads to become excessively large (as confirmed experimentally in \citet{Ford19b}). \citet{Cham22} extended the Ford model to include macrophage proliferation. This introduces an additional means of reducing cell lipid loads as internalised lipid in the parent cell is split between its daughter cells upon division. The model demonstrates that macrophage proliferation can reduce necrotic core formation by enhancing the capacity for necrotic lipid consumption. However, results suggest that proliferation can also be a double-edged sword because the reduction in necrotic core often comes at the expense of a substantially enlarged macrophage population.

\citet{Ford19a} and \citet{Cham22} assume that all plaque macrophage behaviours occur at constant rates independent of internalised lipid content. In this paper, we generalise their lipid-structured framework to include live macrophage behaviour that depends smoothly and continuously on ingested lipid load. For now, we include lipid-dependent behaviour only in macrophage apoptosis, emigration and proliferation. Modelling of lipid-dependent phagocytosis and efferocytosis will be addressed in a future study. By simulating and analysing this new model, we demonstrate that lipid-dependent macrophage behaviour can substantially alter plaque fate by changing both the distribution and the net accumulation of lipid in the system. Note that, while lipid loading is believed to influence plaque fate by modulating the phenotypic profile of the macrophage population \citep{Moor13, Taba16}, we do not explicitly consider this possibility here. The current work can be regarded as a step towards more detailed mathematical models that link macrophage phenotype to internalised lipid load.

The remainder of the manuscript is structured as follows. In Section 2, we outline our methodology. This includes a brief presentation of the model equations; definitions of the functions that characterise lipid-dependent macrophage behaviour; the model non-dimensionalisation; and details of our numerical solution techniques. Section 3 reports results and analysis from an in-depth simulation study that addresses how lipid-dependent apoptosis, emigration and proliferation influence plaque progression. We discuss the implications of our results for both theoretical and experimental atherosclerosis research in Section 4, and end with broad conclusions on the significance of the study in Section 5.

\section{Methods} \label{sMethods}
  
\subsection{Definitions} \label{ssDefs}
The number densities of live and apoptotic plaque macrophages with lipid load $a \geqslant a_0$ at time $t \geqslant 0$ are denoted $m\left(a,t\right)$ and $p\left(a,t\right)$, respectively. The minimum lipid load $a_0$ represents the endogenous lipid contained in the internal structures of each cell. We denote the acellular necrotic lipid content of the plaque at time $t$ by $N\left(t\right)$. For $m\left(a,t\right)$ and $p\left(a,t\right)$, we define the total number of cells in each population by:
\begin{equation}
M\left(t\right)=\int_{a_0}^{\infty} m\left(a,t\right) da, \;\;\; P\left(t\right)=\int_{a_0}^{\infty} p\left(a,t\right) da,
\end{equation}
and the total lipid content of each population by:
\begin{equation}
A_M\left(t\right)=\int_{a_0}^{\infty} a m\left(a,t\right) da, \;\;\; A_P\left(t\right)=\int_{a_0}^{\infty} a p\left(a,t\right) da,
\end{equation}
respectively.

Lipid-dependent cell behaviour is modelled by assuming that the dimensional reference rate of a given behaviour is modulated by a dimensionless factor $g_\diamond\left(a\right)$. The symbol $\diamond$ is a placeholder that represents $\beta$ for apoptosis, $\gamma$ for emigration or $\rho$ for proliferation. For notational convenience, we define the related quantities:
\begin{equation}
G_\diamond\left(t\right) = \int_{a_0}^{\infty} g_\diamond\left(a\right) m\left(a,t\right) \, da, \;\;\; G_{\diamond a}\left(t\right) = \int_{a_0}^{\infty} g_\diamond\left(a\right) a m\left(a,t\right) \, da,
\end{equation}
which will appear in the model equations below. Note that lipid-independence in a given behaviour can be recovered by setting $g_\diamond\left(a\right) \equiv 1$. This, in turn, gives $G_\diamond\left(t\right) = M\left(t\right)$ and $G_{\diamond a}\left(t\right) = A_M\left(t\right)$.

\subsection{Model Statement} \label{ssModel}
A formal derivation of the original lipid-independent model can be found in \citet{Ford19a} and, for the macrophage proliferation terms, in \citet{Cham22}. The interested reader is directed to these works for in-depth explanation. Here, we give a brief summary of the modelling assumptions before stating the equations in full. The model considers the following processes in the plaque:
\begin{enumerate}
  \item LDL and HDL are implicitly assumed to enter the plaque from the bloodstream at fixed rates. Live macrophages consume the LDL and offload lipid to the HDL with net uptake rate $\propto \lambda$. Live macrophages also consume necrotic lipid at rate $\propto \theta$. These two forms of phagocytic lipid uptake are modelled by a continuous advection term in the $m\left(a,t\right)$ equation.
  \item Live macrophages become apoptotic macrophages at rate $\beta g_\beta\left(a\right)$ or emigrate from the plaque at rate $\gamma g_\gamma\left(a\right)$.
  \item Live macrophages consume apoptotic macrophages --- ingesting their entire lipid content --- at rate $\propto \eta$. This process (efferocytosis) contributes a local sink term and a non-local source term to the $m\left(a,t\right)$ equation. The source term takes the form of a convolution integral, which accounts for all possible consumption events that produce live macrophages with lipid content $a$. Note that the value of this integral is to be interpreted as 0 for all $a \leqslant 2a_0$.
  \item Live macrophages proliferate at rate $\rho g_\rho\left(a\right)$. Shortly before division, we assume that the parent cell newly-synthesises the $a_0$ endogenous lipid required to form a second daughter cell \citep{Scag14}. Then, upon division, we assume that the total lipid content of the parent is divided equally between the two daughters. Proliferation contributes a local sink term and a non-local source term to the $m\left(a,t\right)$ equation. The source term accounts for the fact that daughter cells with lipid content $a$ are produced by parent cells with lipid content $2a-a_0$ (i.e.\ the transient, pre-division increase in the parent cell lipid content is not explicitly modelled as it occurs on a timescale much shorter than the timescale of interest).
  \item Apoptotic macrophages undergo post-apoptotic necrosis, producing necrotic lipid at rate $\nu$.
  \item Live macrophages with lipid content $a_0$ are recruited to the plaque from the bloodstream. The rate of cell recruitment is assumed to be a saturating function of the total ingested lipid in the live macrophage population $A_M\left(t\right) - a_0 M\left(t\right)$, with maximal recruitment rate $\alpha$ and half-maximal recruitment when $A_M\left(t\right) - a_0 M\left(t\right) = \kappa$. (Note that this assumption encodes an implicit lipid-dependence because the expression $A_M - a_0 M$ reflects an assumption that macrophages produce recruitment-stimulating cytokines at a rate proportional to their accumulated lipid content $a-a_0$.) Recruitment is modelled by a boundary condition on the $m\left(a,t\right)$ equation.
\end{enumerate}
The equations and boundary condition that reflect the above assumptions are as follows:
\begin{gather}
\begin{split}
& \frac{\partial m\left(a,t\right)}{\partial t} + \bigg[\frac{\lambda}{M\left(t\right)} + \theta N\left(t\right)\bigg] \, \frac{\partial m\left(a,t\right)}{\partial a} = \eta \int_{a_0}^{a-a_0} m(a',t) p(a-a',t)  \, da' \\ & + 4 \rho g_\rho\left(2a-a_0\right) m(2a-a_0, t) - \Big[\beta g_\beta\left(a\right) + \gamma g_\gamma\left(a\right) + \rho g_\rho\left(a\right) + \eta P\left(t\right) \Big] m\left(a,t\right), \label{dim_m_eqn}
\end{split}
\\
\frac{\partial p\left(a,t\right)}{\partial t} = \beta g_\beta\left(a\right) m\left(a,t\right) - \big[\nu M\left(t\right) + \eta\big] p\left(a,t\right), \label{dim_p_eqn} \\
\frac{dN\left(t\right)}{dt} = \nu A_P\left(t\right) - \theta M\left(t\right) N\left(t\right), \label{dim_N_eqn} \\
\bigg[\frac{\lambda}{M\left(t\right)} + \theta N\left(t\right)\bigg]m\left(a_0,t\right) = \frac{\alpha \big[A_M\left(t\right) - a_0 M\left(t\right)\big]}{\kappa + A_M\left(t\right) - a_0 M\left(t\right)}. \label{dim_BC}
\end{gather}

Additional ODEs for $M\left(t\right)$, $A_M\left(t\right)$, $P\left(t\right)$ and $A_P\left(t\right)$ can be generated by integrating (\ref{dim_m_eqn}) and (\ref{dim_p_eqn}) with respect to $a$, both with and without pre-multiplying by $a$. For these calculations, we impose the requirement that $m\left(a,t\right)$, $a m\left(a,t\right)$, $p\left(a,t\right)$ and $a p\left(a,t\right)$ all $\to$ 0 as $a \to \infty$. This leads to the following equations:
\begin{gather}
\frac{dM\left(t\right)}{dt} = \frac{\alpha \big[A_M\left(t\right) - a_0 M\left(t\right)\big]}{\kappa + A_M\left(t\right) - a_0 M\left(t\right)} + \rho G_\rho\left(t\right) - \beta G_\beta\left(t\right) - \gamma G_\gamma\left(t\right), \label{dim_M_eqn} \\
\begin{split}
\frac{dA_M\left(t\right)}{dt} &= \frac{a_0 \alpha \big[A_M\left(t\right) - a_0 M\left(t\right)\big]}{\kappa + A_M\left(t\right) - a_0 M\left(t\right)} + \lambda + \theta M\left(t\right) N\left(t\right) \\ & \quad\quad + \eta A_P\left(t\right) + a_0 \rho G_\rho\left(t\right) - \beta G_{\beta a}\left(t\right) - \gamma G_{\gamma a}\left(t\right), \label{dim_AM_eqn}
\end{split}
\\
\frac{dP\left(t\right)}{dt} = \beta G_\beta\left(t\right) - \big[\nu M\left(t\right) + \eta\big] P\left(t\right), \label{dim_P_eqn} \\
\frac{dA_P\left(t\right)}{dt} = \beta G_{\beta a}\left(t\right) - \big[\nu M\left(t\right) + \eta\big] A_P\left(t\right). \label{dim_AP_eqn}
\end{gather}
Note that, in the presence of lipid-dependent cell behaviour, the ODE equations (\ref{dim_N_eqn}), (\ref{dim_M_eqn})--(\ref{dim_AP_eqn}) and the PDE equations (\ref{dim_m_eqn})--(\ref{dim_p_eqn}) cannot be readily decoupled (in contrast to the simpler models of \citet{Ford19a} and \citet{Cham22}). This limits the opportunity for analytical investigation of the model equations and, hence, this paper will use numerical simulations to study plaque fate and dynamics in a range of pertinent scenarios.
    
The model is closed by assigning appropriate initial conditions to each variable. The PDE variables require initial distributions and the ODE variables require initial values. Generically, we set: 
\begin{equation}
\begin{gathered}
m\left(a,0\right)=m_0\left(a\right), \;\; p\left(a,0\right)=p_0\left(a\right), \\
M\left(0\right)=M_0, \;\; P\left(0\right)=P_0, \;\; A_M\left(0\right)=A_{M0}, \;\; A_P\left(0\right)=A_{P0}, \;\; N\left(0\right)=N_0. \label{dim_ICs}
\end{gathered}
\end{equation}
For $m_0\left(a\right)$ and $p_0\left(a\right)$, we use the following half-normal distributions:
\begin{equation}
m_0\left(a\right) = \frac{2M_0}{a_\sigma \sqrt{2\pi}} \, \exp\left(- \, \frac{\left(a \, - \, a_0\right)^2}{2 {a_\sigma}^2}\right), \;\; 
p_0\left(a\right) = \frac{2P_0}{a_\sigma \sqrt{2\pi}} \, \exp\left(- \, \frac{\left(a \, - \, a_0\right)^2}{2 {a_\sigma}^2}\right), \label{dim_mp_ICs}
\end{equation}
which are scaled such that $\int_{a_0}^{\infty} m_0\left(a\right) \, da = M_0$ and $\int_{a_0}^{\infty} p_0\left(a\right) \, da = P_0$. The parameter $a_\sigma > 0$ defines the shape of the distributions. For $A_{M0}$ and $A_{P0}$, we correspondingly define:
\begin{equation}
A_{M0} = \int_{a_0}^{\infty} a m_0\left(a\right) \, da = M_0\left(a_0 \, + \, \frac{2}{\sqrt{2\pi}} \, a_\sigma\right), \;\;
A_{P0} = \int_{a_0}^{\infty} a p_0\left(a\right) \, da = P_0\left(a_0 \, + \, \frac{2}{\sqrt{2\pi}} \, a_\sigma\right). \label{dim_AMAP_ICs}
\end{equation}
The initial number of live macrophages $M_0$ can be defined in terms of $a_\sigma$ by assuming that the initial distribution $m_0\left(a\right)$ satisfies the boundary condition (\ref{dim_BC}) at $t = 0$. This leads to the relationship:
\begin{equation}
M_0 = \frac{\kappa \lambda \sqrt{2 \pi}}{a_\sigma\Big[a_\sigma \alpha \sqrt{2 \pi} - 2 \lambda\Big]}. \label{dim_M_IC}
\end{equation}
Note that for a valid (positive) $M_0$ value, we require $a_\sigma > \frac{\lambda \sqrt{2}}{\alpha \sqrt{\pi}}$. Finally, we assume that initially there is no necrotic lipid in the system ($N_0 = 0$), and the system contains fewer dead cells than live cells. To satisy the latter assumption, we arbitrarily set $P_0 = 0.5 M_0$.

\subsection{Lipid-Dependent Rate Functions} \label{ssLipidFunc}
In the model, lipid-dependent cell behaviour is described by either a monotonic function $g_\diamond\left(a\right) = g_\diamond^s\left(a\right)$ or a non-monotonic function $g_\diamond\left(a\right) = g_\diamond^r\left(a\right)$. For the monotonic function, we use the following saturating relationship:
\begin{equation}
g_\diamond^s\left(a\right)=\frac{\left(a_\diamond - a_0\right)^{n_\diamond} + \delta_\diamond \left(a - a_0\right)^{n_\diamond}}{\left(a_\diamond - a_0\right)^{n_\diamond} + \left(a - a_0\right)^{n_\diamond}}. \label{dim_g_mono}
\end{equation}
Here, $g_\diamond^s\left(a_0\right) = 1$ and the non-negative parameter $\delta_\diamond = \lim_{a\to\infty} g_\diamond^s\left(a\right)$. Hence, $g_\diamond^s\left(a\right)$ increases from 1 to $\delta_\diamond$ when $\delta_\diamond > 1$ or decreases from 1 to $\delta_\diamond$ when $0 \leqslant \delta_\diamond < 1$. The exponent $n_\diamond \geqslant 1$ determines the shape of the function and the parameter $a_\diamond > a_0$ denotes the lipid content for which $g_\diamond^s\left(a_\diamond\right) = \frac{1 + \delta_\diamond}{2}$. For the non-monotonic function, we use the following scaled relationship:
\begin{equation}
g_\diamond^r\left(a\right)=\epsilon_\diamond + \left(1 - \epsilon_\diamond\right)\left(\frac{q_\diamond {b_\diamond}^{\left(q_\diamond - k_\diamond\right)}}{\sqrt[\uproot{3} q_\diamond]{{k_\diamond}^{k_\diamond} {\left(q_\diamond - k_\diamond\right)}^{\left(q_\diamond - k_\diamond\right)}}} \left[\frac{\left(a - a_0\right)^{k_\diamond}}{{b_\diamond}^{q_\diamond} + \left(a - a_0\right)^{q_\diamond}}\right]\right), \label{dim_g_nonmono}
\end{equation}
where $0 \leqslant \epsilon_\diamond < 1$, $b_\diamond > 0$ and the exponents $q_\diamond$ and $k_\diamond$ satisfy $q_\diamond > k_\diamond \geqslant 1$. This function increases from $g_\diamond^r\left(a_0\right) = \epsilon_\diamond$ to a peak value of 1 at $a = a_0 + b_\diamond \sqrt[\uproot{3} q_\diamond]{\frac{k_\diamond}{q_\diamond - k_\diamond}}$, before decreasing towards $\epsilon_\diamond$ as $a$ tends to infinity. The exponent $k_\diamond$ controls the rate of increase of $g_\diamond^r\left(a\right)$ for $a \approx a_0$, while the larger exponent $q_\diamond$ controls the rate of decline of $g_\diamond^r\left(a\right)$ as $a \to \infty$.

In practice, we shall primarily focus on modulating factors that have a monotonic dependence on $a$ (i.e.\ monotonic increasing for $g_\beta\left(a\right)$, monotonic decreasing for $g_\gamma\left(a\right)$ and $g_\rho\left(a\right)$). However, for $g_\gamma\left(a\right)$, we shall also consider the non-monotonic dependence, since physical arguments suggest that this may be a more realistic assumption.

\subsection{Nondimensionalisation} \label{ssNondim}
Using tildes to denote dimensionless quantities, the independent and dependent variables are nondimensionalised as follows \citep{Cham22}:
\begin{equation}
\begin{gathered}
\tilde{a} = \frac{a}{a_0}, \;\; \tilde{t} = \beta t, \\
\tilde{m}\left(\tilde{a},\tilde{t}\right) = \bigg(\frac{a_0}{M\left(t\right)}\bigg) \, m\left(a,t\right), \;\; \tilde{p}\left(\tilde{a},\tilde{t}\right) = \bigg(\frac{a_0}{P\left(t\right)}\bigg) \, p\left(a,t\right), \\
\tilde{M}\left(\tilde{t}\right) = \bigg(\frac{\beta \, + \, \gamma}{\alpha}\bigg) \, M\left(t\right), \;\; \tilde{P}\left(\tilde{t}\right) = \bigg(\frac{\beta \, + \, \gamma}{\alpha}\bigg) \, P\left(t\right), \\
\tilde{A}_M\left(\tilde{t}\right) = \bigg(\frac{\beta \, + \, \gamma}{a_0 \, \alpha}\bigg) \, A_M\left(t\right), \;\; \tilde{A}_P\left(\tilde{t}\right) = \bigg(\frac{\beta \, + \, \gamma}{a_0 \, \alpha}\bigg) \, A_P\left(t\right), \;\; \tilde{N}\left(\tilde{t}\right) = \bigg(\frac{\beta \, + \, \gamma}{a_0 \, \alpha}\bigg) \, N\left(t\right). \label{nondim}
\end{gathered}
\end{equation}
These scalings are chosen to give $\int_{1}^{\infty} \tilde{m}\left(\tilde{a},\tilde{t}\right) \, d\tilde{a} = \int_{1}^{\infty} \tilde{p}\left(\tilde{a},\tilde{t}\right) \, d\tilde{a} = 1$, such that $\tilde{m}\left(\tilde{a},\tilde{t}\right)$ and $\tilde{p}\left(\tilde{a},\tilde{t}\right)$ may be considered as probability density functions for the live and apoptotic macrophage populations, respectively. We further define the following dimensionless parameters:
\begin{equation}
\begin{gathered}
\tilde{\psi} = \frac{\beta \, + \, \gamma}{\beta}, \;\; \tilde{\kappa} = \frac{\kappa \, \left(\beta \, + \, \gamma\right)}{a_0 \, \alpha}, \;\; \tilde{\rho} = \frac{\rho}{\beta}, \;\; \tilde{\nu} = \frac{\nu}{\beta}, \;\;
\tilde{\lambda} = \frac{\lambda}{a_0 \, \alpha}, \;\; \tilde{\theta} = \frac{\theta \, \alpha}{\beta \, \left(\beta \, + \, \gamma\right)}, \;\; \tilde{\eta} = \frac{\eta \, \alpha}{\beta \, \left(\beta \, + \, \gamma\right)}, \\
\tilde{a}_\diamond = \frac{a_\diamond}{a_0}, \;\;
\tilde{b}_\diamond = \frac{b_\diamond}{a_0}, \;\;  
\tilde{a}_\sigma = \frac{a_\sigma}{a_0}, \;\;
\tilde{n}_\diamond = n_\diamond, \;\; \tilde{\delta}_\diamond = \delta_\diamond, \;\; 
\tilde{k}_\diamond = k_\diamond, \;\; \tilde{q}_\diamond = q_\diamond,
\\
\tilde{M}_0 = \bigg(\frac{\beta \, + \, \gamma}{\alpha}\bigg) \, M_0, \;\; \tilde{P}_0 = \bigg(\frac{\beta \, + \, \gamma}{\alpha}\bigg) \, P_0, \;\;
\tilde{A}_{M0} = \bigg(\frac{\beta \, + \, \gamma}{a_0 \, \alpha}\bigg) \, A_{M0}, \;\; \tilde{A}_{P0} = \bigg(\frac{\beta \, + \, \gamma}{a_0 \, \alpha}\bigg) \, A_{P0}. \label{params}
\end{gathered}
\end{equation}
Using (\ref{nondim}) and (\ref{params}), and dropping tildes for notational convenience, the dimensionless model equations can be expressed as follows:
\begin{gather}
\begin{split}
\frac{\partial m\left(a,t\right)}{\partial t} & + \bigg[\frac{\lambda \psi}{M\left(t\right)} + \theta N\left(t\right)\bigg] \frac{\partial m\left(a,t\right)}{\partial a} \\ & = \eta P\left(t\right) \bigg[\int_{1}^{a-1} m(a',t) p(a-a',t) \, da' - m\left(a,t\right) \bigg] \\ & \quad\quad + \bigg[G_\beta\left(t\right) - g_\beta\left(a\right) + \Big(\psi - 1\Big) \Big(G_\gamma\left(t\right) - g_\gamma\left(a\right)\Big) \bigg] m\left(a,t\right)\\ & \quad\quad\quad + 4 \rho g_\rho\left(2a-1\right) m(2a-1,t) - \Big[G_\rho\left(t\right) + g_\rho\left(a\right)\Big] \rho m\left(a,t\right)\\ & \quad\quad\quad\quad - \bigg[\frac{\psi}{M\left(t\right)} \, \frac{A_M\left(t\right) - M\left(t\right)}{\kappa + A_M\left(t\right) - M\left(t\right)}\bigg] m\left(a,t\right), \label{dimless_m_eqn}
\end{split}
\\
\frac{\partial p\left(a,t\right)}{\partial t} = \frac{M\left(t\right)}{P\left(t\right)} \, \Big[g_\beta\left(a\right) m\left(a,t\right) - G_\beta\left(t\right) p\left(a,t\right)\Big], \label{dimless_p_eqn} \\
\frac{dN\left(t\right)}{dt} = \nu A_P\left(t\right) - \theta M\left(t\right) N\left(t\right), \label{dimless_N_eqn} \\
\frac{dM\left(t\right)}{dt} = \frac{\psi \big[A_M\left(t\right) - M\left(t\right)\big]}{\kappa + A_M\left(t\right) - M\left(t\right)} - \Big[G_\beta\left(t\right) + \big(\psi - 1\big) G_\gamma\left(t\right) - \rho G_\rho\left(t\right)\Big] M\left(t\right), \label{dimless_M_eqn} \\
\begin{split}
\frac{dA_M\left(t\right)}{dt} & = \psi \, \bigg[\frac{A_M\left(t\right) - M\left(t\right)}{\kappa + A_M\left(t\right) - M\left(t\right)} + \lambda\bigg] + \theta M\left(t\right) N\left(t\right)  \\ & \quad + \eta M\left(t\right) A_P\left(t\right) - \Big[G_{\beta a}\left(t\right) + \big(\psi - 1\big) G_{\gamma a}\left(t\right) - \rho G_\rho\left(t\right)\Big] M\left(t\right), \label{dimless_AM_eqn} 
\end{split}
\\
\frac{dP\left(t\right)}{dt} = G_\beta\left(t\right) M\left(t\right) - \Big[\nu + \eta M\left(t\right)\Big] P\left(t\right), \label{dimless_P_eqn} \\
\frac{dA_P\left(t\right)}{dt} = G_{\beta a}\left(t\right) M\left(t\right) - \Big[\nu + \eta M\left(t\right)\Big] A_P\left(t\right). \label{dimless_AP_eqn}
\end{gather}
The boundary condition (\ref{dim_BC}) becomes:
\begin{equation}
\bigg[\frac{\lambda \psi}{M\left(t\right)} + \theta N\left(t\right)\bigg] m\left(1,t\right) = \frac{\psi}{M\left(t\right)} \, \frac{A_M\left(t\right) - M\left(t\right)}{\kappa + A_M\left(t\right) - M\left(t\right)}, \label{dimless_BC}
\end{equation}
and the initial conditions become: 
\begin{equation}
\begin{gathered}
m_0\left(a\right) = p_0\left(a\right) = \frac{2}{a_\sigma \sqrt{2\pi}} \, \exp\left(- \, \frac{\left(a - 1\right)^2}{2 {a_\sigma}^2}\right), \\
N_0 = 0, \\
M_0 = \frac{\kappa \lambda \sqrt{2 \pi}}{a_\sigma\Big[a_\sigma \sqrt{2 \pi} - 2 \lambda\Big]}, \;\;
P_0 = 0.5M_0, \\
A_{M0} = M_0\left(1 + \frac{2}{\sqrt{2\pi}} \, a_\sigma\right), \;\;
A_{P0} = P_0\left(1 + \frac{2}{\sqrt{2\pi}} \, a_\sigma\right). \label{dimless_ICs}
\end{gathered}
\end{equation}
Note that for all simulations in this paper we set $a_\sigma = 0.5$.

Based on a similar argument to that presented in \citet{Cham22}, we note that equations (\ref{dimless_M_eqn}) and (\ref{dimless_AM_eqn}) will lead to unbounded growth of $M\left(t\right)$ and $A_M\left(t\right)$ if:
\begin{equation*}
G_\beta\left(t\right) + \big(\psi - 1\big) G_\gamma\left(t\right) - \rho G_\rho\left(t\right) \leqslant 0.
\end{equation*}
To avoid this eventuality in the current study, we ensure that the condition:
\begin{equation*}
g_\beta\left(a\right) + \big(\psi - 1\big) g_\gamma\left(a\right) - \rho g_\rho\left(a\right) > 0,
\end{equation*}
is satisfied for all $a \in [1, \infty)$.

The functions that control lipid-dependent cell behaviour are recast in dimensionless quantities as follows:
\begin{gather}
g_\diamond^s\left(a\right)=\frac{\left(a_\diamond - 1\right)^{n_\diamond} + \delta_\diamond \left(a - 1\right)^{n_\diamond}}{\left(a_\diamond - 1\right)^{n_\diamond} + \left(a - 1\right)^{n_\diamond}}, \label{dimless_g_mono} \\ 
g_\diamond^r\left(a\right)=\epsilon_\diamond + \left(1 - \epsilon_\diamond\right)\left(\frac{q_\diamond {b_\diamond}^{\left(q_\diamond - k_\diamond\right)}}{\sqrt[\uproot{3} q_\diamond]{{k_\diamond}^{k_\diamond} {\left(q_\diamond - k_\diamond\right)}^{\left(q_\diamond - k_\diamond\right)}}} \left[\frac{\left(a - 1\right)^{k_\diamond}}{{b_\diamond}^{q_\diamond} + \left(a - 1\right)^{q_\diamond}}\right]\right). \label{dimless_g_nonmono}
\end{gather}
The dimensionless integral terms are now defined as $G_\diamond\left(t\right) = \int_{1}^{\infty} g_\diamond\left(a\right) m\left(a,t\right) \, da$ and $G_{\diamond a}\left(t\right) = \int_{1}^{\infty} g_\diamond\left(a\right) a m\left(a,t\right) \, da$. Lipid-dependence in a given behaviour can again be removed by setting $g_\diamond\left(a\right) \equiv 1$. In dimensionless terms, this leads to $G_\diamond\left(t\right) = 1$ and $G_{\diamond a}\left(t\right) = \frac{A_M\left(t\right)}{M\left(t\right)}$.

\subsection{Numerical Solution of Equations} \label{ssNumer}
The model equations are solved by reformulating (\ref{dimless_m_eqn})--(\ref{dimless_AP_eqn}) as a large system of coupled ODEs (method of lines) and integrating with the MATLAB routine \emph{ode15s}. This approach requires the infinite $a$ domain for equations (\ref{dimless_m_eqn}) and (\ref{dimless_p_eqn}) to be capped at a finite upper limit $a_{max}$ and appropriately discretised. In choosing $a_{max}$, we aim to minimise numerical error associated with the loss of live cells (and their ingested lipid) across the upper domain boundary. For the scenarios considered in this paper, we find that this typically requires $a_{max}$ values on the order $10^2$--$10^3$. Since uniform discretisation of such large domains would require a substantial (and potentially unfeasible) number of grid points, we hereby adopt a non-uniform gridding strategy to reduce the number of points required.

We discretise the $a$ domain into $I$ points $a_i$ ($i = 1,2,...,I$) by assuming that the spacing between adjacent points $\Delta a_i = a_{i+1} - a_i$ increases linearly from $\Delta a_1 = \Delta a_{min}$ to $\Delta a_{I-1} = c\Delta a_{min}$, where $c > 1$ is a constant. (Note that this equates to adding the constant increment $\frac{\Delta a_{min} \left(c-1\right)}{\left(I-2\right)}$ to the width of each subsequent grid spacing). The positions of individual grid points are thus given by the expression:
\begin{equation}
	a_i = 1 + \Delta a_{min} \left(i-1\right) \left[1 + \frac{\left(i-2\right) \left(c-1\right)}{2\left(I-2\right)}\right],
\end{equation}
where $a_{max} = a_I = 1 + \frac{1}{2} \Delta a_{min} \left(c + 1\right) \left(I - 1\right)$. We stress that this non-uniform gridding approach is by no means optimised for the problem. However, we find that it is well suited to the model equations because it allows for the use of a high grid point density near $a = 1$ (where solutions can vary rapidly with $a$) and a low grid point density for $a \gg 1$ (where solutions have relatively little variation with $a$). For the simulations in this paper, we use two different domain sizes with the following discretisations:
\begin{enumerate}
  \item $\Delta a_{min} = 0.005$, $c = 126$, $I = 1258$ $\implies a_{max} = 400.0975$; 
  \item $\Delta a_{min} = 0.005$, $c = 200$, $I = 2001$ $\implies a_{max} = 1006$.
\end{enumerate}
Note that the ratios $\frac{\Delta a_{min} \left(c-1\right)}{\left(I-2\right)}$ are very similar for the two discretisations, meaning that corresponding points $a_i$ have almost identical positions for all $i \leqslant 1258$. Most simulations in this study use the smaller domain. However, for cases that consider lipid-dependent emigration, calculations are performed on the larger domain. The larger domain is required in this case because the reduced rate of emigration for lipid-loaded cells tends to drive an increase in the proportion of cells with very large lipid quantities.

Using the above discretisation, we define the quantities $m_i\left(t\right) = m\left(a_i,t\right)$ and $p_i\left(t\right) = p\left(a_i,t\right)$ for $i = 1,...,I$. Equations (\ref{dimless_m_eqn})--(\ref{dimless_AP_eqn}) are then recast as a system of $2I + 4$ ODEs for $m_2\left(t\right),...,$ $m_I\left(t\right)$, $p_1\left(t\right),...,$ $p_I\left(t\right)$, $N\left(t\right)$, $M\left(t\right)$, $A_M\left(t\right)$, $P\left(t\right)$ and $A_P\left(t\right)$ coupled to an algebraic constraint for $m_1\left(t\right)$ as given by the boundary condition (\ref{dimless_BC}). All integral terms in these ODEs are approximated in terms of the $m_i\left(t\right)$ and $p_i\left(t\right)$ using the trapezoidal rule. The derivative in the structural variable is approximated by the non-uniform second-order upwind scheme:
\begin{equation}
\frac{\partial m}{\partial a}\left(a_i,t\right) \approx \frac{3m_i\left(t\right) - 4m_{i-1}\left(t\right) + m_{i-2}\left(t\right)}{3\Delta a_{i-1} - \Delta a_{i-2}},
\end{equation}
for $3 \leqslant i \leqslant I$, and by the non-uniform second-order centered scheme:
\begin{equation}
\frac{\partial m}{\partial a}\left(a_2,t\right) \approx \frac{m_3\left(t\right) - m_1\left(t\right)}{\Delta a_2 + \Delta a_1}
\end{equation}
for $i=2$.

The proliferation and efferocytosis terms in equation (\ref{dimless_m_eqn}) pose additional challenges to the numerical solution of the ODE system. The following provides further details on our handling of these non-local terms:
\begin{enumerate}
\item  The proliferation source term $4 \rho g_\rho\left(2a-1\right) m(2a-1,t)$ is not well-defined for $2a-1 > a_{max}$ (i.e.\ proliferation of cells with lipid loads greater than $a_{max}$ cannot be quantified). We therefore omit this term from all $m_i\left(t\right)$ equations for which $a_i > \frac{1}{2}\left(a_{max}+1\right)$. For the large simulation domains that we use, the numerical error associated with these omissions is negligible (indeed, in the case that $g_\rho\left(a\right)$ tends rapidly to 0, the error is vanishingly small).  
\item To evaluate the proliferation and efferocytosis source terms in the $m_i\left(t\right)$ equations, we often need to approximate $m$ and $p$ values at positions that are not coincident with grid points (e.g.\ $m\left(2a-1,t\right)$ in the proliferation source term, $p\left(a-a',t\right)$ in the integrand of the efferocytosis source term). Whenever such an approximation is required, we perform a linear interpolation using the appropriate pair of $m_i\left(t\right)$ or $p_i\left(t\right)$ values from adjacent grid points.
\item Upon each evaluation of an efferocytosis source integral, we divide the result by the numerical approximation to the following integral:
\begin{equation*}
	\int_{1}^{\infty}\int_{1}^{a-1} m(a',t) p(a-a',t) \, da' \, da = \left(\int_{1}^{\infty} m\left(a,t\right) \, da\right) \cdot \left(\int_{1}^{\infty} p\left(a,t\right) \, da\right).
\end{equation*}
This re-normalises the result and acts to prevent the growth of small errors that arise in the numerical calculation of the efferocytosis integral. Note that no such intervention is required for the other integral terms in the model.
\end{enumerate}

\subsection{Total System Lipid and Average Lipid per Cell} \label{ssTotAv}
Before presenting numerical results, we pause to draw attention to three other dimensionless quantities of interest. Specifically, the total amount of lipid in the plaque $L\left(t\right) = A_M\left(t\right) + A_P\left(t\right) + N\left(t\right)$, the average lipid content per live cell ${\bar{A}}_M\left(t\right) = \frac{A_M\left(t\right)}{M\left(t\right)}$, and the average lipid content per apoptotic cell ${\bar{A}}_P\left(t\right) = \frac{A_P\left(t\right)}{P\left(t\right)}$. Using the ODE system (\ref{dimless_N_eqn})-(\ref{dimless_AP_eqn}), additional ODEs for $L\left(t\right)$, ${\bar{A}}_M\left(t\right)$ and ${\bar{A}}_P\left(t\right)$ can be derived. These ODEs give insight into how $L\left(t\right)$, ${\bar{A}}_M\left(t\right)$ and ${\bar{A}}_P\left(t\right)$ evolve with time, and this proves to be useful for interpreting the simulation outcomes. We therefore present the equations below. Note that, in what follows, we make the definition $F\left(t\right) = \frac{A_M\left(t\right) - M\left(t\right)}{\kappa + A_M\left(t\right) - M\left(t\right)}$, where $F\left(t\right)$ is proportional to the dimensionless macrophage recruitment rate.

For total system lipid $L\left(t\right)$, we have the equation:
\begin{equation}
\frac{dL\left(t\right)}{dt} = \psi F\left(t\right) + \psi\lambda + \rho G_\rho\left(t\right) M\left(t\right) - \left(\psi - 1\right) G_{\gamma a}\left(t\right) M\left(t\right). \label{dimless_L_eqn}
\end{equation}
Accordingly, the mechanisms that add lipid to the system are monocyte recruitment, LDL consumption, and local macrophage proliferation (first, second and third terms on the right-hand side, respectively). Contrastingly, the sole mechanism of lipid removal from the system is macrophage emigration (final term on the right-hand side). Note that the time evolution of $L\left(t\right)$ is driven entirely by the behaviour of live macrophages, while explicit lipid-dependent effects appear only in the proliferation and emigration terms.

For average lipid content per live cell, we have the equation:
\begin{equation}
\begin{split}
\frac{d{\bar{A}}_M\left(t\right)}{dt} & = \frac{\psi \lambda}{M\left(t\right)} + \theta N\left(t\right) + \eta A_P\left(t\right) - \frac{\psi F\left(t\right)}{M\left(t\right)} \Big[{\bar{A}}_M\left(t\right) - 1\,\Big] 
- \rho G_\rho\left(t\right) \Big[{\bar{A}}_M\left(t\right) - 1\,\Big] \\ & \quad\quad + \Big[G_\beta\left(t\right) {\bar{A}}_M\left(t\right) - G_{\beta a}\left(t\right)\Big] 
+ \left(\psi - 1\right)\Big[G_\gamma\left(t\right) {\bar{A}}_M\left(t\right) - G_{\gamma a}\left(t\right)\Big]. \label{dimless_AMbar_eqn} 
\end{split}
\end{equation}
Here, the first three terms on the right-hand side demonstrate that LDL consumption, necrotic lipid consumption and efferocytosis, respectively, all act to increase ${\bar{A}}_M\left(t\right)$. On the other hand, both monocyte recruitment and macrophage proliferation (terms four and five, respectively) always act to reduce ${\bar{A}}_M\left(t\right)$ (note that ${\bar{A}}_M\left(t\right) > 1$ in all but the extreme case where $m\left(a,t\right)$ is a delta distribution). The final two terms, which relate to apoptosis (term six) and emigration (term seven), can act to either increase or decrease ${\bar{A}}_M\left(t\right)$ depending on their signs. Letting $\diamond$ denote $\beta$ or $\gamma$, we see that these terms act to increase ${\bar{A}}_M\left(t\right)$ when $\frac{G_{\diamond a}\left(t\right)}{G_\diamond\left(t\right)} < {\bar{A}}_M\left(t\right)$ and act to decrease ${\bar{A}}_M\left(t\right)$ when $\frac{G_{\diamond a}\left(t\right)}{G_\diamond\left(t\right)} > {\bar{A}}_M\left(t\right)$. Here, the ratio $\frac{G_{\diamond a}\left(t\right)}{G_\diamond\left(t\right)}$ can be interpreted as the average lipid load of cells leaving the live cell population via apoptosis or emigration at time $t$. Thus, intuitively, when the cells leaving the live cell population have a smaller (larger) average lipid load than the live cells that remain, the average lipid load of the live cell population tends to increase (decrease). Note that the impact of apoptosis and emigration on ${\bar{A}}_M\left(t\right)$ observed here is entirely due to the lipid-dependent terms in the model. In the absence of lipid-dependent apoptosis and emigration
, the final two terms on the right-hand side of (\ref{dimless_AMbar_eqn}) equate to zero.

Finally, the ODE for average lipid content per apoptotic cell is:
\begin{equation}
\frac{d{\bar{A}}_P\left(t\right)}{dt} = \frac{M\left(t\right)}{P\left(t\right)}\Big[G_{\beta a}\left(t\right) - G_\beta\left(t\right) {\bar{A}}_P\left(t\right)\Big]. \label{dimless_APbar_eqn} 
\end{equation}
This equation is noticeably simpler than that for the live cells and contains only a term relating to macrophage apoptosis. Here, we see that ${\bar{A}}_P\left(t\right)$ increases when $\frac{G_{\beta a}\left(t\right)}{G_\beta\left(t\right)} > {\bar{A}}_P\left(t\right)$ and decreases when $\frac{G_{\beta a}\left(t\right)}{G_\beta\left(t\right)} < {\bar{A}}_P\left(t\right)$. Again, this makes intuitive sense because ${\bar{A}}_P\left(t\right)$ increases (decreases) when the average lipid load of dying cells is larger (smaller) than the average lipid load of those already dead. Note that, unlike the corresponding term in the ${\bar{A}}_M\left(t\right)$ equation, the right-hand side here does not vanish in the absence of lipid-dependent apoptosis. Rather, the term in brackets reduces to ${\bar{A}}_M\left(t\right) - {\bar{A}}_P\left(t\right)$ and the ODE acts to equilibrate the average apoptotic cell lipid content to the average live cell lipid content. 

\section{Results} \label{sResults}
The model outlined in Section \ref{sMethods} includes lipid-dependent terms for macrophage apoptosis, emigration and proliferation. We stress, however, that it is not our intention to apply the model in its full generality. Rather, we shall consider lipid-dependence in each behaviour individually to investigate how each lipid-dependent behaviour can influence plaque progression. To maintain our focus on the role of lipid-dependence, we fix all parameter values not related to the functions $g_\diamond\left(a\right)$ (see Table \ref{LIparams}). For the functions $g_\diamond\left(a\right)$, we consider a range of parameterisations, both in unscaled and scaled formats. For unscaled simulations, we apply the functions $g_\diamond^s\left(a\right)$ or $g_\diamond^r\left(a\right)$ exactly as defined in Section \ref{ssNondim}. For scaled simulations, we pre-multiply the relevant function by a scaling value that we have found (by simulation) to give $G_\diamond\left(t\right) \approx 1$ at steady-state (specifically, we accept $G_\diamond\left(\infty\right) = 1 \pm 0.01$). Scaling the functions in this way allows for a consistent comparison of steady-state results from both lipid-dependent \emph{and} lipid-independent cases because the net (population level) rate of the behaviour of interest is conserved. Note that we focus mainly on steady state results because the time to reach steady state (typically around 100 macrophage lifetimes) is considerably shorter than the lifespan of a plaque. Table \ref{mono_params} (monotonic functions) and Table \ref{nonmono_params} (non-monotonic functions) summarise the cases that we consider in the following sections. Each table reports function parameterisations, $G_\diamond\left(\infty\right)$ values from unscaled simulations and corresponding scaling values required to give $G_\diamond\left(\infty\right) \approx 1$. 
\begin{table}[hp]
	\centering
	\footnotesize
	\begin{tabular}{| l | l | l |}
		\hline
		Parameter & Description & Value \\ \hline
		$\psi$ & Reference dimensionless live cell loss rate & 1.2\\ \hline
		$\kappa$ & Dimensionless live cell accumulated lipid content for half-maximal recruitment & 5\\ \hline
		$\rho$ & Reference dimensionless live cell proliferation rate & 0 (Section \ref{ssNoProlif})\\
		& & 0.5 (Section \ref{ssProlif})\\ \hline
		$\nu$ & Dimensionless post-apoptotic necrosis rate & 1\\ \hline
		$\lambda$ & Dimensionless net LDL consumption/HDL offloading rate & 0.1\\ \hline
		$\theta$ & Dimensionless necrotic lipid consumption rate & 0.5\\ \hline
		$\eta$ & Dimensionless efferocytosis rate & 8\\ \hline
	\end{tabular}
	\caption{Base case lipid-independent parameter values. These values, which are consistent with the dimensional parameter estimates reported in \citet{Ford19a}, are used for all simulation results in Section \ref{sResults}. Note that $\rho$ has two values because we neglect proliferation for the results in Section \ref{ssNoProlif}, but include it for the results in Section \ref{ssProlif}.} \label{LIparams}
\end{table}
\begin{table}[hp]
	\centering
	\footnotesize
	\begin{tabular}{| c | c | >{\centering}p{0.52cm} | >{\centering}p{0.52cm} | >{\centering}p{0.52cm} || >{\centering}p{1.88cm} | >{\centering}p{1.88cm} || >{\centering}p{1.88cm} | c |}
		\hline
		\multirow{3}*{Cell Behaviour} & \multirow{3}*{$g_\diamond\!\left(a\right)$} & \multicolumn{3}{ c ||}{Parameter Values} & \multicolumn{2}{ c ||}{No Proliferation ($\rho = 0$)} & \multicolumn{2}{ c |}{Proliferation ($\rho = 0.5$)}\\ \cline{3-9}
		& & \multirow{2}*{$a_\diamond$} & \multirow{2}*{$\delta_\diamond$} & \multirow{2}*{$n_\diamond$} & $G_\diamond\!\left(\infty\right)$ & Scaling Values & $G_\diamond\!\left(\infty\right)$ & Scaling Values\\
		& & & & & (unscaled) & ($G_\diamond\!\left(\infty\right) \approx 1$) & (unscaled) & ($G_\diamond\!\left(\infty\right) \approx 1$)\\ \hhline{|=|=|=|=|=#=|=#=|=|}
		Apoptosis & $g_\beta^s\left(a\right)$ & 15 & 2 & 2 & 1.188 & 0.86 & \multicolumn{2}{ c |}{Not considered}\\ \hline
		Apoptosis & $g_\beta^s\left(a\right)$ & 12 & 3 & 2 & 1.576 & 0.72 & 1.445 & 0.76\\ \hline
		Apoptosis & $g_\beta^s\left(a\right)$ & 9 & 4 & 2 & 2.367 & 0.565 & \multicolumn{2}{ c |}{Not considered}\\ \hline
		Emigration & $g_\gamma^s\left(a\right)$ & 12 & 0.1 & 1.5 & 0.6902 & 1.38* & \multicolumn{2}{ c |}{Not considered}\\ \hline
		Emigration & $g_\gamma^s\left(a\right)$ & 18 & 0.1 & 1.5 & 0.7812 & 1.25* & 0.7650 & 1.25*\\ \hline
		Emigration & $g_\gamma^s\left(a\right)$ & 24 & 0.1 & 1.5 & 0.8340 & 1.18* & \multicolumn{2}{ c |}{Not considered}\\ \hline
		Proliferation & $g_\rho^s\left(a\right)$ & 4 & 0 & 2 & \multicolumn{2}{ c ||}{Not applicable} & 0.6377 & 1.47\\ \hline
		Proliferation & $g_\rho^s\left(a\right)$ & 9 & 0 & 2 & \multicolumn{2}{ c ||}{Not applicable} & 0.8070 & 1.22\\ \hline
		Proliferation & $g_\rho^s\left(a\right)$ & 14 & 0 & 2 & \multicolumn{2}{ c ||}{Not applicable} & 0.8675 & 1.15\\ \hline
	\end{tabular}
	\caption{Parameterisations, unscaled $G_\diamond\!\left(\infty\right)$ values and corresponding scaling values for simulations with the monotonic rate modulating function (\ref{dimless_g_mono}). Asterisks denote that $\delta_\gamma$ is divided by the scaling value when the scaling is applied. This ensures that both the scaled and unscaled functions tend towards the same value as $a \to \infty$.} \label{mono_params}
\end{table}
\begin{table}[hp]
	\centering
	\footnotesize
	\begin{tabular}{| c | c | >{\centering}p{0.52cm} | >{\centering}p{0.52cm} | >{\centering}p{0.52cm} | >{\centering}p{0.52cm} | c | c |}
		\hline
		\multirow{2}*{Cell Behaviour} & \multirow{2}*{$g_\diamond\!\left(a\right)$} & \multicolumn{4}{ c |}{Parameter Values} & $G_\diamond\!\left(\infty\right)$ & Scaling Values\\ \cline{3-6}
		& & $\epsilon_\diamond$ & $b_\diamond$ & $k_\diamond$ & $q_\diamond$ & (unscaled) & ($G_\diamond\!\left(\infty\right) \approx 1$)\\ \hhline{|=|=|=|=|=|=|=|=|}
		Emigration & $g_\gamma^r\left(a\right)$ & 0.1 & 3 & 1 & 2 & 0.5736 & 1.9*\\ \hline
		Emigration & $g_\gamma^r\left(a\right)$ & 0.1 & 6 & 1 & 2 & 0.5418 & 1.99*\\ \hline
		Emigration & $g_\gamma^r\left(a\right)$ & 0.1 & 9 & 1 & 2 & 0.4964 & 2.08*\\ \hline
	\end{tabular}
	\caption{Parameterisations, unscaled $G_\diamond\!\left(\infty\right)$ values and corresponding scaling values for simulations with the non-monotonic rate modulating function (\ref{dimless_g_nonmono}). Note that these simulations do not consider macrophage proliferation (i.e.\ $\rho = 0$). Asterisks denote that scaling values multiply only the second term on the right-hand side of (\ref{dimless_g_nonmono}). This ensures that both the scaled and unscaled functions tend towards $\epsilon_\gamma$ as $a \to \infty$.} \label{nonmono_params}
\end{table}

\subsection{Lipid-Independent Base Case Simulations} \label{ssBase}
\begin{figure}
	\centering
	\begin{subfigure}[b]{0.49\textwidth}
		\centering
		\includegraphics[height=5.5cm]{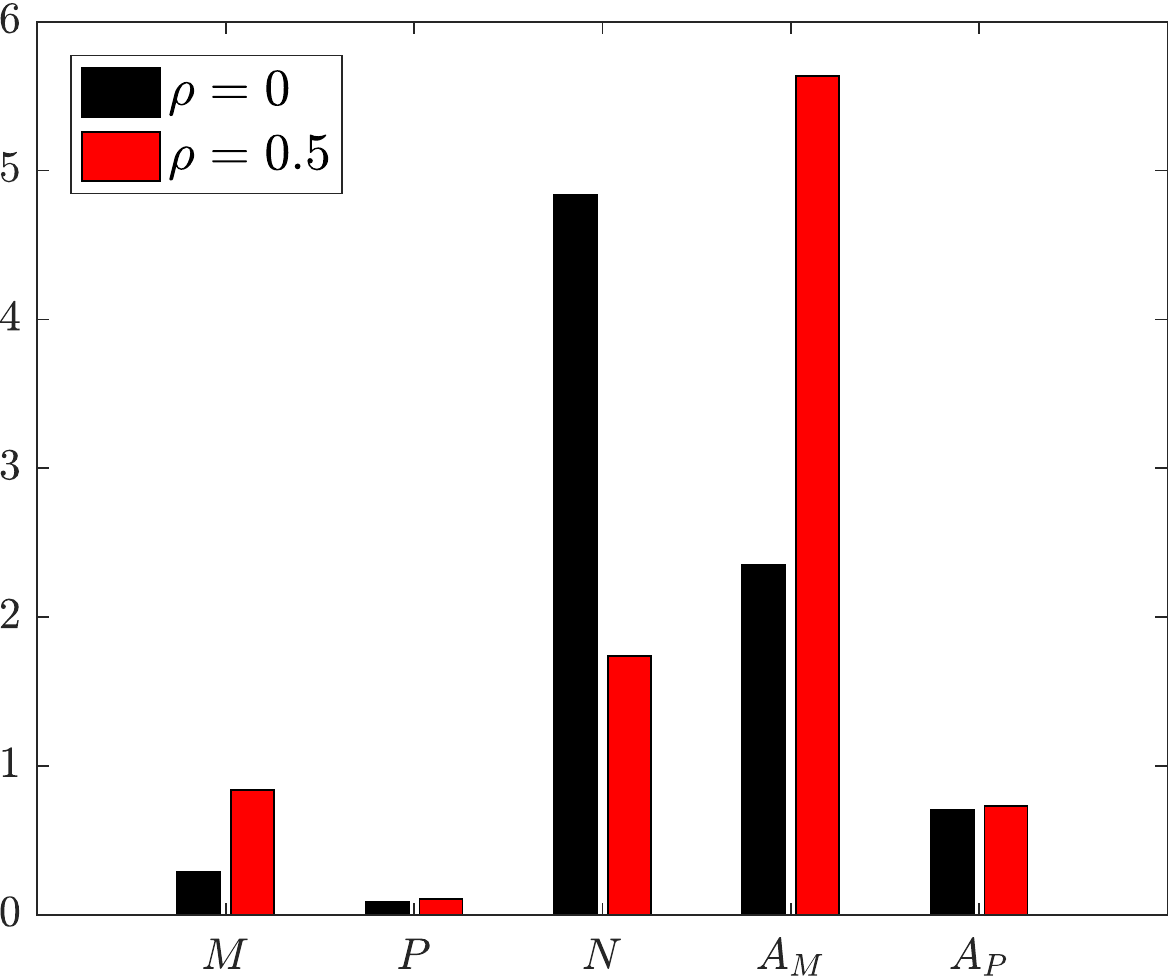}
		\caption{} \label{BaseCase_ODEs}
	\end{subfigure}
	\begin{subfigure}[b]{0.49\textwidth}
		\centering
		\includegraphics[height=5.5cm]{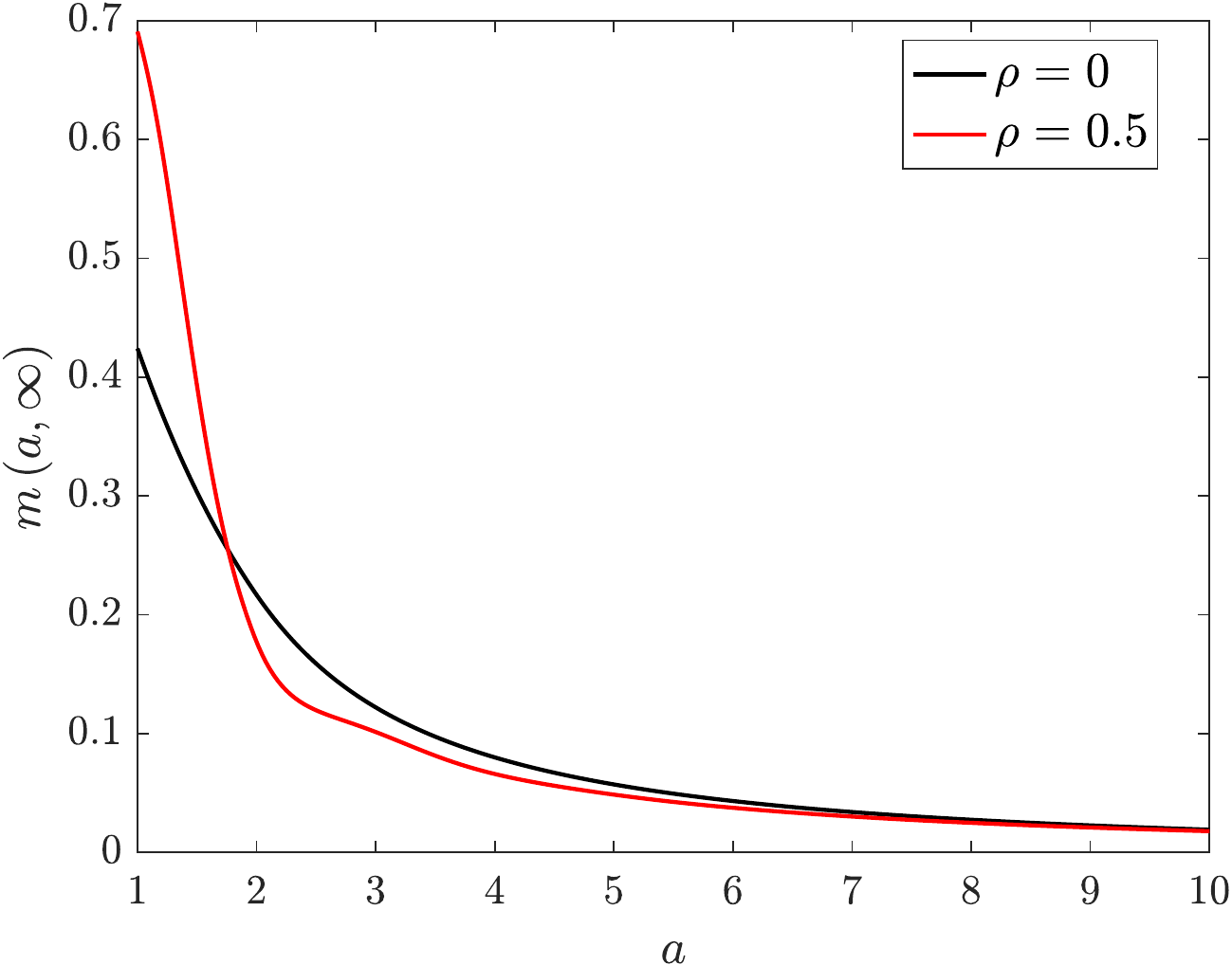}
		\caption{} \label{BaseCase_m}
	\end{subfigure}
	\caption{Steady state solutions for (a) the ODE variables and (b) the live cell distribution $m\left(a,t\right)$ from lipid-independent simulations both without macrophage proliferation ($\rho = 0$; black bars/lines) and with macrophage proliferation ($\rho = 0.5$; red bars/lines). Note that the apoptotic cell distributions $p\left(a,\infty\right)$ are effectively identical to the corresponding $m\left(a,\infty\right)$ plots in each case.} \label{BaseCase}
\end{figure}

Intra-plaque macrophage proliferation has only recently been established as an important contributor to plaque progression \citep{Robb13}. The extent of this proliferation is not well characterised but it is understood to vary over the lifetime of a plaque \citep{Lhot16}. Given this uncertainty, we shall perform numerical investigations of lipid-dependent cell behaviour both in the absence (Section \ref{ssNoProlif}) and the presence (Section \ref{ssProlif}) of macrophage proliferation. To provide a reference point for our lipid-dependent simulations, we first generate base case results where macrophage behaviour is independent of internalised lipid (all $g_\diamond\left(a\right) = 1$). Steady-state results for these cases, which use only the parameter values in Table \ref{LIparams}, are shown in Figure \ref{BaseCase}. For an in-depth understanding of these results, interested readers are referred to the simulations and analysis in \citet{Ford19a} and \citet{Cham22}. Here, we provide only a brief summary of some key features of the results.

In the absence of proliferation ($\rho = 0$), the model elicits a relatively mild immune response and forms a moderately sized necrotic core ($M \approx 0.29$, $N \approx 4.84$; Figure \ref{BaseCase_ODEs} black bars). This reflects our conservative model parameterisation, since we wish to allow for stronger immune responses in lipid-dependent cases where poorer outcomes are anticipated (e.g.\ when apoptosis increases with $a$, or when emigration decreases with $a$). In the case with proliferation ($\rho = 0.5$), we observe an almost 3-fold increase in the live cell population ($M \approx 0.84$; Figure \ref{BaseCase_ODEs} red bar). This is partly due to proliferation itself, but also due to enhanced recruitment courtesy of the substantial increase in $A_M$ (5.64 vs.\ 2.35). The increased cell population leads to a reduction in both the necrotic core size ($N \approx 1.74$) and the average lipid per cell (${\bar{A}}_M = {\bar{A}}_P \approx 6.7$, down from 8.1). The reduction in ${\bar{A}}_M$ due to proliferation can be seen qualitatively by comparing the steady state $m\left(a,t\right)$ distributions in Figure \ref{BaseCase_m}. The case with proliferation (red line) has a considerably larger proportion of cells with very small lipid loads.

Overall, it is prudent to note that we have chosen a lipid-independent parameterisation where no single macrophage behaviour dominates plaque formation. Given that these lipid-independent parameters are fixed for the entirety of the study, we argue that avoiding a single dominant cell behaviour is the most appropriate way to gain a general appreciation of the impact of lipid-dependence on plaque fate and dynamics.

\subsection{Lipid-Dependent Simulations Without Proliferation} \label{ssNoProlif}
In this section, we neglect macrophage proliferation and investigate, in turn, the role of lipid-dependence in macrophage apoptosis and emigration.    

\subsubsection{Apoptosis Only} \label{sssApo}
\begin{figure}
	\centering
	\begin{subfigure}[b]{0.49\textwidth}
		\centering
		\includegraphics[height=5.5cm]{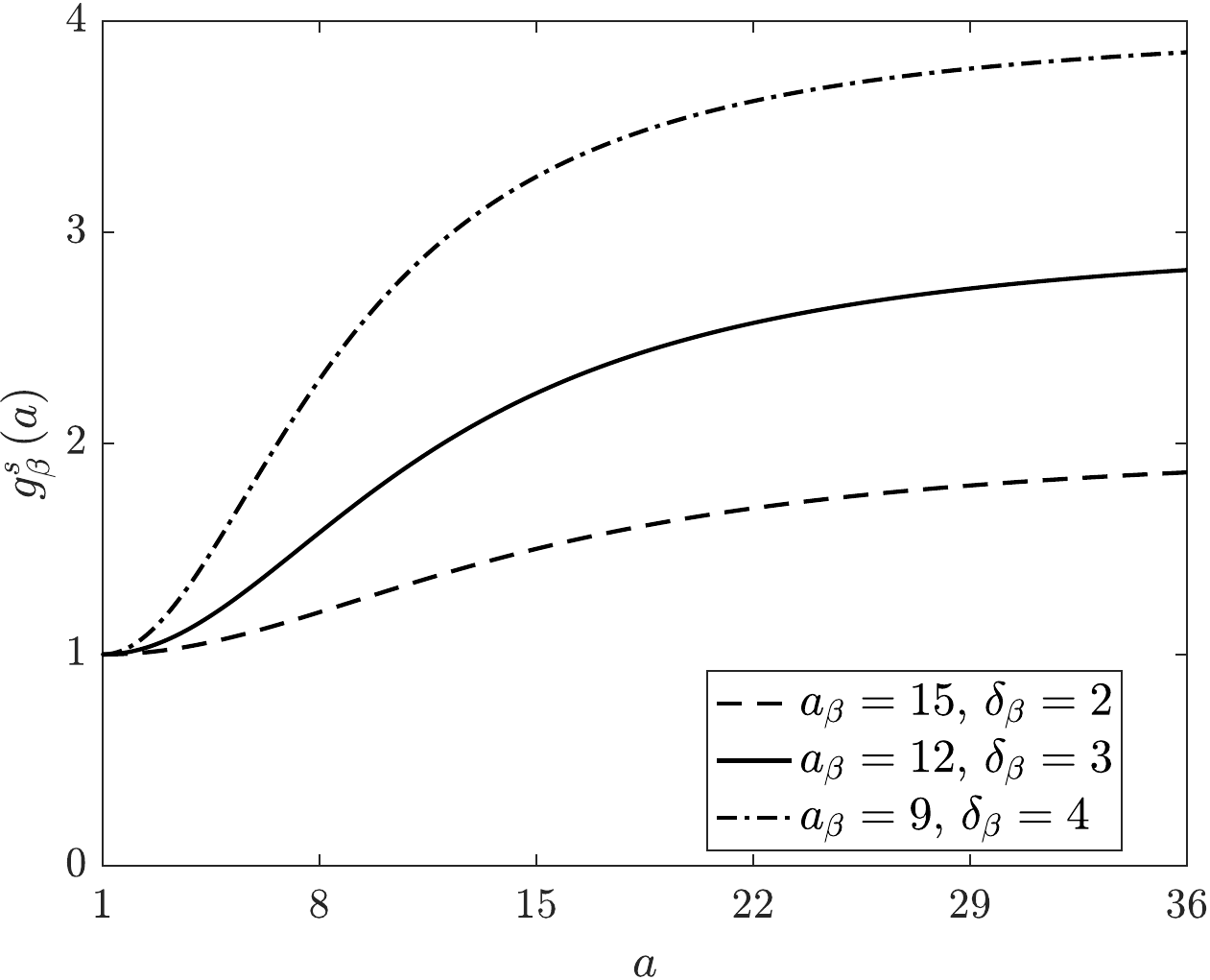}
		\caption{} \label{g_beta_s_unscaled}
	\end{subfigure}
	\begin{subfigure}[b]{0.49\textwidth}
		\centering
		\includegraphics[height=5.5cm]{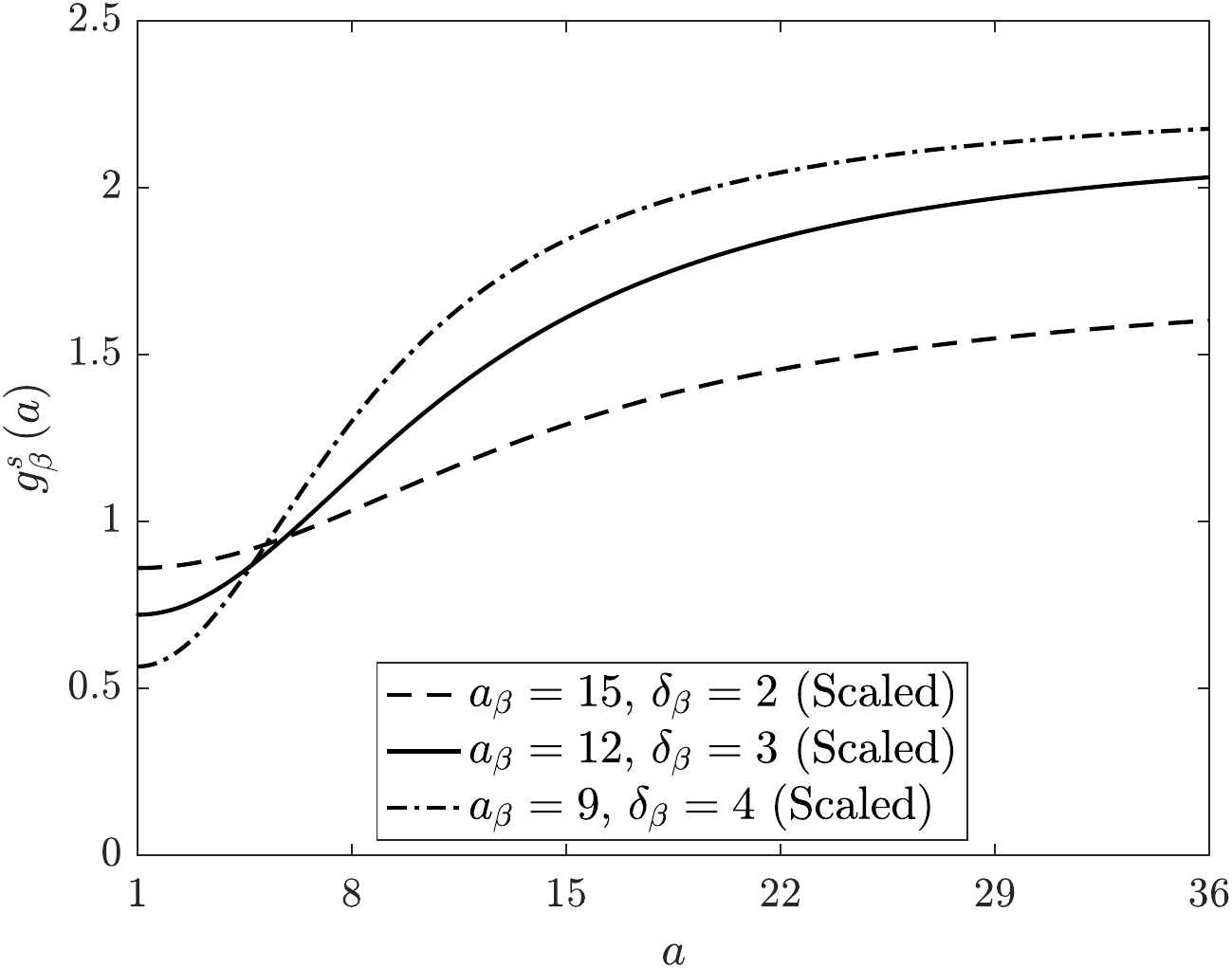}
		\caption{} \label{g_beta_s_scaled}
	\end{subfigure}
	\caption{(a) Unscaled and (b) scaled rate modulating functions for macrophage apoptosis $g_\beta\left(a\right) = g_\beta^s\left(a\right)$. Plots correspond to equation (\ref{dimless_g_mono}) with parameter values $n_\beta = 2$ and $a_\beta = 15$, $\delta_\beta = 2$ (dashed lines), $a_\beta = 12$, $\delta_\beta = 3$ (solid lines) or $a_\beta = 9$, $\delta_\beta = 4$ (dot-dashed lines). Scaling values for the plots in (b) are 0.86, 0.72 and 0.565, respectively.} \label{g_beta_s}
\end{figure}

To investigate lipid-dependent apoptosis, we set $g_\gamma\left(a\right) = 1$ and $g_\beta\left(a\right) = g_\beta^s\left(a\right)$. We consider three different unscaled forms for $g_\beta^s$, each of which has $\delta_\beta > 1$ and $n_\beta = 2$ (Figure \ref{g_beta_s_unscaled}). Each function reflects an assumption that the likelihood of macrophage apoptosis increases with increasing ingested lipid content \citep{Taba02, Feng03}. By varying the values of both $a_\beta$ and $\delta_\beta$, we investigate the impact of a mild ($a_\beta = 15$, $\delta_\beta = 2$), moderate ($a_\beta = 12$, $\delta_\beta = 3$) or severe ($a_\beta = 9$, $\delta_\beta = 4$) increase in the apoptosis rate (beyond the reference value) with increasing $a$. (Note that by changing the parameter values in this way, we anticipate that \emph{all} scenarios with $a_\beta \in \left[9, 15\right]$ and $\delta_\beta \in \left[2, 4\right]$ will have solutions that lie within the bounds of those reported below.)

\begin{figure}
	\centering		
	\includegraphics[height=15.3cm]{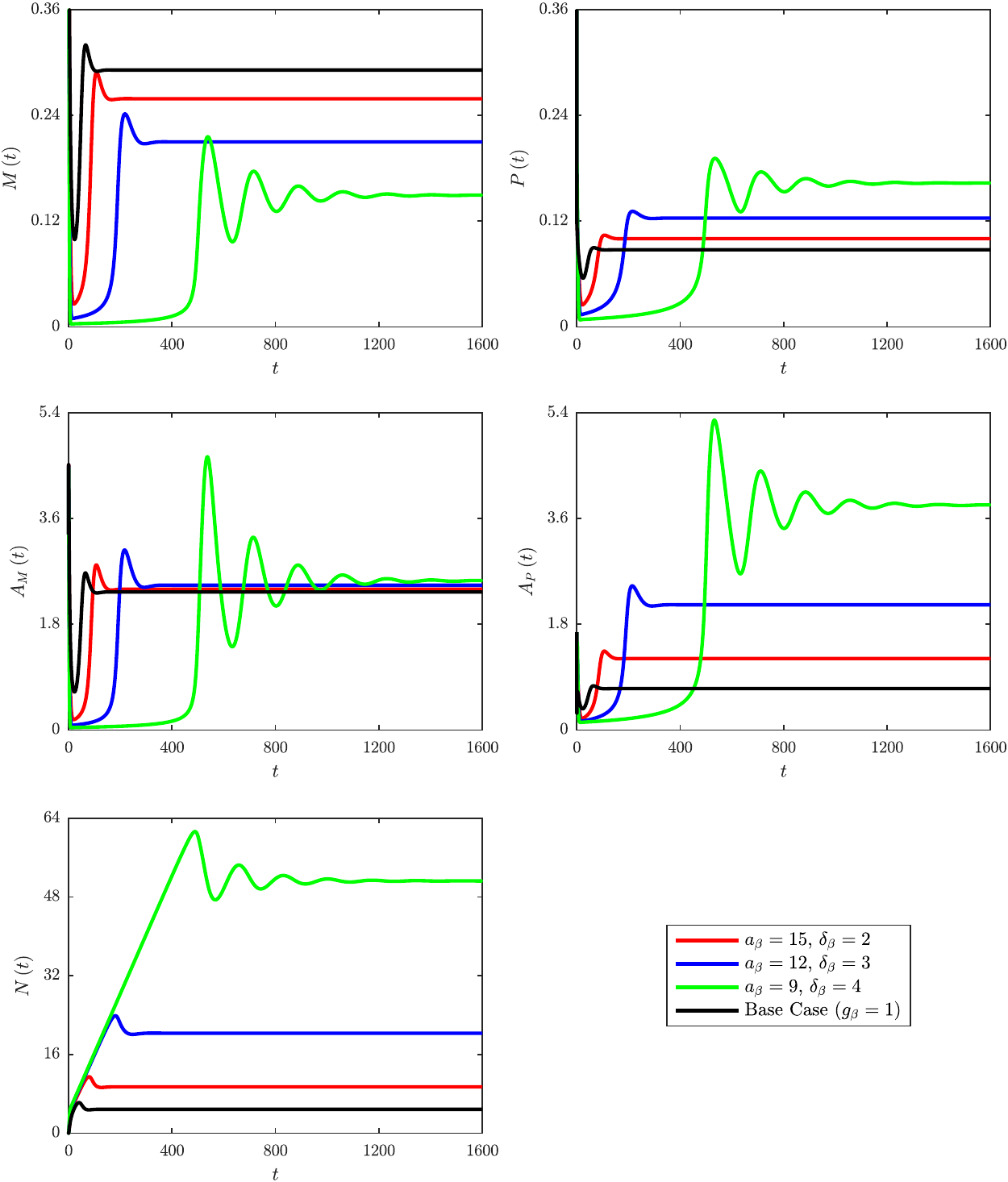}
	\caption{Time dependent solutions of the ODE variables $M\left(t\right)$, $P\left(t\right)$, $A_M\left(t\right)$, $A_P\left(t\right)$ and $N\left(t\right)$ for the lipid-independent base case ($g_\beta = 1$; black lines) and for three cases with unscaled lipid-dependent apoptosis ($g_\beta = g_\beta^s$). Lipid-dependent cases have parameter values $n_\beta = 2$ and $a_\beta = 15$, $\delta_\beta = 2$ (red lines), $a_\beta = 12$, $\delta_\beta = 3$ (blue lines) or $a_\beta = 9$, $\delta_\beta = 4$ (green lines).} \label{ODEs_time_apo}
\end{figure}

\begin{figure}
	\centering		
	\includegraphics[height=7cm]{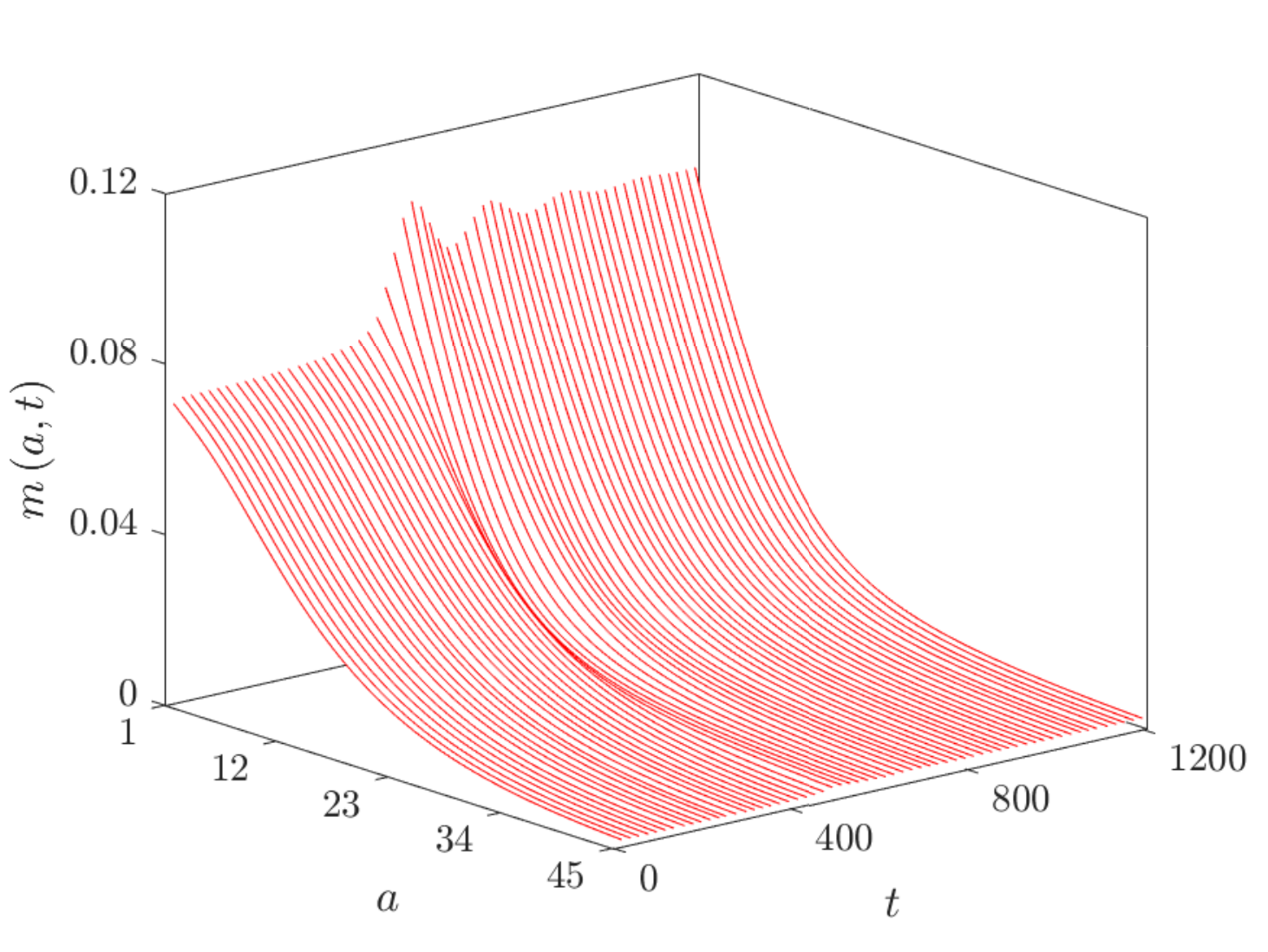}
	\caption{Surface plot showing the temporal evolution of $m\left(a,t\right)$ for $1 \leqslant a \leqslant 45$ and $0 \leqslant t \leqslant 1200$ in a simulation with unscaled lipid-dependent apoptosis ($g_\beta = g_\beta^s$ with $n_\beta = 2$,  $a_\beta = 9$ and $\delta_\beta = 4$). Note the onset of temporal oscillations in $m\left(a,t\right)$ near $a=1$ at $t \approx 400$. These oscillations reflect repeated bursts of macrophage recruitment that gradually diminish in intensity.} \label{m_surf}
\end{figure}

Time dependent solutions of the ODE variables for these cases are compared with those for the base case simulation ($g_\beta\left(a\right) = 1$) in Figure \ref{ODEs_time_apo}. An immediate observation is the interesting behaviour of the case with $a_\beta = 9$ and $\delta_\beta = 4$. While the other cases all display similar dynamics, this case elicits oscillations that eventually decay towards steady state. Noting that this is the case with the highest net apoptosis rate (dimensionless value $G_\beta\left(t\right) = 2.367$ at steady state; see Table \ref{mono_params}), a likely explanation for the oscillations is as follows. Initially, live macrophage numbers $M\left(t\right)$ drop to a very low level because cells that ingest even moderate quantities of lipid quickly die to become apoptotic macrophages $P\left(t\right)$. The associated conversion of live cell lipid $A_M\left(t\right)$ to apoptotic cell lipid $A_P\left(t\right)$ reduces the rate of macrophage recruitment. Low cell numbers and sustained apoptotic lipid generation lead to rapid accumulation of necrotic lipid $N\left(t\right)$. Eventually, however, the necrotic lipid pool becomes so vast that the live macrophages begin to ingest lipid at a rate that outstrips the death rate. This allows $A_M\left(t\right)$ to rise, which stimulates macrophage recruitment and enhances lipid ingestion to shrink the necrotic core. Of course, as more live cells now attain higher quantities of ingested lipid, their apoptosis rates increase further and $M\left(t\right)$ drops once again. A new period of growth in $N\left(t\right)$ is then initiated until a further (smaller) wave of macrophage recruitment is triggered. This cycle repeats until the magnitude of the oscillations in each variable finally tend to zero. Figure \ref{m_surf} presents a corresponding surface plot of the live macrophage distribution $m\left(a,t\right)$ in this case. The solution shows temporal oscillations near $a = 1$, which are associated with the repeated waves of macrophage recruitment. 

\begin{figure}
	\centering
	\begin{subfigure}[b]{\textwidth}
		\centering	
		\includegraphics[height=6cm]{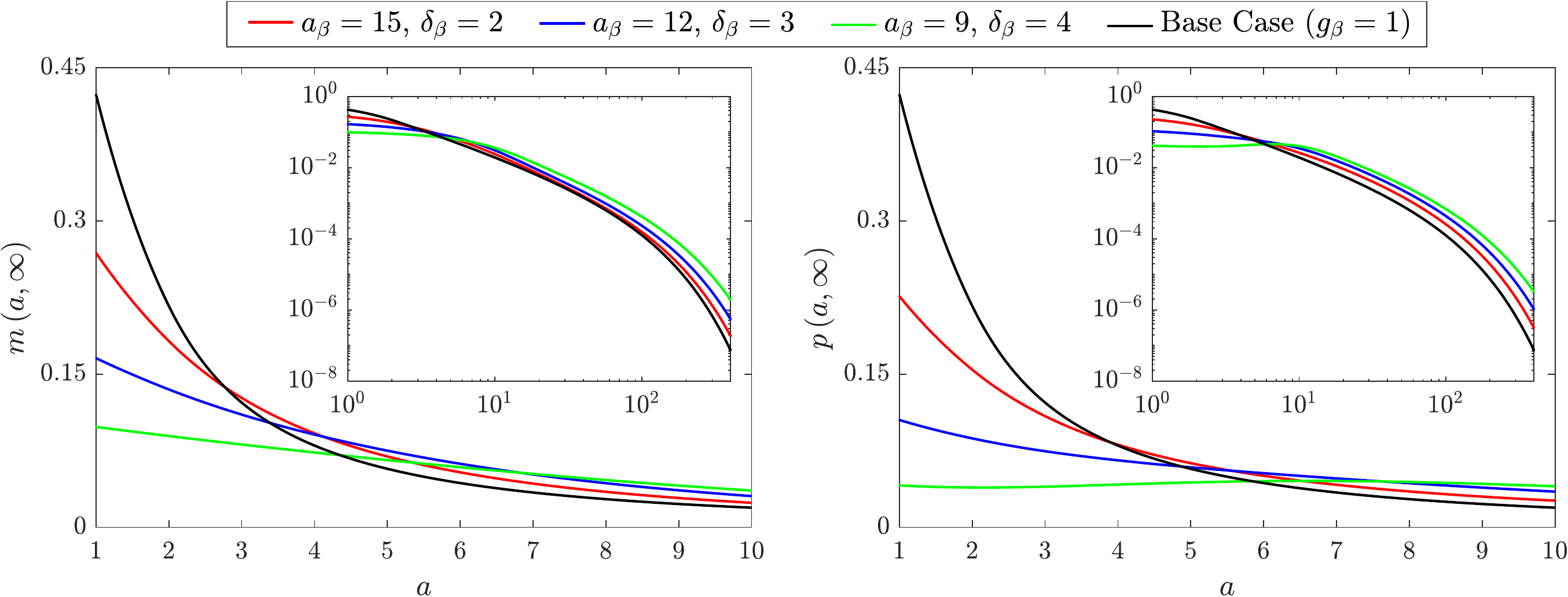}
		\caption{} \label{unsc_m_p_SS_apo}
	\end{subfigure}
	\par\medskip
	\begin{subfigure}[b]{\textwidth}
		\centering	
		\includegraphics[height=6cm]{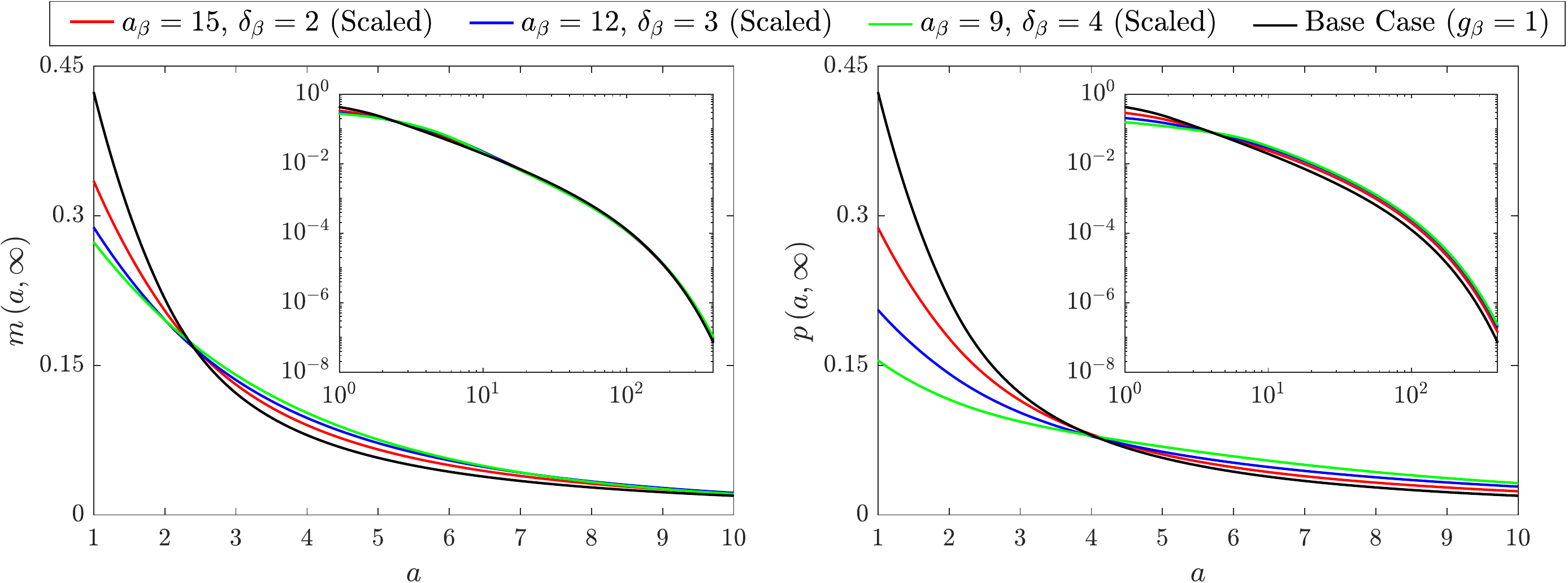}
		\caption{} \label{sc_m_p_SS_apo}
	\end{subfigure}
	\caption{Steady state $m\left(a,t\right)$ distributions (left panels) and $p\left(a,t\right)$ distributions (right panels) for the lipid-independent base case ($g_\beta = 1$; black lines) and for three cases with (a) unscaled or (b) scaled lipid-dependent apoptosis ($g_\beta = g_\beta^s$). Lipid-dependent cases have parameter values $n_\beta = 2$ and $a_\beta = 15$, $\delta_\beta = 2$ (red lines), $a_\beta = 12$, $\delta_\beta = 3$ (blue lines) or $a_\beta = 9$, $\delta_\beta = 4$ (green lines). Primary plots show the results on the interval $a \in \left[1,10\right]$ and inset log-log plots show the results on the entire $a$ domain.} \label{m_p_SS_apo}
\end{figure}

Long-time solutions of the ODE variables for each simulation indicate that the severity of the increase in $g_\beta^s$ correlates with a decrease in $M$ and an increase in each of the other variables (ranging from a marginal rise in $A_M$ to a substantial rise in $N$). The severity of the change in $g_\beta^s$ also appears to be correlated to the time required to reach steady state, which increases from $t \approx 100$ in the base case to $t \approx 180$ ($a_\beta = 15$, $\delta_\beta = 2$), $t \approx 350$ ($a_\beta = 12$, $\delta_\beta = 3$) or $t \approx 2000$ ($a_\beta = 9$, $\delta_\beta = 4$). The steady state distributions of live macrophages $m\left(a,t\right)$ and apoptotic macrophages $p\left(a,t\right)$ for each simulation are presented in Figure \ref{unsc_m_p_SS_apo} (note, from equation (\ref{dimless_p_eqn}), that $m\left(a,t\right)$ and $p\left(a,t\right)$ differ at steady state only in the case of lipid-dependent apoptosis). For both $m$ and $p$, we see that increasing the severity of the change in $g_\beta^s$ skews the distributions towards larger lipid loads. This seems counter-intuitive (shouldn't the increase in apoptosis rate with increasing $a$ produce a \emph{smaller} proportion of cells with large lipid loads?), but it appears that the elevated lipid ingestion rate (due to the increase in $N$ and $A_P$) produces more cells with large lipid loads than are lost through apoptosis. 
At steady state, the $m$ and $p$ distributions are related by the expression $p\left(a,t\right) = \frac{g_\beta\left(a\right) m\left(a,t\right)}{G_\beta\left(t\right)}$ (see equation (\ref{dimless_p_eqn})). In the simulation with $a_\beta = 9$ and $\delta_\beta = 4$, we note that this gives a non-monotonic profile for $p\left(a,\infty\right)$ with a shallow peak near $a = 8$.

\begin{figure}
	\centering		
	\includegraphics[width=5.5cm]{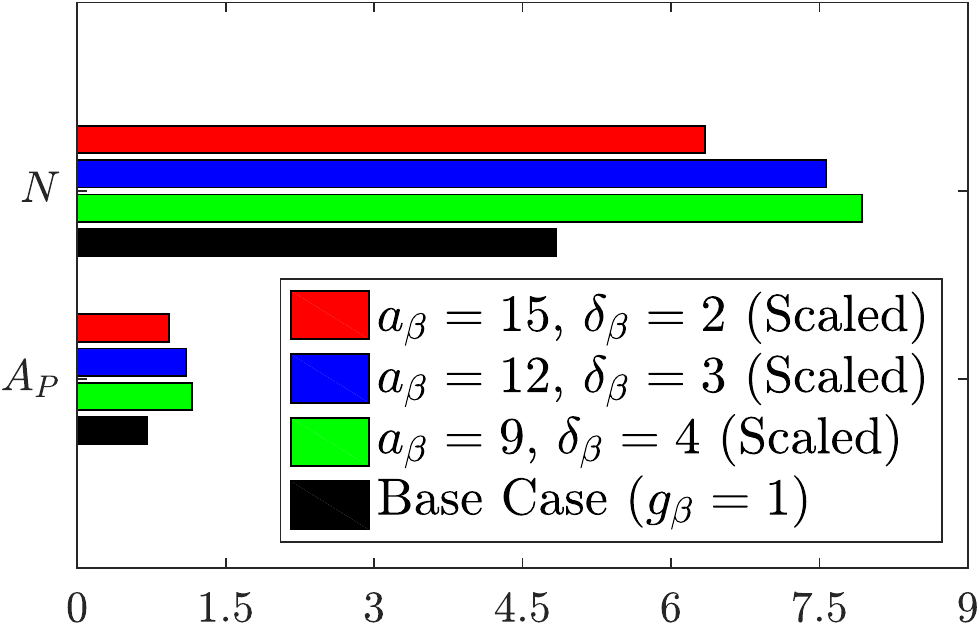}
	\caption{Steady state solutions of the ODE variables $A_P\left(t\right)$ and $N\left(t\right)$ for the lipid-independent base case ($g_\beta = 1$; black bars) and for three cases with scaled lipid-dependent apoptosis ($g_\beta = g_\beta^s$). Lipid-dependent cases have parameter values $n_\beta = 2$ and $a_\beta = 15$, $\delta_\beta = 2$ (red bars), $a_\beta = 12$, $\delta_\beta = 3$ (blue bars) or $a_\beta = 9$, $\delta_\beta = 4$ (green bars). Solutions for $M\left(t\right)$, $P\left(t\right)$ and $A_M\left(t\right)$ are omitted from the plot as their values are unchanged across cases.} \label{sc_ODEs_apo}
\end{figure}

When interpreting the above results, it is important to consider that the net apoptosis rates $G_\beta\left(t\right)$ vary considerably between the different simulations (from 1 in the base case to 2.367 at steady state in the case with $a_\beta = 9$ and $\delta_\beta = 4$). To correct for this difference, we repeat the lipid-dependent simulations with appropriately scaled functions $g_\beta^s\left(a\right)$ that give $G_\beta\left(t\right) \approx 1$ at steady state (Figure \ref{g_beta_s_scaled}). For these simulations, we find that the steady state values of $M$, $P$ and $A_M$ effectively remain fixed, and only $A_P$ and $N$ vary across the different cases. (Note that the oscillatory dynamics observed above do not occur in this case, and, while the trend of increasing time to steady state remains, it is much less pronounced than before). Figure \ref{sc_ODEs_apo} compares the steady state solutions for $A_P$ and $N$ from the base case with those from the scaled lipid-dependent simulations. While the absolute variation in these quantities is much smaller than in the unscaled scenario, the trend of increasing $A_P$ and $N$ with increasing severity of change in $g_\beta^s$ remains intact.

Plots of the corresponding steady state PDE solutions are shown in Figure \ref{sc_m_p_SS_apo}. These demonstrate that the trends in the $m\left(a,\infty\right)$ and $p\left(a,\infty\right)$ distributions with scaled $g_\beta^s\left(a\right)$ are largely conserved from the unscaled cases (Figure \ref{unsc_m_p_SS_apo}), albeit the differences relative to the base case results are somewhat less pronounced. Note that each of the $p\left(a,\infty\right)$ distributions presented in Figure \ref{sc_m_p_SS_apo} satisfies the relationship $p\left(a,\infty\right) = g_\beta\left(a\right) m\left(a,\infty\right)$. Accordingly, the $p\left(a,\infty\right)$ distribution in each case corresponds exactly to the distribution of apoptosis events at steady state. Moreover, equation (\ref{dimless_APbar_eqn}) shows that, in each case, the steady state average lipid per apoptotic cell ${\bar{A}}_P = \frac{A_P}{P}$ is given by the steady state $G_{\beta a}$ value. Given that the steady state $P$ values are essentially identical across all cases, this indicates that the steady state $A_P$ values are exactly proportional to the steady state $G_{\beta a}$ values. The trend observed in Figure \ref{sc_ODEs_apo} for increasing steady state $A_P$ (and $N$) with increasing severity of change in $g_\beta^s$ is therefore related to a trend for increasing steady state $G_{\beta a}$. The precise reason for this trend of increasing steady state $G_{\beta a}$ is, however, not entirely clear. It may be correlated to the increase in the (maximum) steepness of $g_\beta^s$, or it may reflect the increase in the limiting value of $g_\beta^s$ (c.f.\ Figure \ref{g_beta_s_scaled}).

\begin{figure}
	\centering
	\includegraphics[height=6cm]{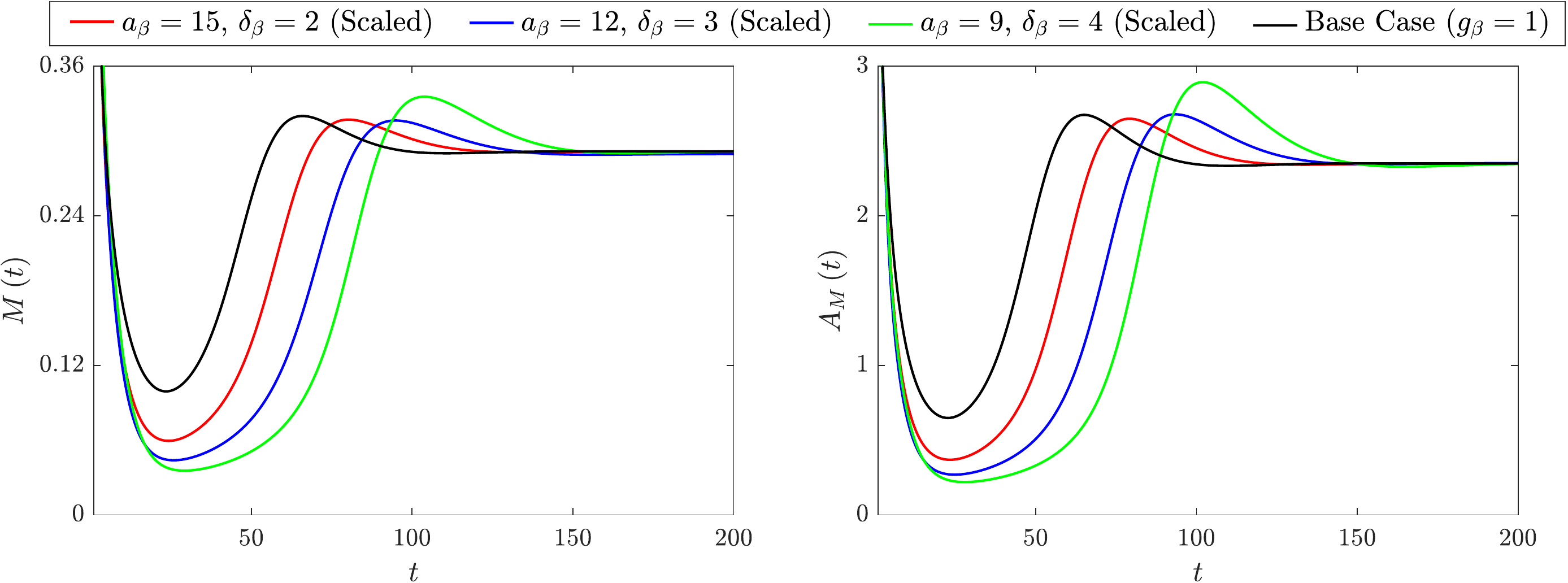}
	\caption{Time dependent solutions of the ODE variables $M\left(t\right)$ and $A_M\left(t\right)$ for the lipid-independent base case ($g_\beta = 1$; black lines) and for three cases with scaled lipid-dependent apoptosis ($g_\beta = g_\beta^s$). Lipid-dependent cases have parameter values $n_\beta = 2$ and $a_\beta = 15$, $\delta_\beta = 2$ (red lines), $a_\beta = 12$, $\delta_\beta = 3$ (blue lines) or $a_\beta = 9$, $\delta_\beta = 4$ (green lines). Results show that, while all simulations reach similar steady state $M$ and $A_M$ values, the paths taken to get there vary considerably.} \label{M_AM_time_apo_scaled}
\end{figure}

The fact that the steady state $M$ and $A_M$ values are essentially identical across these scaled simulations has interesting consequences when viewed through the lens of equations (\ref{dimless_L_eqn}) and (\ref{dimless_AMbar_eqn}). For the scenarios simulated here, equation (\ref{dimless_L_eqn}) reduces to:
\begin{equation}
\frac{dL\left(t\right)}{dt} = \psi F\left(t\right) + \psi\lambda - \left(\psi - 1\right) A_M\left(t\right). \label{dimless_L_apo_only}
\end{equation}
The total system lipid $L\left(t\right)$ has explicit dependence only on $M\left(t\right)$ and $A_M\left(t\right)$. Interestingly, however, our simulation results indicate that the steady state $M$ and $A_M$ values do not uniquely define the steady state $L$ (i.e.\ steady state $M$ and $A_M$ remain essentially fixed across simulations, but steady state $A_P$ and $N$ do not, leading to overall change in $L = A_M + A_P + N$). This absence of uniqueness at steady state demonstrates that, in terms of total system lipid, the model plaque can carry the weight of its history. That is, the observed increase in steady state $L$ with increasing severity of change in $g_\beta^s$ can be explained by the historical accumulation of lipid due to periods of reduced lipid removal by emigration (i.e.\ reduced $A_M\left(t\right)$) or increased lipid addition by recruitment (i.e.\ increased $F\left(t\right)$ or, equivalently, increased $A_M\left(t\right) - M\left(t\right)$). The time courses of the relevant $M\left(t\right)$ and $A_M\left(t\right)$ solutions (Figure \ref{M_AM_time_apo_scaled}) suggest that reduced lipid removal by emigration is the predominant mechanism in this case.

Unlike $L\left(t\right)$, the steady state solution for ${\bar{A}}_M\left(t\right)$ remains fixed across all four simulations. In each of these cases, the steady state solution to equation (\ref{dimless_AMbar_eqn}) can be expressed as:
\begin{equation}
\frac{\psi \lambda}{M} + \theta N + \eta A_P - \frac{\psi F}{M} \Big[{\bar{A}}_M - 1\,\Big] + {\bar{A}}_M - G_{\beta a} = 0, \label{dimless_AMbar_apo_only}
\end{equation}
where all time-dependent variables take their steady state values. As steady state $M$ and ${\bar{A}}_M$ have fixed values for all simulations, only terms two, three and six on the left-hand side of (\ref{dimless_AMbar_apo_only}) change in each case. Thus, steady state ${\bar{A}}_M$ is unchanged across all scaled simulations because any reduction in average lipid per cell due to lipid-dependent apoptosis (term six) is exactly compensated by an increase in necrotic and efferocytic lipid consumption (terms two and three, respectively). This finding supports our earlier interpretation of the unscaled simulation results, where we found that the contribution of lipid-dependent apoptosis to reducing ${\bar{A}}_M\left(t\right)$ was outstripped by other factors.

\subsubsection{Emigration Only} \label{sssEmi}
To study lipid-dependent emigration, we set $g_\beta\left(a\right) = 1$ and consider that $g_\gamma\left(a\right)$ takes either the monotonic form $g_\gamma^s\left(a\right)$ with $\delta_\gamma < 1$, or the non-monotonic form $g_\gamma^r\left(a\right)$. In the monotonic case, we assume that the rate at which macrophages leave the plaque decreases with increasing lipid load \citep{Gils12, Wans13, Chen19}. In the non-monotonic case, we retain this argument. However, we additionally assume a substantially reduced emigration rate for macrophages with little or no accumulated lipid. This assumption reflects a range of considerations, including the innate propensity for macrophages to pursue foreign bodies, and the fact that macrophages usually traverse the plaque before exiting to the media \citep{Llod04}. Either way, we anticipate that newly-recruited macrophages are unlikely to leave the plaque without first ingesting at least some lipid. Combined with the assumption of reduced emigration for macrophages with large lipid loads, this leads us to consider that the lipid-dependent emigration rate may peak at some intermediate lipid quantity. (Note that the argument regarding the scarcity of emigration of macrophages with small lipid loads may not hold if the lipid load can be reduced by proliferation. As such, we will only consider the function $g_\gamma^r\left(a\right)$ in the current section where we neglect macrophage proliferation.)    

\begin{figure}
	\centering
	\begin{subfigure}[b]{0.49\textwidth}
		\centering
		\includegraphics[height=5.5cm]{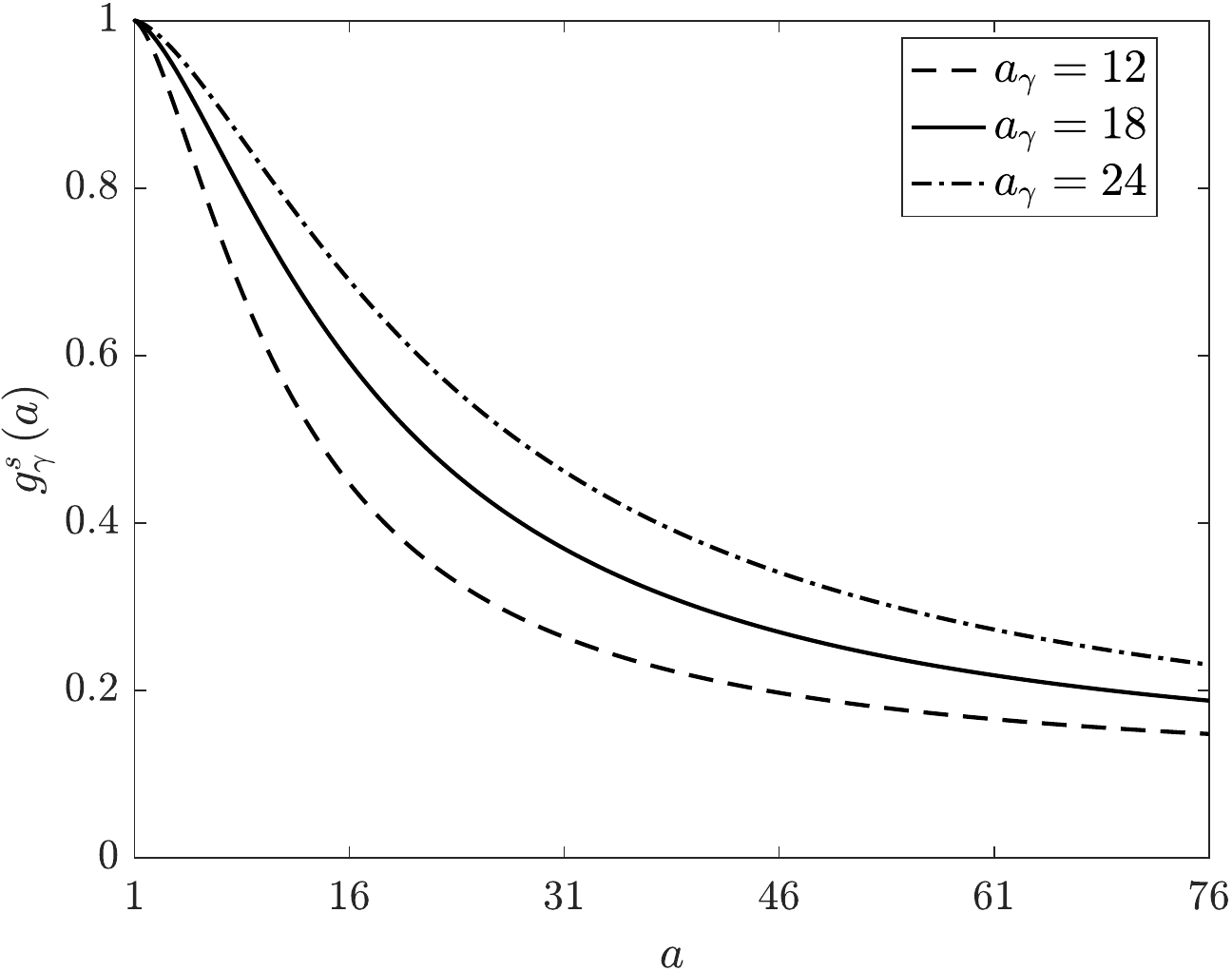}
		\caption{} \label{g_gamma_s_unscaled}
	\end{subfigure}
	\begin{subfigure}[b]{0.49\textwidth}
		\centering
		\includegraphics[height=5.5cm]{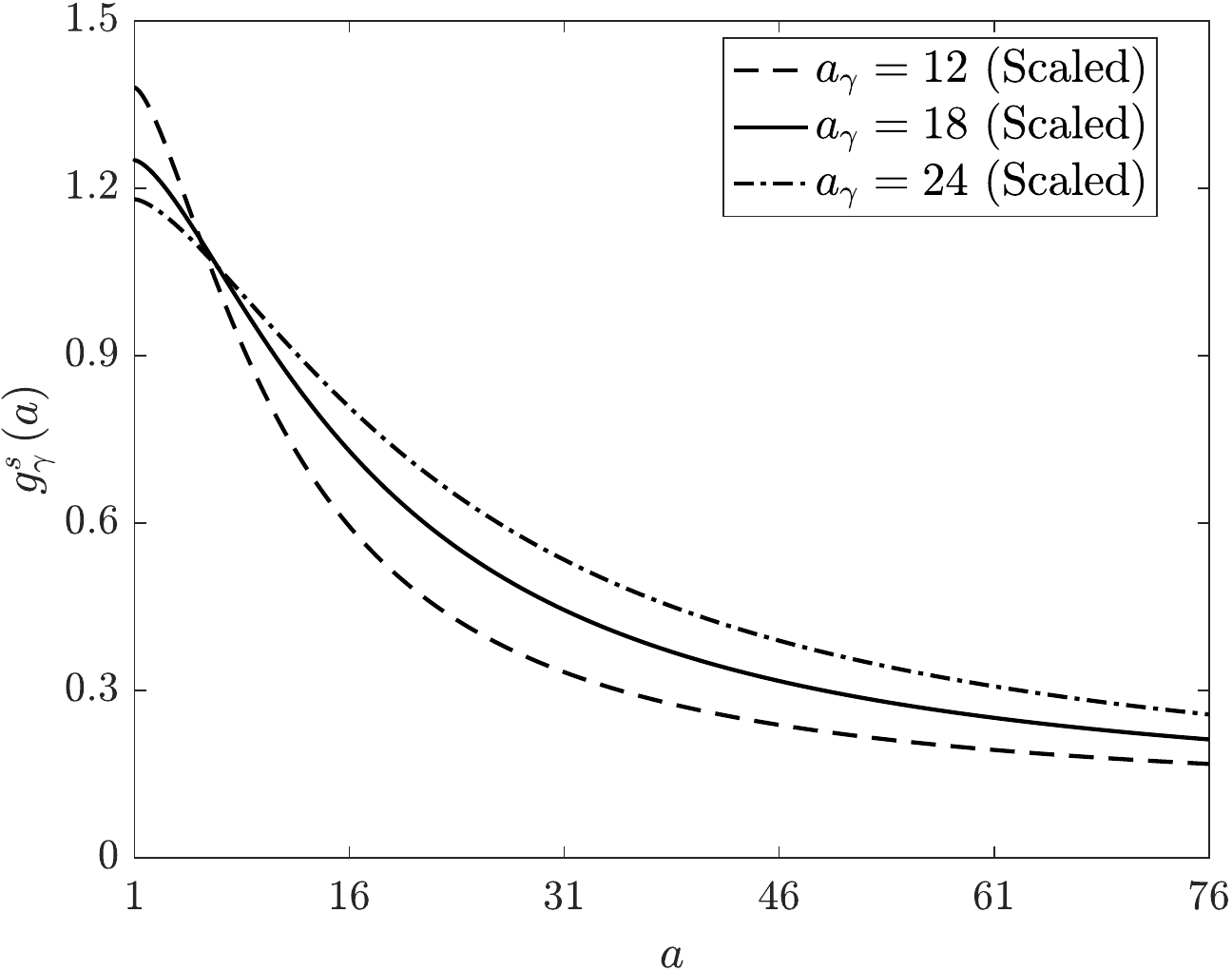}
		\caption{} \label{g_gamma_s_scaled}
	\end{subfigure}
	\caption{(a) Unscaled and (b) scaled rate modulating functions for macrophage emigration $g_\gamma\left(a\right) = g_\gamma^s\left(a\right)$. Plots correspond to equation (\ref{dimless_g_mono}) with parameter values $n_\gamma = 1.5$, $\delta_\gamma = 0.1$ and $a_\gamma = 12$ (dashed lines), $a_\gamma = 18$ (solid lines) or $a_\gamma = 24$ (dot-dashed lines). Scaling values for the plots in (b) are 1.38, 1.25 and 1.18, respectively. Note that, in the scaled cases, $\delta_\gamma$ is divided by the scaling value so that each $g_\gamma^s$ retains the limiting value of the unscaled functions.} \label{g_gamma_s}
\end{figure}

We first investigate monotonic lipid-dependence in macrophage emigration and consider three alternative unscaled forms for $g_\gamma^s\left(a\right)$ (Figure \ref{g_gamma_s_unscaled}). Each function has $\delta_\gamma = 0.1$ and $n_\gamma = 1.5$, while we vary $a_\gamma$ to study the impact of a gentle ($a_\gamma = 24$), moderate ($a_\gamma = 18$) or steep ($a_\gamma = 12$) initial reduction in the emigration rate with increasing $a$. Note that we deliberately choose a small $n_\gamma$ value and a non-zero $\delta_\gamma$ value to maintain feasibility in the numerical solution of the equations. When the functions $g_\gamma^s$ decline rapidly and/or tend towards zero as $a \to \infty$, a non-negligible proportion of cells can acquire extremely large lipid loads ($a \gg 1000$). For numerical accuracy, such cases need to be solved on extended domains at significant computational expense. However, as this increase in computational effort is unlikely to produce additional physical insight, we avoid such scenarios here. (While physically we may expect effectively zero emigration of macrophages with very large lipid loads, note that our parameterisation ensures cells remain fifty times more likely to die than to emigrate as $a \to \infty$).

\begin{figure}
	\centering
	\begin{subfigure}[b]{0.49\textwidth}
		\centering
		\includegraphics[height=6.05cm]{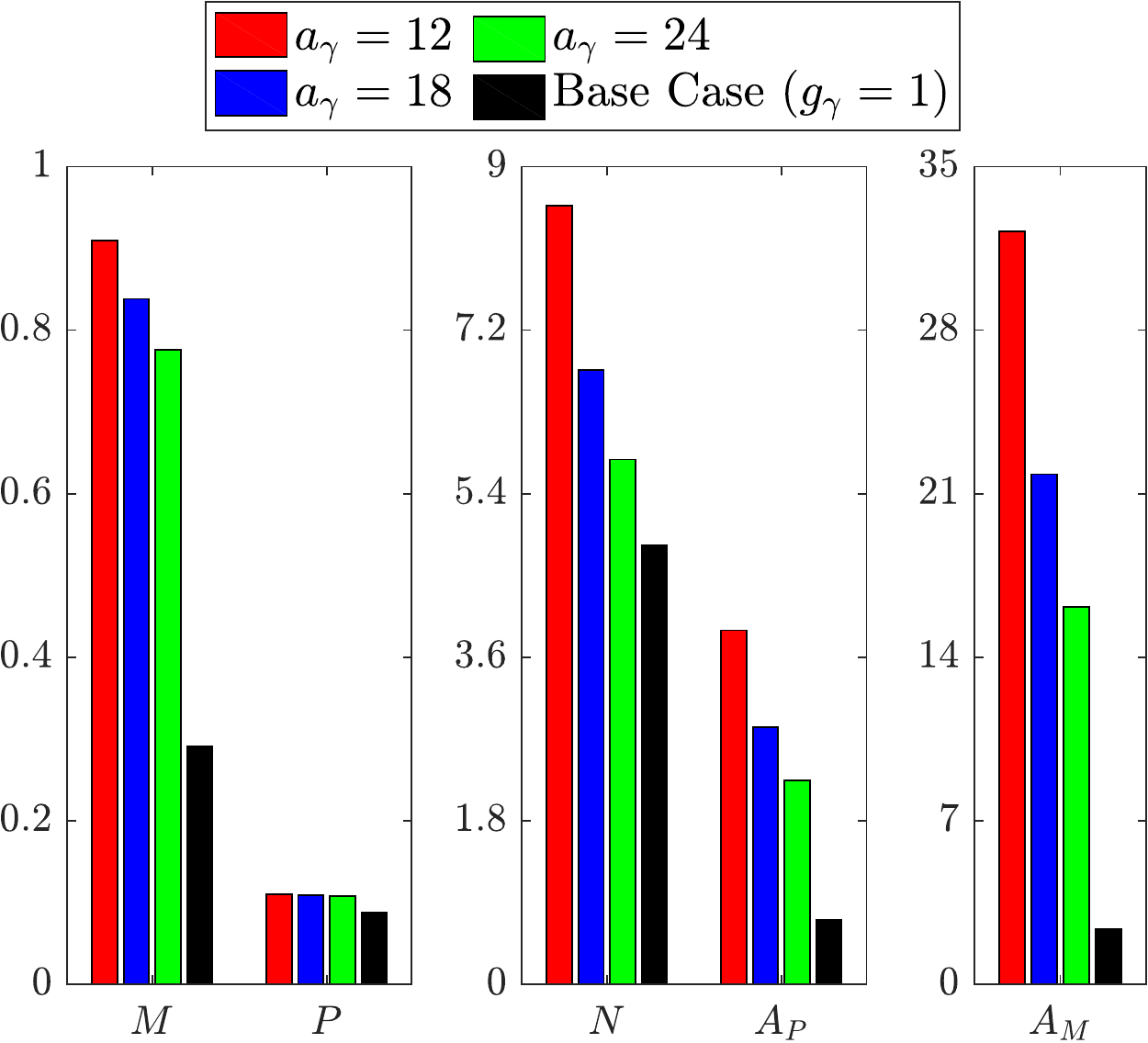}
		\caption{} \label{unsc_ODEs_emi_mono}
	\end{subfigure}
	\begin{subfigure}[b]{0.49\textwidth}
		\centering
		\includegraphics[height=6.05cm]{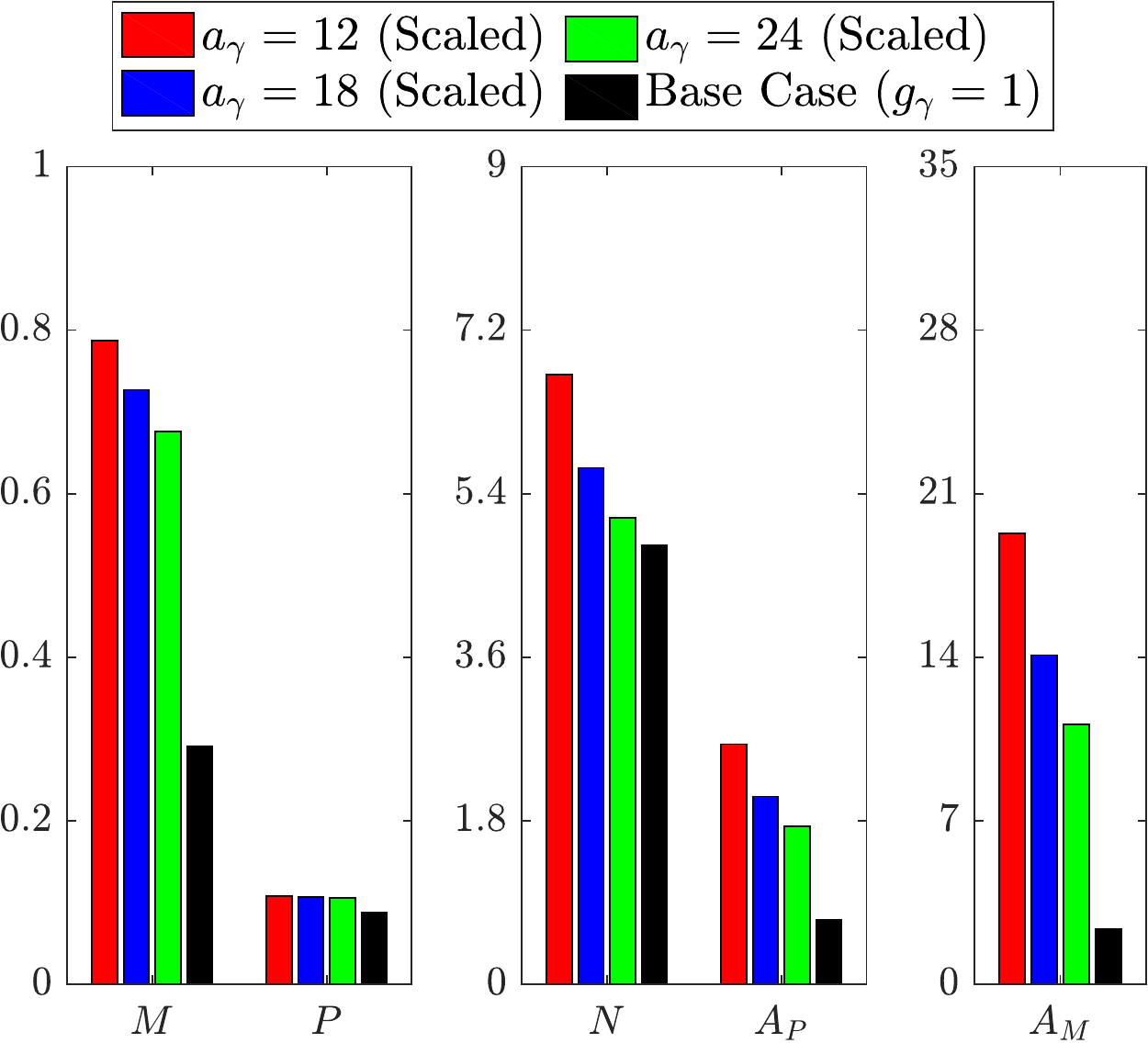}
		\caption{} \label{sc_ODEs_emi_mono}
	\end{subfigure}
	\par\medskip
	\begin{subfigure}[b]{0.49\textwidth}
		\centering
		\includegraphics[height=6.05cm]{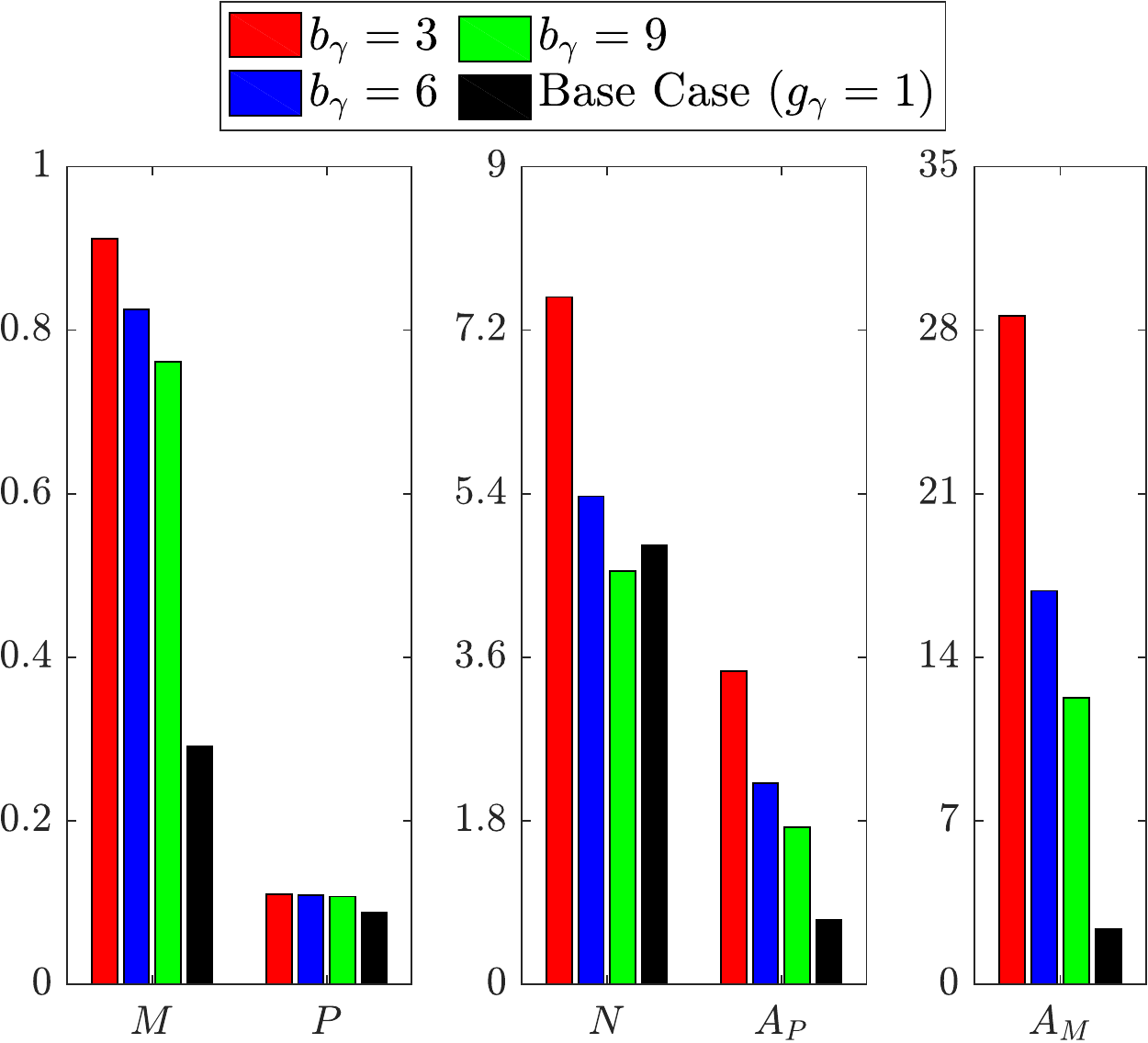}
		\caption{} \label{unsc_ODEs_emi_nonmono}
	\end{subfigure}
	\begin{subfigure}[b]{0.49\textwidth}
		\centering
		\includegraphics[height=6.05cm]{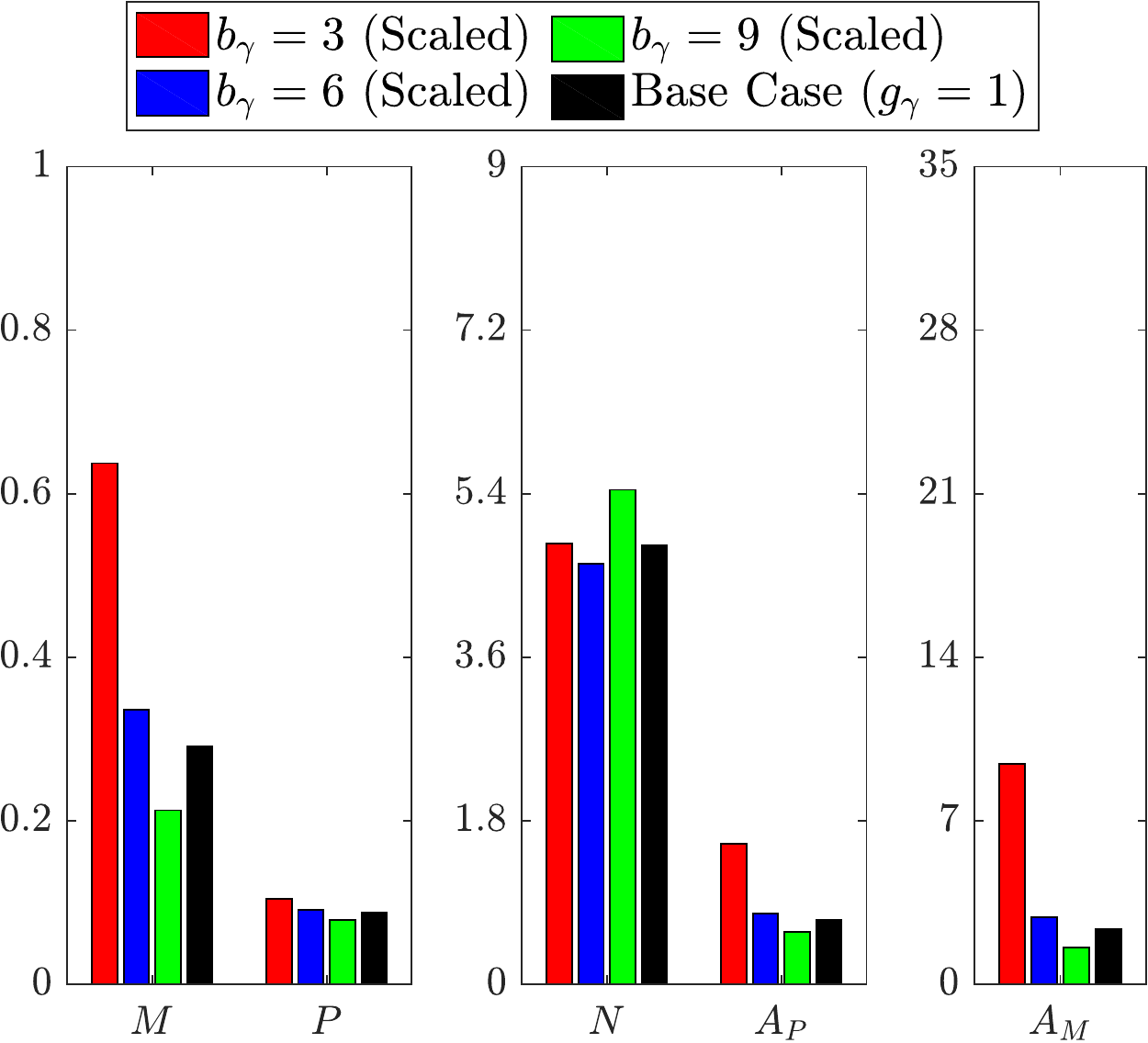}
		\caption{} \label{sc_ODEs_emi_nonmono}
	\end{subfigure}
	\caption{Steady state solutions of the ODE variables $M\left(t\right)$, $P\left(t\right)$, $N\left(t\right)$, $A_P\left(t\right)$ and $A_M\left(t\right)$ for the lipid-independent base case ($g_\gamma = 1$; black bars) and for several scenarios with lipid-dependent emigration. Lipid-dependent simulations use (a) unscaled or (b) scaled monotonic lipid-dependence ($g_\gamma = g_\gamma^s$) and (c) unscaled or (d) scaled non-monotonic lipid-dependence ($g_\gamma = g_\gamma^r$). Monotonic lipid-dependent cases have parameter values $n_\gamma = 1.5$, $\delta_\gamma = 0.1$ and $a_\gamma = 12$ (red bars), $a_\gamma = 18$ (blue bars) or $a_\gamma = 24$ (green bars). Non-monotonic lipid-dependent cases have parameter values $\epsilon_\gamma = 0.1$, $k_\gamma = 1$, $q_\gamma = 2$ and $b_\gamma = 3$ (red bars), $b_\gamma = 6$ (blue bars) or $b_\gamma = 9$ (green bars).} \label{ODEs_emi}
\end{figure}

\begin{figure}
	\centering
	\begin{subfigure}[b]{0.49\textwidth}
		\centering
		\includegraphics[height=6.5cm]{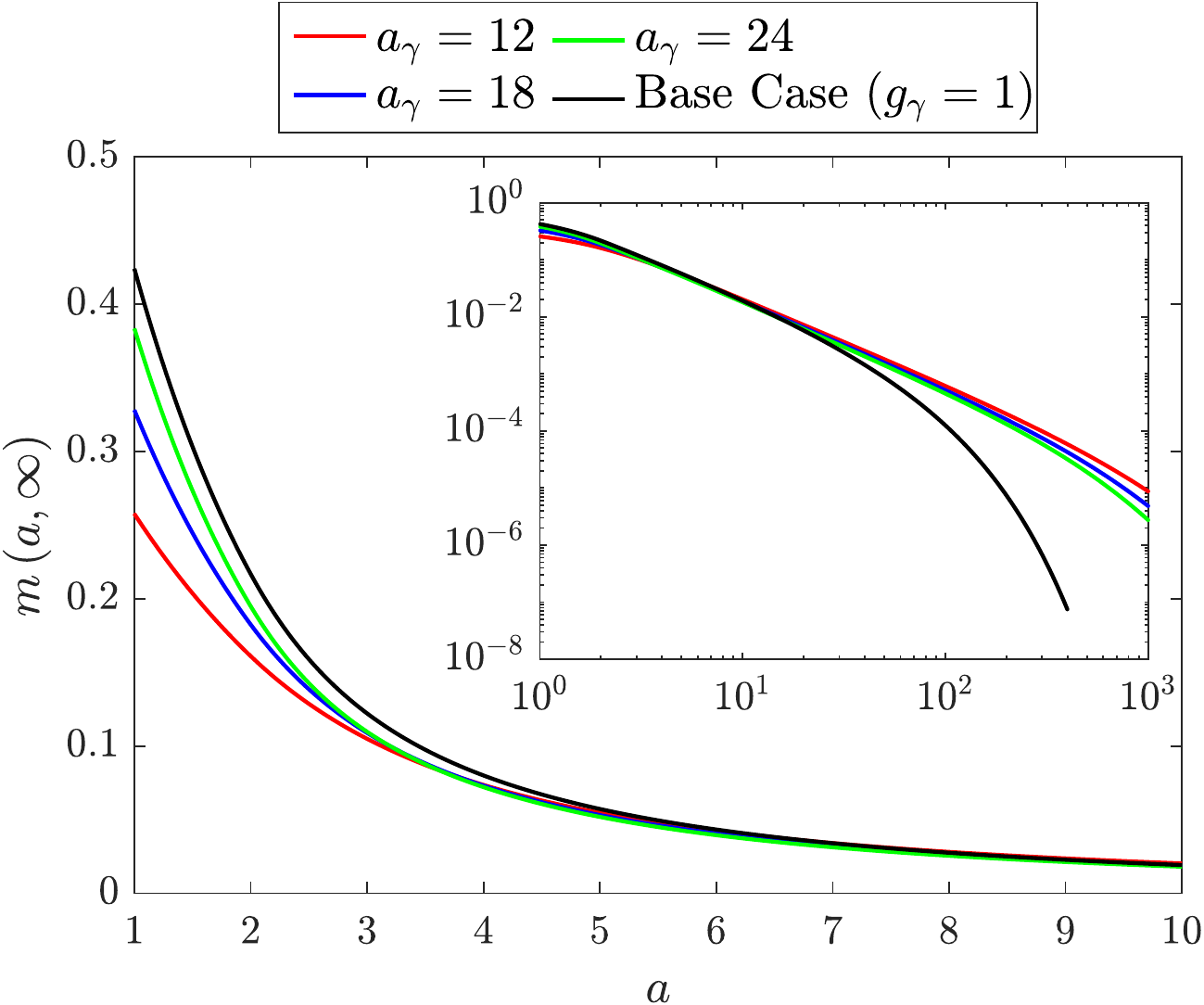}
		\caption{} \label{unsc_m_SS_emi_mono}
	\end{subfigure}
	\begin{subfigure}[b]{0.49\textwidth}
		\centering
		\includegraphics[height=6.5cm]{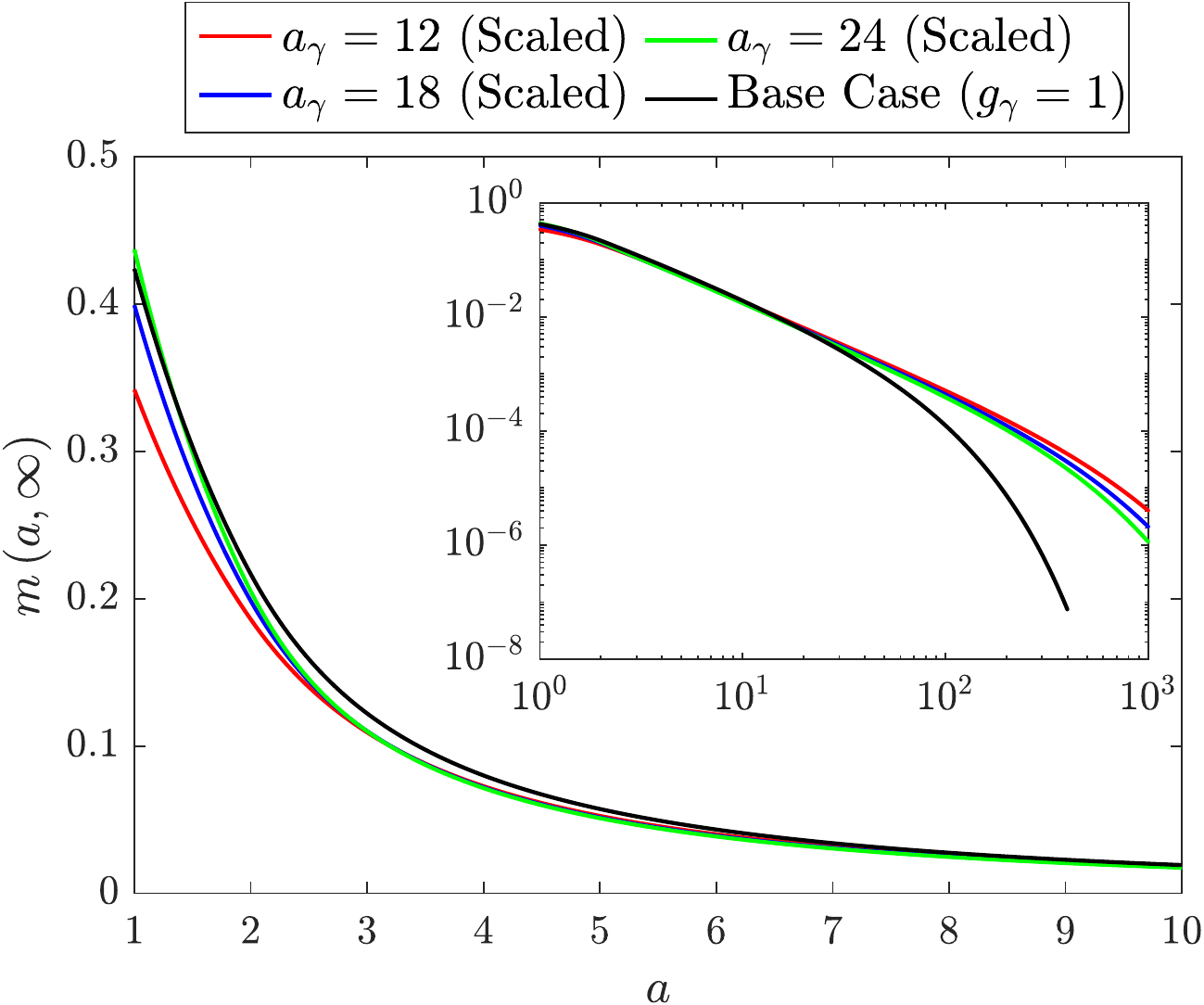}
		\caption{} \label{sc_m_SS_emi_mono}
	\end{subfigure}
	\par\medskip
	\begin{subfigure}[b]{0.49\textwidth}
		\centering
		\includegraphics[height=6.5cm]{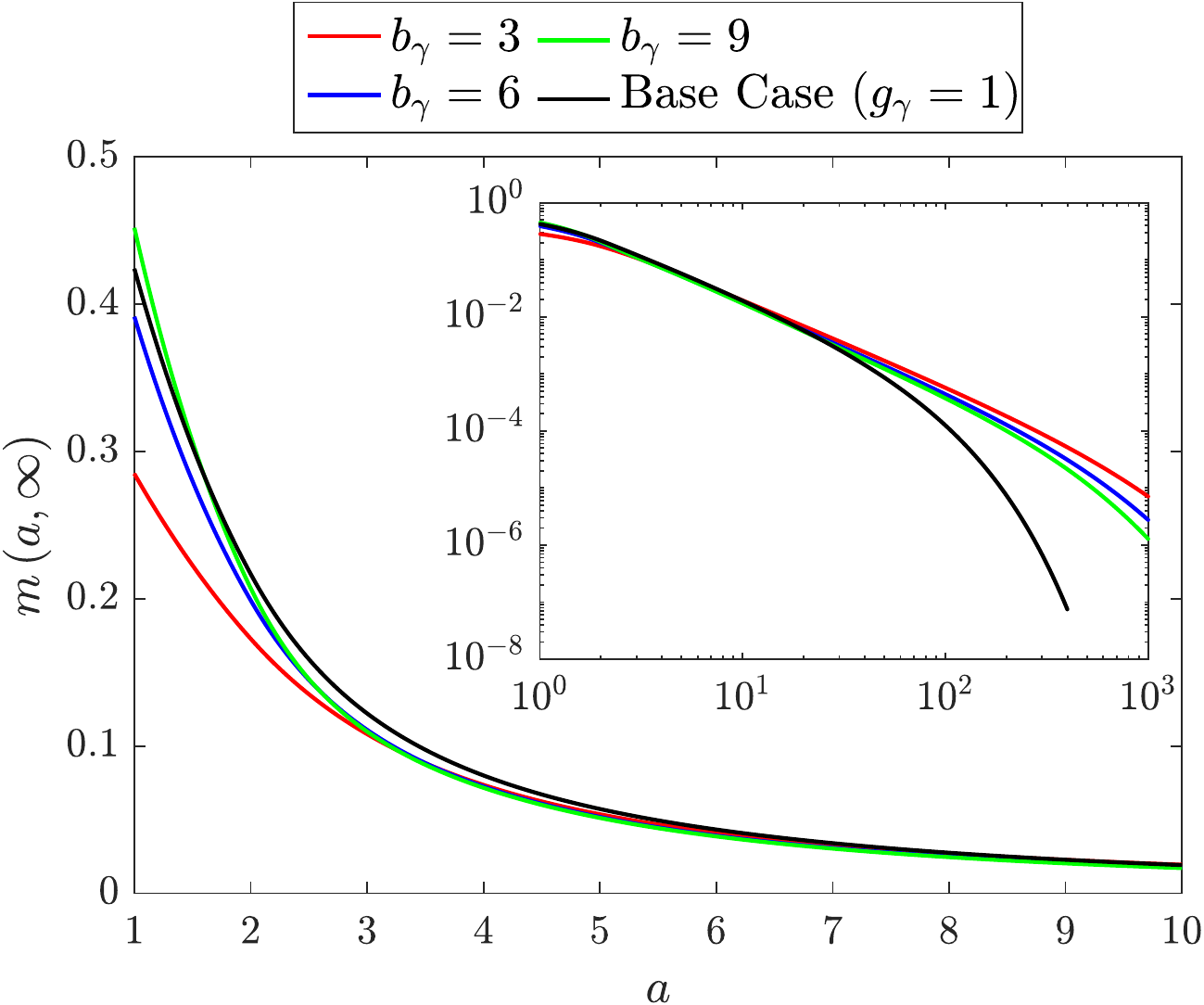}
		\caption{} \label{unsc_m_SS_emi_nonmono}
	\end{subfigure}
	\begin{subfigure}[b]{0.49\textwidth}
		\centering
		\includegraphics[height=6.5cm]{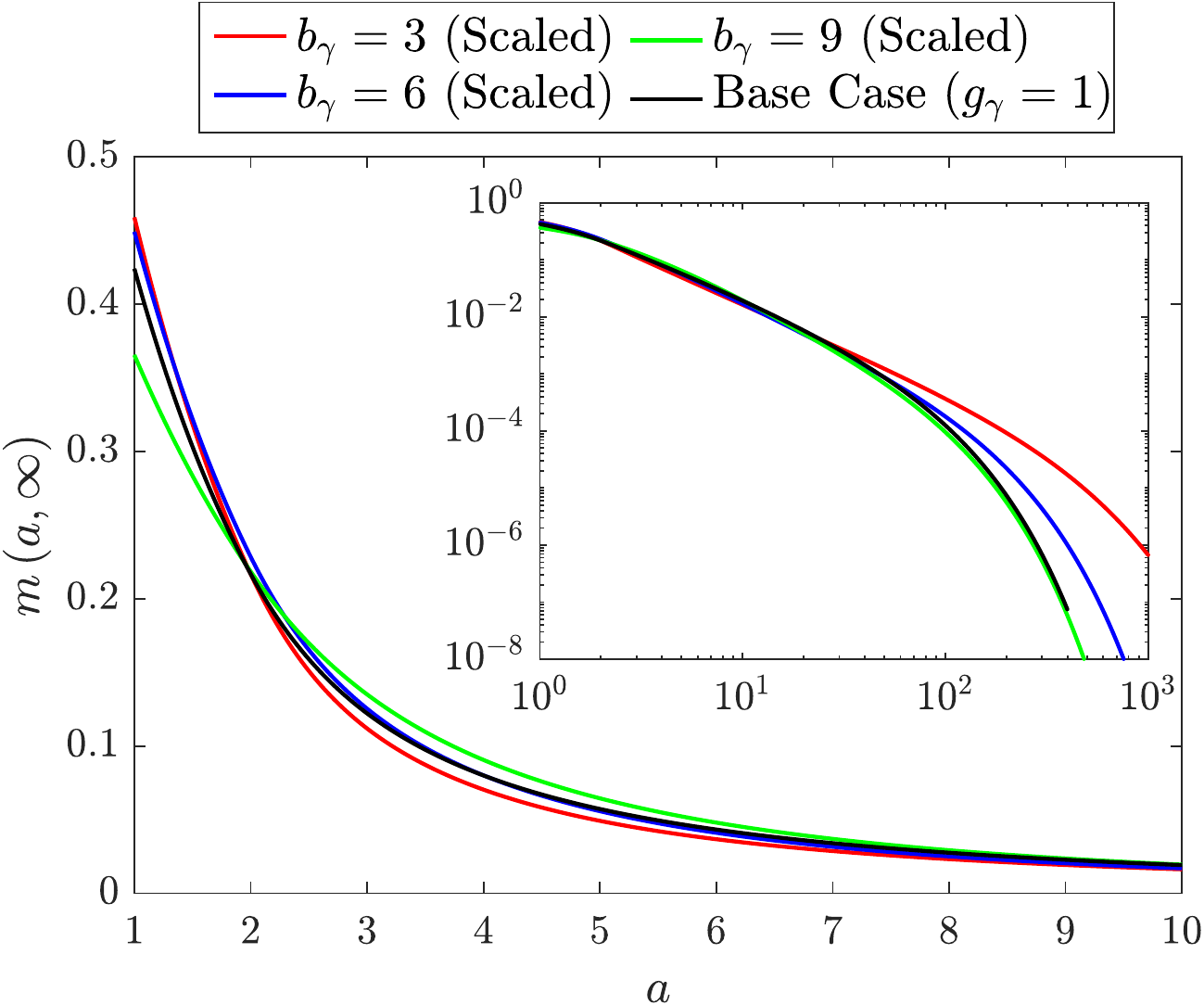}
		\caption{} \label{sc_m_SS_emi_nonmono}
	\end{subfigure}
	\caption{Steady state $m\left(a,t\right)$ distributions for the lipid-independent base case ($g_\gamma = 1$; black lines) and for several scenarios with lipid-dependent emigration. Lipid-dependent simulations use (a) unscaled or (b) scaled monotonic lipid-dependence ($g_\gamma = g_\gamma^s$) and (c) unscaled or (d) scaled non-monotonic lipid-dependence ($g_\gamma = g_\gamma^r$). Monotonic lipid-dependent cases have parameter values $n_\gamma = 1.5$, $\delta_\gamma = 0.1$ and $a_\gamma = 12$ (red lines), $a_\gamma = 18$ (blue lines) or $a_\gamma = 24$ (green lines). Non-monotonic lipid-dependent cases have parameter values $\epsilon_\gamma = 0.1$, $k_\gamma = 1$, $q_\gamma = 2$ and $b_\gamma = 3$ (red lines), $b_\gamma = 6$ (blue lines) or $b_\gamma = 9$ (green lines). Primary plots show the results on the interval $a \in \left[1,10\right]$ and inset log-log plots show the results on the entire $a$ domain.} \label{}
\end{figure}

Steady state results for the ODE variables in each lipid-dependent case are compared against those from the base case ($g_\gamma = 1$) in Figure \ref{unsc_ODEs_emi_mono} (note the use of several vertical axis scales for ease of visualisation). These results show all five variables trending upwards with decreasing $a_\gamma$ (or, equivalently, with decreasing net emigration rate $G_\gamma\left(\infty\right)$). These upward trends have varying degrees, with the case for $a_\gamma = 12$ showing approximately 1.3-, 1.8-, 3.1-, 5.5- and 13.7-fold increases in $P$, $N$, $M$, $A_P$ and $A_M$, respectively, versus the base case. The increase in steady state $M$ in each lipid-dependent case is partly due to reduced net emigration, but mainly due to increased cell recruitment as a consequence of the substantial rise in $A_M$. The steady state $A_M$ values increase so significantly because as cells accumulate more and more lipid they become less and less likely to emigrate. This is shown in the inset of Figure \ref{unsc_m_SS_emi_mono} by the (relatively) large proportions of cells with very large lipid loads in the steady state $m\left(a,t\right)$ distributions. As these heavily lipid-loaded cells are highly likely to undergo apoptosis, $A_P$ is increased and this ultimately fuels an increase in $N$. Note that these enlarged necrotic lipid pools emerge despite the significant increase in lipid consumption afforded by the increase in $M$.

To correct for the variation in the net steady state emigration rates in the above lipid-dependent cases, we repeat our study with appropriately scaled $g_\gamma^s\left(a\right)$ functions. Note that the functions are scaled in such a way that the limiting value of each $g_\gamma^s$ remains at 0.1 (in practice, this involves dividing $\delta_\gamma$ by the scaling value when the scaling is applied). We make this assumption because the model results are particularly sensitive to the limiting value of $g_\gamma^s$ and we wish to retain consistency for comparison with the unscaled cases. Steady state solutions of the ODE variables are presented in Figure \ref{sc_ODEs_emi_mono}. Unlike lipid-dependent apoptosis (where the steady state values of $M$, $P$ and $A_M$ were all unchanged from their base case values upon scaling), here we see that all five variables continue to differ from the base case. Indeed, the trends in the ODE solutions are identical to those seen in the unscaled cases, albeit with slightly less pronounced variation from the base case values. This observation correlates with the steady state $m\left(a,t\right)$ distributions in Figure \ref{sc_m_SS_emi_mono}, which become slightly closer to the base case distribution than those in the unscaled case (Figure \ref{unsc_m_SS_emi_mono}). These results with scaled $g_\gamma^s$ demonstrate that the particular form of the lipid-dependent emigration function (parameterised here by $a_\gamma$ and an appropriate scaling value) can have considerable influence on the long-term plaque composition.

\begin{figure}
	\centering
	\begin{subfigure}[b]{0.49\textwidth}
		\centering
		\includegraphics[height=5.5cm]{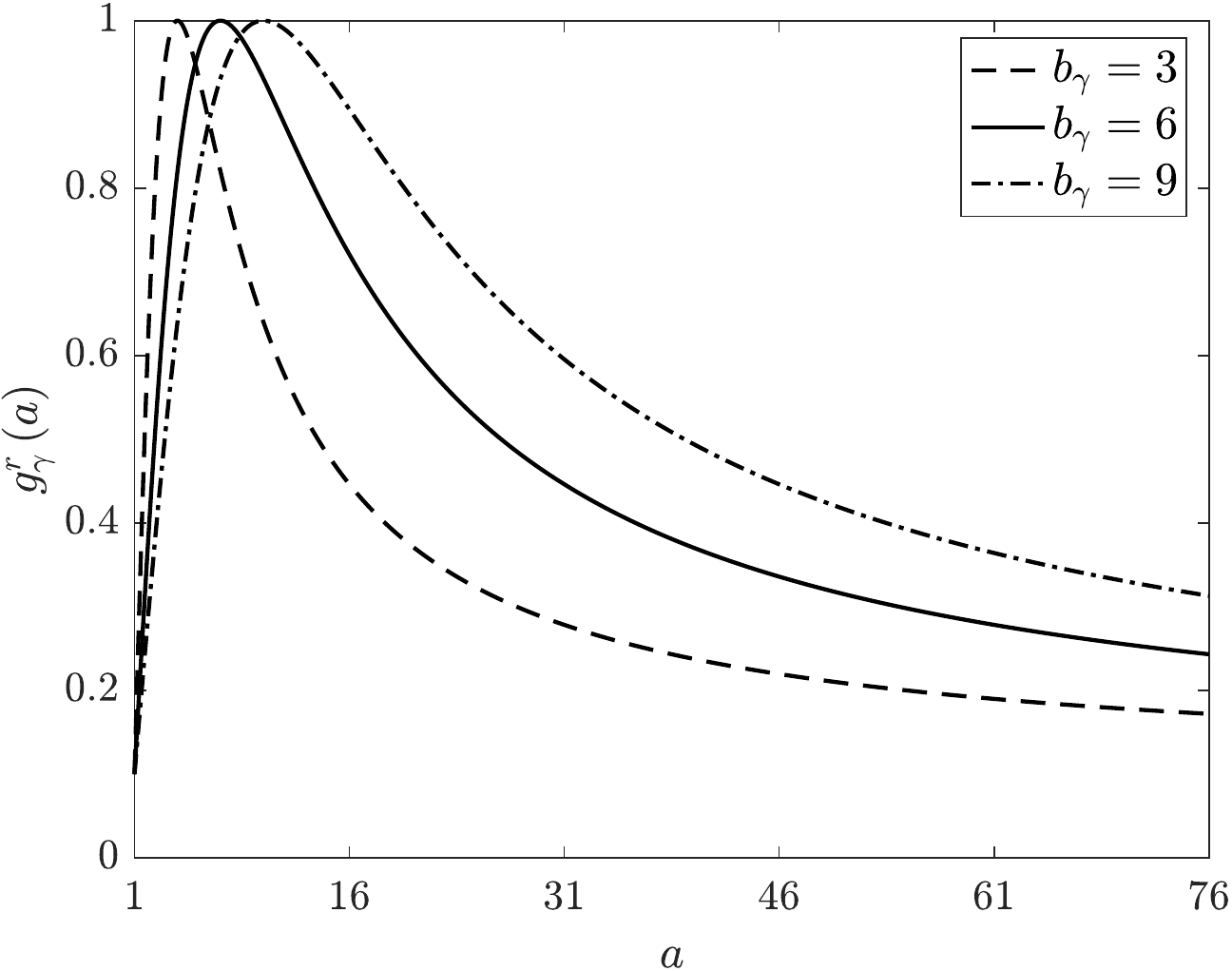}
		\caption{} \label{g_gamma_r_unscaled}
	\end{subfigure}
	\begin{subfigure}[b]{0.49\textwidth}
		\centering
		\includegraphics[height=5.5cm]{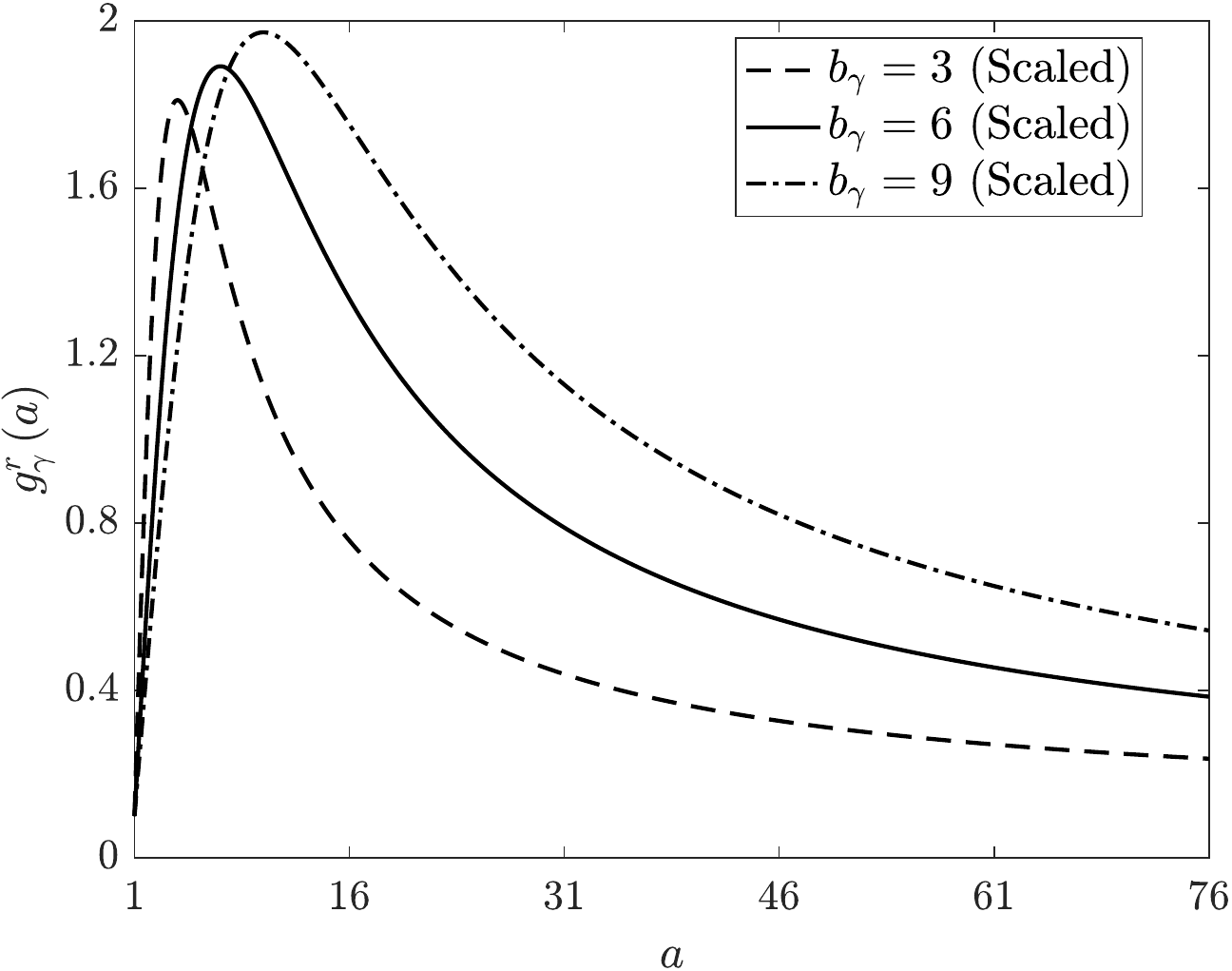}
		\caption{} \label{g_gamma_r_scaled}
	\end{subfigure}
	\caption{(a) Unscaled and (b) scaled rate modulating functions for macrophage emigration $g_\gamma\left(a\right) = g_\gamma^r\left(a\right)$. Plots correspond to equation (\ref{dimless_g_nonmono}) with parameter values $\epsilon_\gamma = 0.1$, $k_\gamma = 1$, $q_\gamma = 2$ and $b_\gamma = 3$ (dashed lines), $b_\gamma = 6$ (solid lines) or $b_\gamma = 9$ (dot-dashed lines). Scaling values for the plots in (b) are 1.9, 1.99 and 2.08, respectively. Note that these scaling values multiply only the second term in equation (\ref{dimless_g_nonmono}) so that each $g_\gamma^r$ retains the limiting value of the unscaled functions.} \label{g_gamma_r}
\end{figure}

\begin{figure}
	\centering		
	\includegraphics[height=5.5cm]{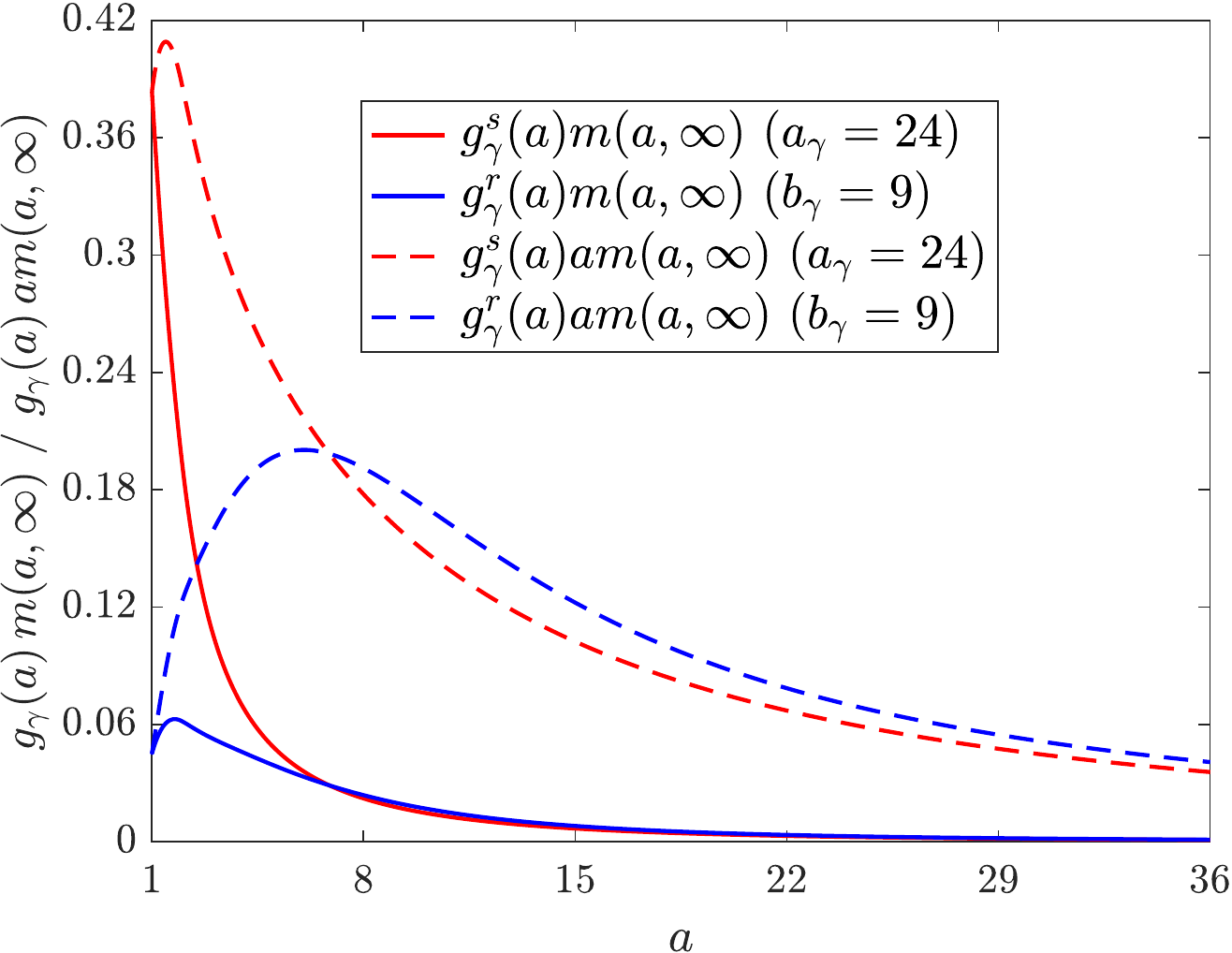}
	\caption{Plot comparing how emigration events $g_\gamma\left(a\right)m\left(a,\infty\right)$ (solid lines) and associated quantities of removed lipid $g_\gamma\left(a\right)am\left(a,\infty\right)$ (dashed lines) are distributed with respect to $a$ at steady state in simulations with monotonic lipid-dependent emigration $g_\gamma = g_\gamma^s$ ($a_\gamma = 24$; red lines) and non-monotonic lipid-dependent emigration $g_\gamma = g_\gamma^r$ ($b_\gamma = 9$; blue lines). The net emigration rate for the non-monotonic case is considerably smaller than that for the monotonic case ($G_\gamma = \int_1^\infty g_\gamma m \, da \approx 0.50$ vs.\ 0.83), but the increased proportion of emigration events with larger lipid loads produces a similar net lipid removal rate ($G_{\gamma a} = \int_1^\infty g_\gamma am \, da \approx 6.29$ vs.\ 6.61). The average lipid removed per cell $\frac{G_{\gamma a}}{G_\gamma}$ is approximately 8.0 for the monotonic case and approximately 12.6 for the non-monotonic case.} \label{SS_emi_events}
\end{figure}

We now consider the case of non-monotonic lipid-dependent emigration, where the function $g_\gamma^r\left(a\right)$ encodes a reduced emigration rate for macrophages with small lipid loads. We fix the parameter values $k_\gamma = 1$, $q_\gamma = 2$ and $\epsilon_\gamma = 0.1$ (note the consistency in the limiting value of $g_\gamma$), and vary $b_\gamma$ such that $g_\gamma^r$ has its peak at $a = 4$ ($b_\gamma = 3$), $a = 7$ ($b_\gamma = 6$) or $a = 10$ ($b_\gamma = 9$). Plots of these three unscaled $g_\gamma^r$ are shown in Figure \ref{g_gamma_r_unscaled}, where we note that each rightward shift in the peak reduces the rate of decline of the function with increasing $a$. The steady state ODE solutions and $m\left(a,t\right)$ distributions generated with these functions are shown in Figure \ref{unsc_ODEs_emi_nonmono} and Figure \ref{unsc_m_SS_emi_nonmono}, respectively. An immediate observation is that the plot of the ODE results is visually very similar to that for the unscaled monotonic case (Figure \ref{unsc_ODEs_emi_mono}). This, however, is coincidental and should not be regarded as significant. Indeed, the lipid-dependent case with the best outcome (smallest $N$) in Figure \ref{unsc_ODEs_emi_mono} is the one with the highest net emigration rate at steady state ($a_\gamma = 24$, $G_\gamma\left(\infty\right) \approx 0.83$), whereas the lipid-dependent case with the best outcome in Figure \ref{unsc_ODEs_emi_nonmono} is the one with the \emph{lowest} net emigration rate at steady state ($b_\gamma = 9$, $G_\gamma\left(\infty\right) \approx 0.50$). This observation demonstrates that when emigration events are skewed towards macrophages with larger lipid loads, the efficiency of lipid removal from the system can be substantially improved (Figure \ref{SS_emi_events}). Note that the case with $b_\gamma = 9$ has a smaller steady state $N$ value than the base case (where $G_\gamma = 1$). This outcome presumably reflects a preferential balance in the steady state $M$ and $A_P$ values. However, the exact mechanism that underlies this result is difficult to pinpoint.

Given that non-monotonic lipid-dependent emigration can remove substantial quantities of lipid from the plaque even at reduced net emigration rates, it seems reasonable to expect an improvement in outcomes when the net emigration rate is scaled to the reference value. (See Figure \ref{g_gamma_r_scaled} for plots of the scaled lipid-dependent functions $g_\gamma^r$. Note that we again preserve the limiting value of 0.1 by multiplying only the non-constant part of (\ref{dimless_g_nonmono}) by the scaling value.) This is indeed true in some cases, but the steady state ODE results in Figure \ref{sc_ODEs_emi_nonmono} portray a greater subtlety in the model's response to this scaling. Relative to their corresponding unscaled cases (Figure \ref{unsc_ODEs_emi_nonmono}), each scaled case shows a reduction in the steady state $M$, $P$, $A_M$ and $A_P$ values. Steady state $N$ also decreases for $b_\gamma = 3$ and $b_\gamma = 6$ (here falling below the base case value), but for $b_\gamma = 9$ the core size \emph{increases}. It appears that, in this case, the extent of lipid removal from the plaque is so substantial that the recruitment response is blunted and too few macrophages remain present in the plaque to resolve the necrotic core. This explanation is corroborated by the steady state $m\left(a,t\right)$ distribution in Figure \ref{sc_m_SS_emi_nonmono} (green line). Compared to the corresponding result for the unscaled case (Figure \ref{unsc_m_SS_emi_nonmono}), we see sizeable reductions in both the proportion of cells with very large lipid loads (see inset) and the proportion of cells with very small lipid loads ($a \approx 1$). The cases for $b_\gamma = 3$ and $b_\gamma = 6$ (red and blue lines, respectively) show similarly large reductions in $m$ at the top end of their distributions, but at the lower end both distributions experience an increase.

\subsection{Lipid-Dependent Simulations With Proliferation} \label{ssProlif}
In this section, we introduce macrophage proliferation into the system by setting $\rho = 0.5$. First we assume that proliferation is independent of lipid load. We briefly revisit the scenarios from Section \ref{ssNoProlif} to study the manner in which proliferation (with its tendency to reduce average lipid loads) interacts with lipid-dependence in the other cell behaviours. We then revert to lipid-independent apoptosis and emigration, and conclude the results by investigating the impact of lipid-dependence in proliferation itself.

\subsubsection{Apoptosis Only or Emigration Only} \label{sssApoEmi}
We set $g_\rho\left(a\right) = 1$, and select one lipid-dependent case from each of Sections \ref{sssApo} and \ref{sssEmi}. These are $g_\beta = g_\beta^s$ with $a_\beta = 12$, $\delta_\beta = 3$ and $g_\gamma = g_\gamma^s$ with $a_\gamma = 18$, respectively. As before, we allow lipid-dependence in only one cell behaviour at a time, and we investigate each lipid-dependent case using both scaled and unscaled functions. (Note that $g_\beta$ and $g_\gamma$ generally require new scaling values to account for the impact of proliferation; see Table \ref{mono_params}). Here, we shall primarily focus on the scaled cases since preserving the net steady state apoptosis or emigration rate facilitates comparison against the earlier scaled cases without proliferation. The unscaled cases do, however, provide some interest and we comment on these below.

For the unscaled lipid-dependent apoptosis case, Table \ref{mono_params} shows that the inclusion of proliferation acts to reduce the net steady state apoptosis rate $G_\beta$ by approximately 8\% from 1.576 to 1.445. This reflects the fact that proliferation acts to reduce individual cell lipid loads, thereby reducing the likelihood of apoptosis (i.e.\ daughter cell apoptosis rates $g_\beta^s\left(a\right)$ are always less than parent cell apoptosis rates $g_\beta^s\left(2a-1\right)$). Note, however, that a tangible reduction in the daughter cell apoptosis rate requires $g_\beta^s\left(a\right) \ll g_\beta^s\left(2a-1\right)$, and this occurs only for daughter cells with $a$ sufficiently small. The unscaled lipid-dependent emigration case produces the rather curious result that proliferation actually \emph{reduces} the net steady state emigration rate ($G_\gamma\left(\infty\right)$ drops from 0.7812 to 0.7650; see Table \ref{mono_params}). This is counter-intuitive because the associated reduction of lipid loads would be expected to \emph{increase} the likelihood of emigration (i.e.\ $g_\gamma^s\left(a\right) > g_\gamma^s\left(2a-1\right)$ for all $a$). This quirk appears to arise due to the proliferation of cells with very large lipid loads. Proliferation of such cells fails to significantly elevate daughter cell emigration rates because $g_\gamma^s\left(a\right)$ remains similar to $g_\gamma^s\left(2a-1\right)$ for large $a$. Thus, lipid loaded cells with very low emigration rates perpetuate in the system and skew the $G_\gamma$ value upwards. Despite this unexpected increase in $G_\gamma$, proliferation does, as expected, reduce the average lipid per cell at steady state.

\begin{figure}
	\centering
	\begin{subfigure}[b]{0.49\textwidth}
		\centering
		\includegraphics[height=6.35cm]{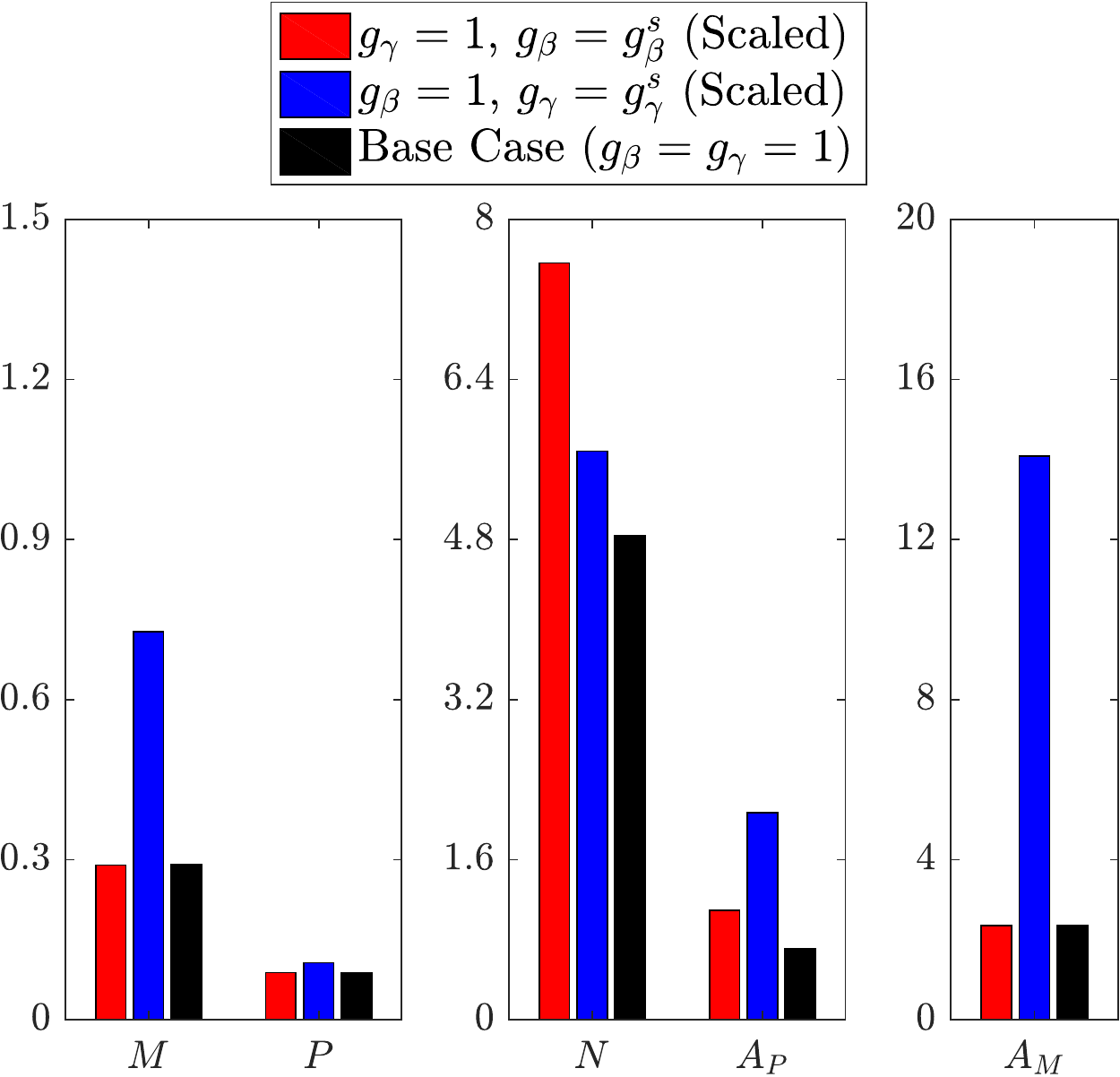}
		\caption{} \label{ODEs_SS_apo_emi_rho0}
	\end{subfigure}
	\begin{subfigure}[b]{0.49\textwidth}
		\centering
		\includegraphics[height=6.35cm]{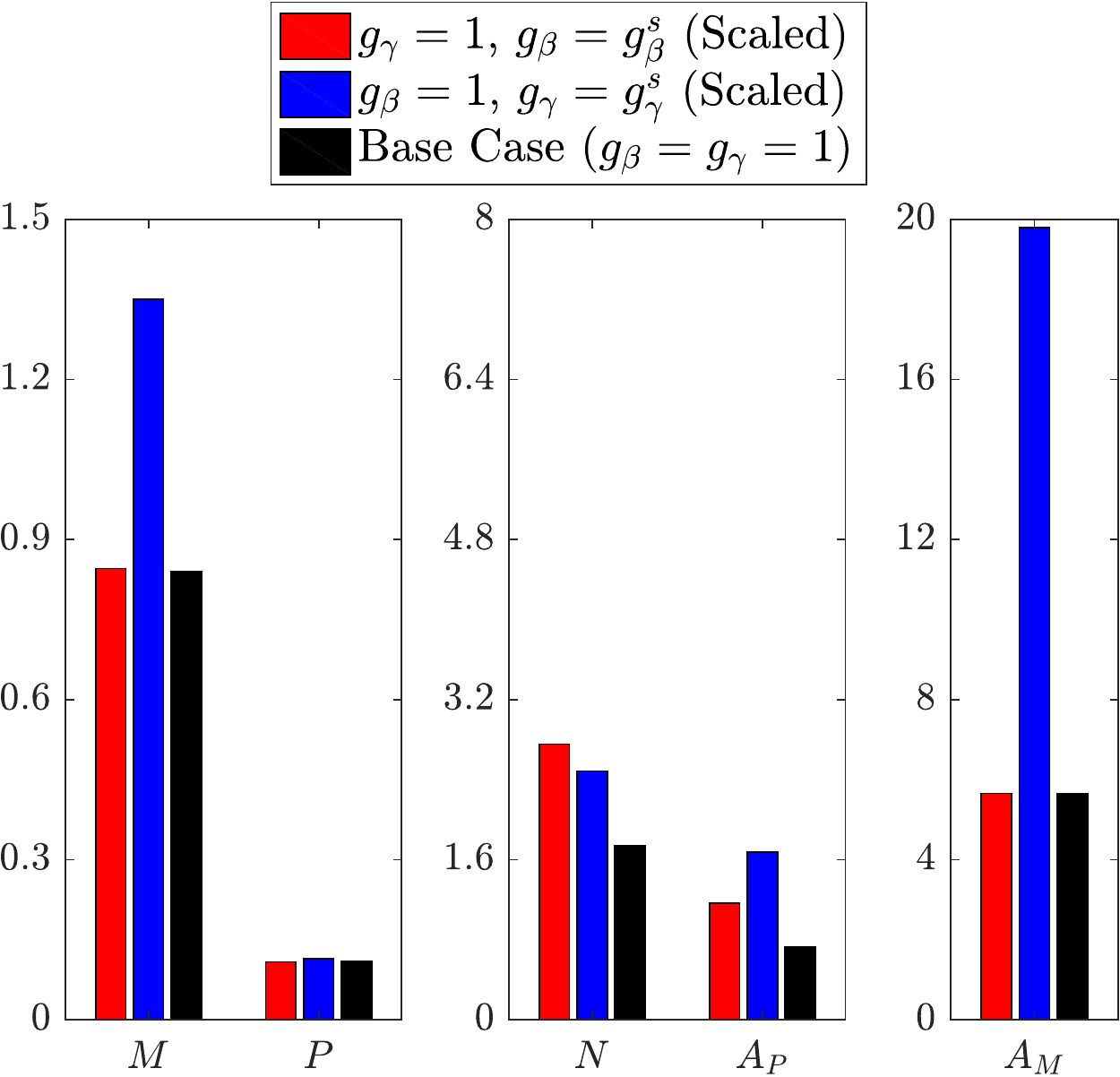}
		\caption{} \label{ODEs_SS_apo_emi_rho5}
	\end{subfigure}
	\caption{Steady state solutions of the ODE variables $M\left(t\right)$, $P\left(t\right)$, $N\left(t\right)$, $A_P\left(t\right)$ and $A_M\left(t\right)$ for lipid-independent ($g_\beta = g_\gamma = 1$; black bars) and scaled lipid-dependent cases (a) without proliferation ($\rho = 0$) and (b) with proliferation ($\rho = 0.5$). Red bars show results with lipid-dependent apoptosis ($g_\beta = g_\beta^s$, $g_\gamma = 1$) using parameter values $n_\beta = 2$, $\delta_\beta = 3$ and $a_\beta = 12$. Blue bars show results with lipid-dependent emigration ($g_\gamma = g_\gamma^s$, $g_\beta = 1$) using parameter values $n_\gamma = 1.5$, $\delta_\gamma = 0.1$ and $a_\gamma = 18$.} \label{ODEs_SS_apo_emi_rho}
\end{figure}

\begin{figure}
	\centering
	\begin{subfigure}[b]{0.49\textwidth}
		\centering
		\includegraphics[height=6.8cm]{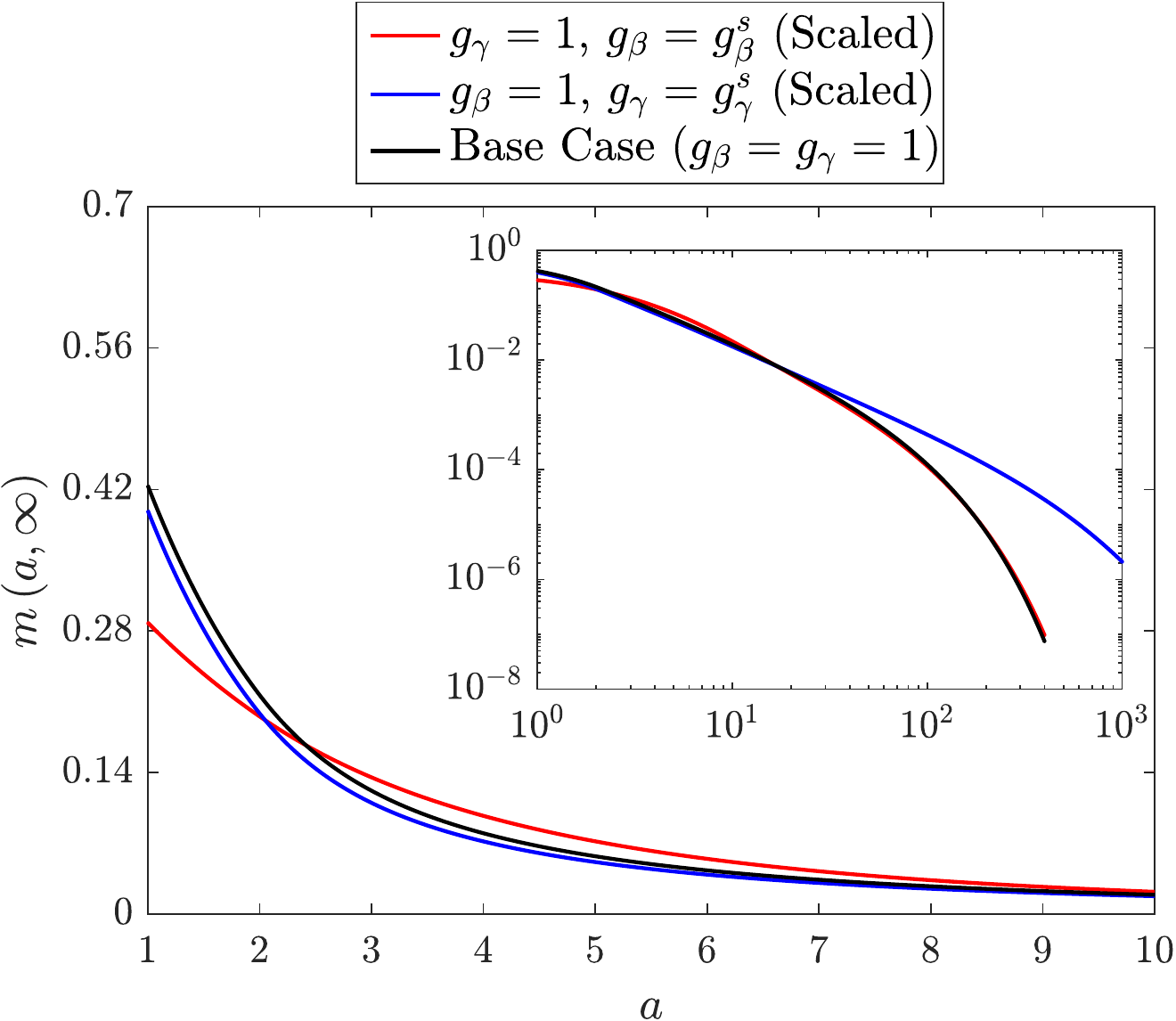}
		\caption{} \label{m_SS_apo_emi_rho0}
	\end{subfigure}
	\begin{subfigure}[b]{0.49\textwidth}
		\centering
		\includegraphics[height=6.8cm]{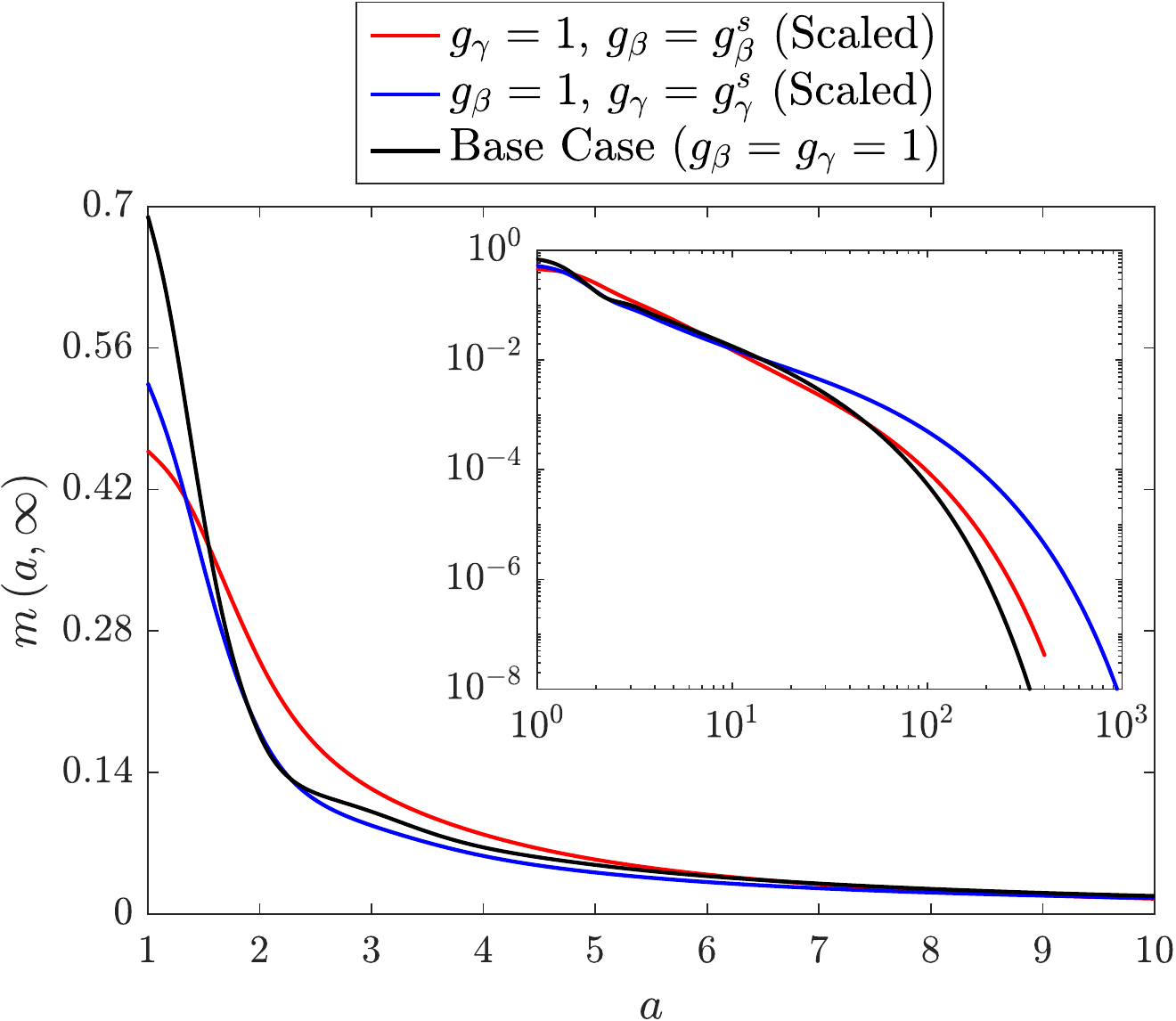}
		\caption{} \label{m_SS_apo_emi_rho5}
	\end{subfigure}
	\caption{Steady state $m\left(a,t\right)$ distributions for lipid-independent ($g_\beta = g_\gamma = 1$; black bars) and scaled lipid-dependent cases (a) without proliferation ($\rho = 0$) and (b) with proliferation ($\rho = 0.5$). Red lines show results with lipid-dependent apoptosis ($g_\beta = g_\beta^s$, $g_\gamma = 1$) using parameter values $n_\beta = 2$, $\delta_\beta = 3$ and $a_\beta = 12$. Blue lines show results with lipid-dependent emigration ($g_\gamma = g_\gamma^s$, $g_\beta = 1$) using parameter values $n_\gamma = 1.5$, $\delta_\gamma = 0.1$ and $a_\gamma = 18$. Primary plots show the results on the interval $a \in \left[1,10\right]$ and inset log-log plots show the results on the entire $a$ domain.} \label{m_SS_apo_emi_rho}
\end{figure}

Turning attention back to the scaled lipid-dependent functions, Figure \ref{ODEs_SS_apo_emi_rho} presents the steady state ODE solutions for the lipid-dependent and base case simulations both in the absence (Figure \ref{ODEs_SS_apo_emi_rho0}) and presence (Figure \ref{ODEs_SS_apo_emi_rho5}) of macrophage proliferation. Although the two solution sets are quantitatively different, we note that the lipid-dependent results display the same qualitative trends regardless of the particular proliferation rate. Lipid-dependent apoptosis increases $N$ and $A_P$ ($M$, $P$ and $A_M$ remain fixed), while lipid-dependent emigration increases the solutions of all five ODE variables. Figure \ref{m_SS_apo_emi_rho} shows the corresponding steady-state $m\left(a,t\right)$ distributions for the cases without proliferation (Figure \ref{m_SS_apo_emi_rho0}) and with proliferation (Figure \ref{m_SS_apo_emi_rho5}). As proliferation adds a second non-local effect into the model, the precise features of the plots in Figure \ref{m_SS_apo_emi_rho5} are difficult to interpret. However, as expected based on equation (\ref{dimless_AMbar_eqn}), proliferation tends to shift the $m\left(a,t\right)$ distributions towards lower accumulated lipid loads. For lipid-dependent apoptosis, proliferation produces a relatively large increase in the proportion of cells with small lipid loads. However, for large $a$, the impact of proliferation seems relatively minimal. This likely reflects the substantial increase in the apoptosis rate with increasing $a$, which effectively acts to suppress proliferation events.     

\subsubsection{Proliferation Only} \label{sssPro}
\begin{figure}
	\centering
	\begin{subfigure}[b]{0.49\textwidth}
		\centering
		\includegraphics[height=5.5cm]{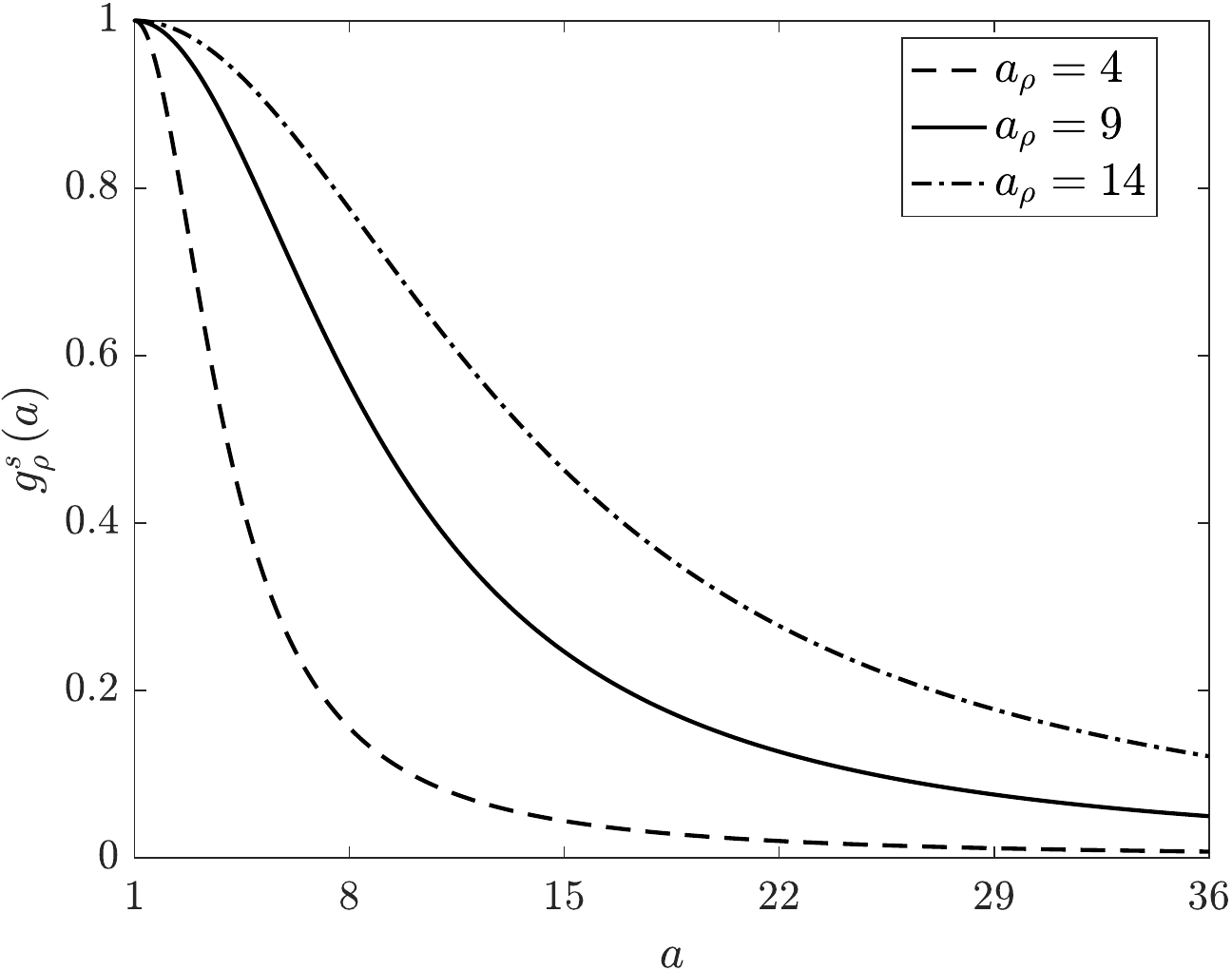}
		\caption{} \label{g_rho_s_unscaled}
	\end{subfigure}
	\begin{subfigure}[b]{0.49\textwidth}
		\centering
		\includegraphics[height=5.5cm]{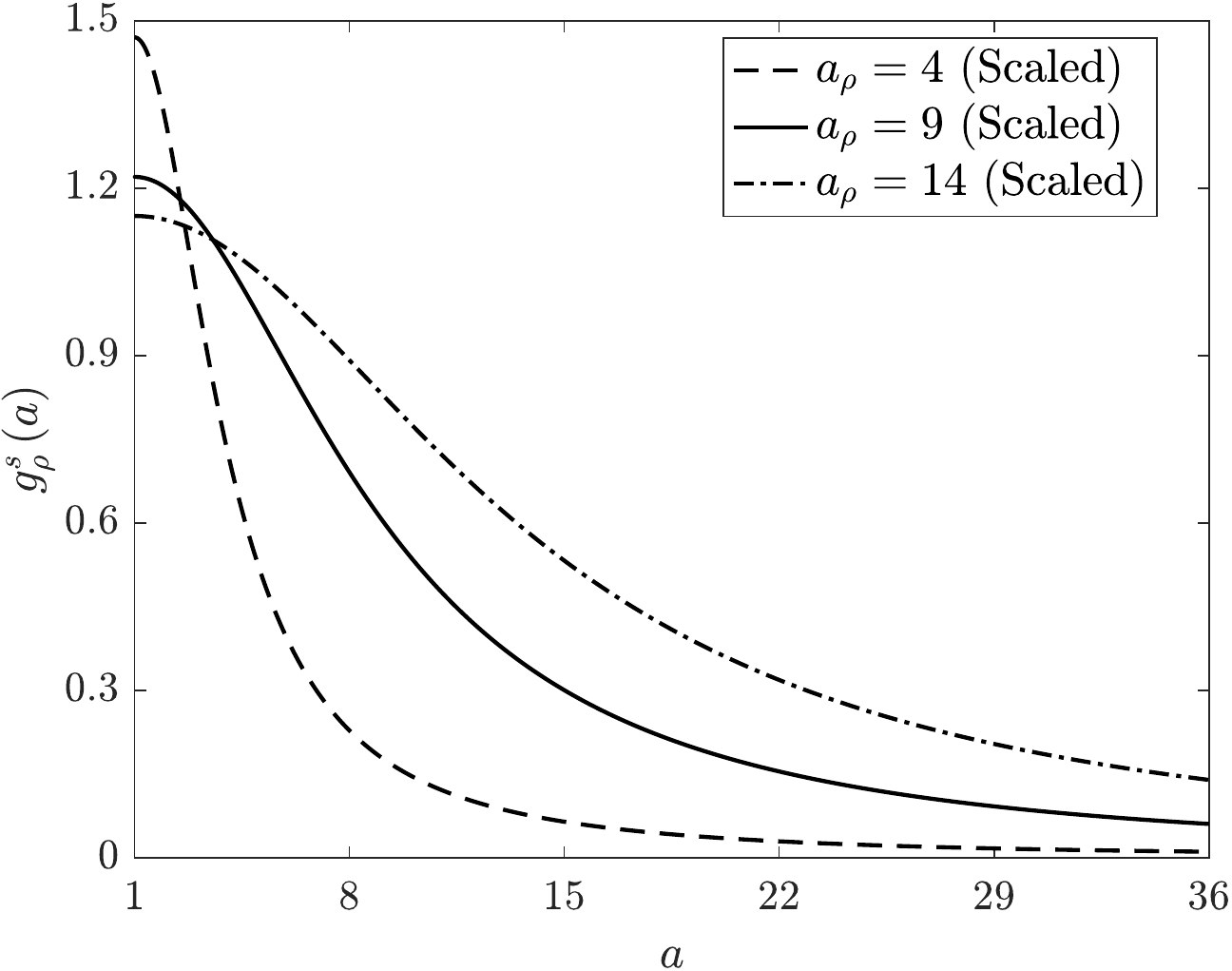}
		\caption{} \label{g_rho_s_scaled}
	\end{subfigure}
	\caption{(a) Unscaled and (b) scaled rate modulating functions for macrophage proliferation $g_\rho\left(a\right) = g_\rho^s\left(a\right)$. Plots correspond to equation (\ref{dimless_g_mono}) with parameter values $n_\rho = 2$, $\delta_\rho = 0$ and $a_\rho = 4$ (dashed lines), $a_\rho = 9$ (solid lines) or $a_\rho = 14$ (dot-dashed lines). Scaling values for the plots in (b) are 1.47, 1.22 and 1.15, respectively.} \label{g_rho_s}
\end{figure}

Finally, we investigate the role of lipid-dependent macrophage proliferation. In this section, we set $g_\rho\left(a\right) = g_\rho^s\left(a\right)$ and fix $g_\beta\left(a\right) = g_\gamma\left(a\right) = 1$. Once again, we consider three different forms for the unscaled lipid dependent functions (Figure \ref{g_rho_s_unscaled}). These functions have $n_\rho = 2$, $\delta_\rho = 0$ and $a_\rho = 4$, 9 or 14. By varying $a_\rho$, we study the effect of different rates of decline in the macrophage proliferation rate with increased lipid loading \citep{Kim18}.

\begin{figure}
	\centering		
	\includegraphics[height=5.5cm]{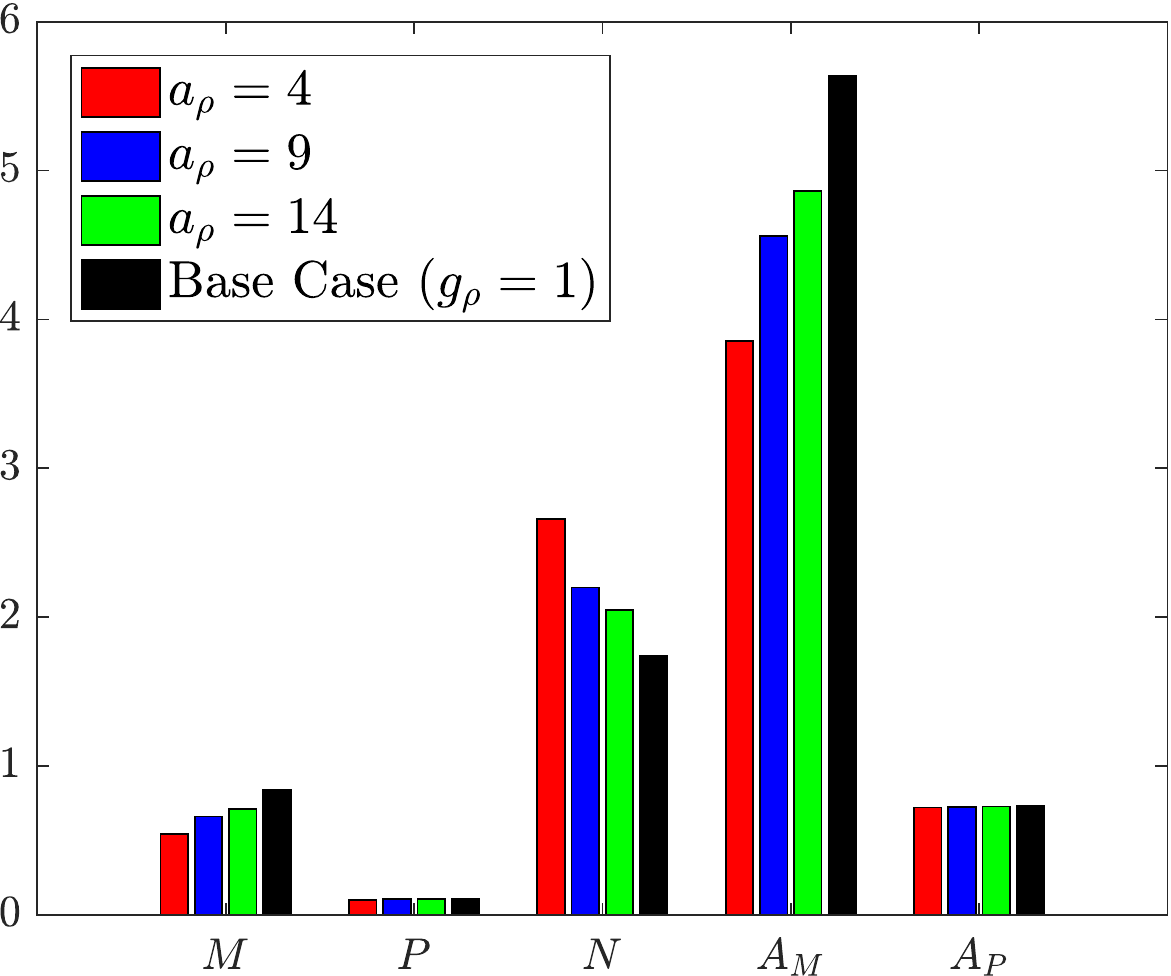}
	\caption{Steady state solutions of the ODE variables $M\left(t\right)$, $P\left(t\right)$, $N\left(t\right)$, $A_P\left(t\right)$ and $A_M\left(t\right)$ for the lipid-independent base case ($g_\rho = 1$; black bars) and for three cases with unscaled lipid-dependent proliferation ($g_\rho = g_\rho^s$). Lipid-dependent cases have parameter values $n_\rho = 2$, $\delta_\rho = 0$ and $a_\rho = 4$ (red bars), $a_\rho = 9$ (blue bars) or $a_\rho = 14$ (green bars).} \label{unsc_ODEs_SS_pro}
\end{figure}

The steady state ODE solutions generated using these functions are presented in Figure \ref{unsc_ODEs_SS_pro}. Here, we observe marginal decreases in $P$ and $A_P$, marked decreases in $M$ and $A_M$, and a marked increase in $N$ with decreasing $a_\rho$. What becomes apparent, however, is that these variations in the steady state solutions across the lipid-dependent cases are not explicitly related to the particular form of $g_\rho$, but simply reflect the associated changes in the net proliferation rates $\rho G_\rho\left(\infty\right)$. The case with $a_\rho = 4$, for example, has net steady state proliferation rate $\rho G_\rho \approx \left(0.5\right)\left(0.6377\right) \approx 0.319$. A lipid-independent simulation ($G_\rho = 1$) with proliferation rate $\rho = 0.319$ would therefore give almost identical steady state ODE solutions, albeit with different temporal dynamics and different steady state PDE solutions.

It is interesting to observe that the $G_\rho$ values in these lipid-dependent simulations remain fairly close to 1, even when $g_\rho^s$ declines rapidly with $a$. This suggests that the model is relatively insensitive to the value of the parameter $a_\rho$, provided it is not too close to 1. This makes intuitive sense because proliferation generally acts to keep cell lipid loads close to $a=1$, and, provided $a_\rho$ is not too small, the lipid-dependent proliferation rate remains high throughout this region (i.e.\ $g_\rho^s \approx 1$). Contrastingly, we note that, while lipid-independent proliferation may allow some cells to acquire small lipid loads from large lipid loads through multiple proliferation events, this phenomenon will be suppressed in the lipid-dependent cases due to the decline in $g_\rho^s$ for large $a$.

\begin{figure}
	\centering		
	\includegraphics[height=6.5cm]{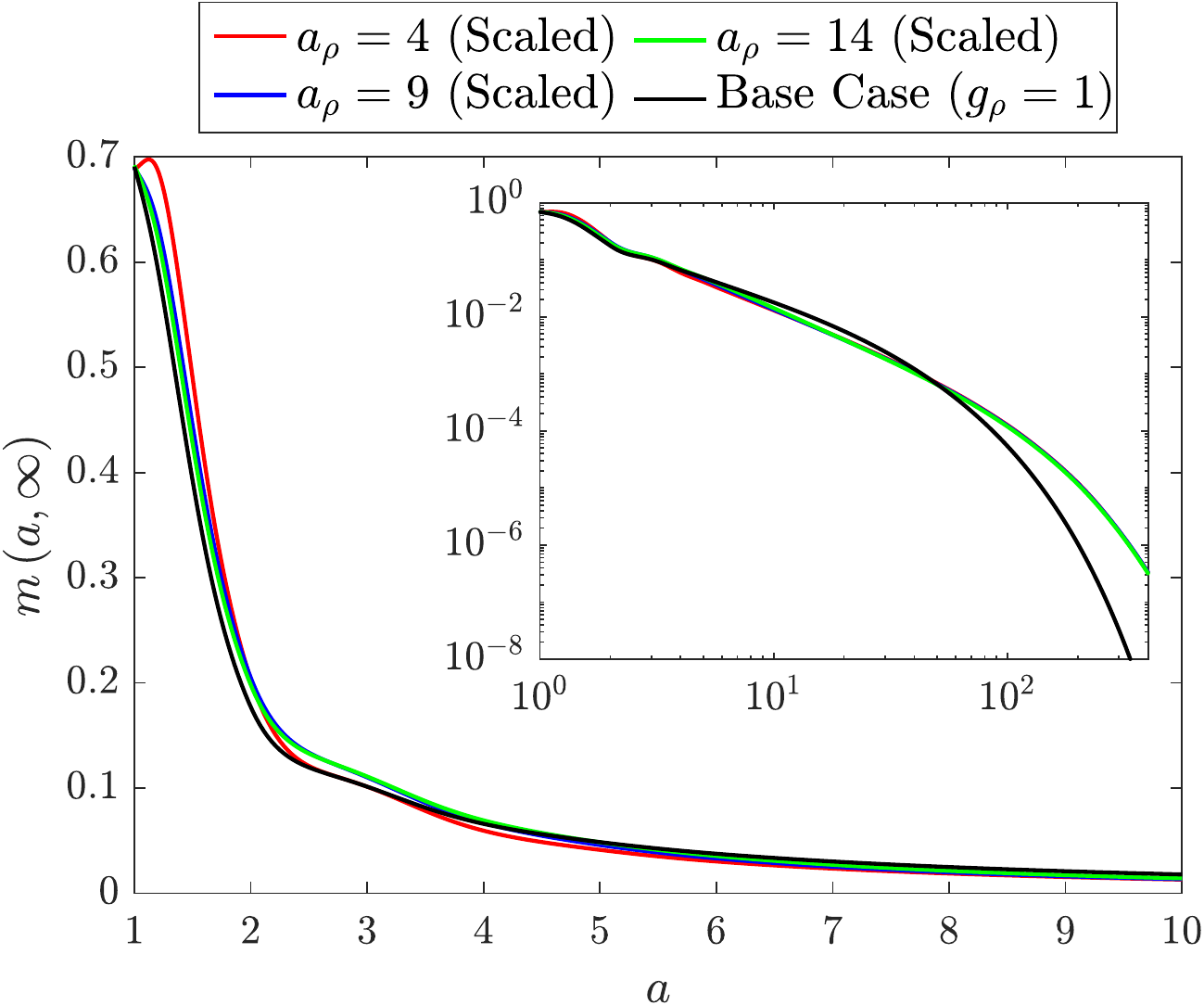}
	\caption{Steady state $m\left(a,t\right)$ distributions for the lipid-independent base case ($g_\rho = 1$; black bars) and for three cases with scaled lipid-dependent proliferation ($g_\rho = g_\rho^s$). Lipid-dependent cases have parameter values $n_\rho = 2$, $\delta_\rho = 0$ and $a_\rho = 4$ (red bars), $a_\rho = 9$ (blue bars) or $a_\rho = 14$ (green bars). The primary plot shows the results on the interval $a \in \left[1,10\right]$ and the inset log-log plot shows the results on the entire $a$ domain. Note that all four distributions coincide at $a = 1$ because the values of the ODE variables in the boundary condition (\ref{dimless_BC}) are almost identical in each case.} \label{sc_m_SS_pro}
\end{figure}

Consistent with their definition, simulations using the scaled $g_\rho^s$ (Figure \ref{g_rho_s_scaled}) lead to net proliferation rates approximately equal to $\rho$ at steady state. Accordingly, each of these simulations produces steady state ODE results that are indistinguishable from the base case. We therefore omit these ODE results and focus instead on the corresponding steady state $m\left(a,t\right)$ distributions (Figure \ref{sc_m_SS_pro}). (Note that we also omit the $m\left(a,t\right)$ distributions generated using the unscaled $g_\rho^s$. These differ slightly from the scaled $g_\rho^s$ results for small $a$, but the distributions become increasingly similar as $a$ increases.) Figure \ref{sc_m_SS_pro} shows that there are few discernible differences between the base case steady state $m\left(a,t\right)$ distribution and the distributions generated using the scaled $g_\rho^s$. One exception is that the lipid-dependent cases all have a larger proportion of cells with lipid load $a>50$. This is because these cells have substantially reduced proliferation rates and cannot readily divide to reduce their lipid load. In the case with $a_\rho = 4$, we note a subtle change in the form of the distribution with the emergence of a small peak near $a=1$. This is likely due to the locally elevated proliferation rate, which alters the relative balance between proliferation and efferocytosis (see Figure 7 in \citet{Cham22}). Overall, we conclude from the results in this section that the outcome of plaque formation is probably more sensitive to the net population-level proliferation rate than to the particular distribution of lipid-dependent rates across the population.

\section{Discussion} \label{sDiscuss}
Lipid consumption is known to alter macrophage behaviour but the implications for atherosclerotic plaque formation are not well understood. In this paper, we establish a novel modelling framework with which to study these implications. We develop a structured population model of plaque macrophage lipid accumulation in which macrophages are classified by their internalised lipid load $a$ and behave in a lipid-dependent manner. This work builds upon the partial integro-differential equation models recently developed in \citet{Ford19a} and \citet{Cham22}. As in these earlier works, we model how live macrophage behaviour feeds into apoptotic macrophage population dynamics and necrotic core formation through mechanisms such as efferocytosis, post-apoptotic necrosis and necrotic core consumption. A key difference in the current model is that the lipid-averaged ODE subsystem cannot be readily decoupled from the governing PDEs.

We focus our investigation on how lipid-dependent macrophage apoptosis, emigration and proliferation can influence plaque progression. This work is predicated on results that show dysfunctionality in macrophages with large ingested lipid loads \citep{Taba02, Moor13, Yin21}. Consequently, we assume that increasing lipid loads cause increased rates of apoptosis and reduced rates of both emigration and proliferation. Note that we do not consider lipid-induced dysfunction in macrophage phagocytic or efferocytic capacity. Thus, heavily lipid-loaded macrophages in our simulations consume extracellular lipid and engulf apoptotic bodies at the same rate as cells with small ingested lipid loads. This is a limitation of the current study that we will address in future by including lipid-dependent phagocytosis and efferocytosis terms in the model. We anticipate that the inclusion of these terms will have dual benefit. Not only will the model become more realistic, but lipid consumption rates that decrease with increasing lipid load will reduce the domain sizes required for accurate numerical solutions. Model simulations will therefore be less computationally demanding and numerical techniques such as the non-uniform gridding strategy implemented here may no longer be required.      

Lipid-dependent macrophage behaviour is encoded in the model through the dimensionless functions $g_\beta\left(a\right)$, $g_\gamma\left(a\right)$ and $g_\rho\left(a\right)$, which modulate the reference rates of apoptosis, emigration and proliferation, respectively. These functions typically take a monotonic form, but for emigration we also consider a non-monotonic lipid-dependence. In the absence of appropriate data to parameterise these functions, we simulate a range of scenarios with different rates of approach to their (finite) limiting values. This is a reasonable approach as the lipid-dependence of each behaviour \emph{in vivo} is likely controlled by several different factors including the activation or disruption of unique signalling pathways \citep{Feng03, Gils12, Robb13}. Thus, the rate of each individual macrophage behaviour may be altered in different ways, at different times and at different internalised lipid loads. In our analysis, we have made the simplifying assumption that only one of apoptosis, emigration or proliferation may be lipid-dependent at any one time. While this may be unrealistic in practice, the benefit of this assumption is that we can unpick the influence of each lipid-dependent behaviour on plaque progression. Of course, simulations with more than one lipid-dependent behaviour can be easily performed within our model framework. However, the increased difficulty in interpretation of such results tends to limit the insight gained.

For each scenario that we model in this paper, we fix the lipid-independent parameter values and perform several simulations using both unscaled and scaled functions $g_\diamond\left(a\right)$. Unscaled simulations use the exact functional forms (\ref{dimless_g_mono}) or (\ref{dimless_g_nonmono}) with an appropriate set of lipid-dependent parameter values. Each corresponding scaled simulation uses the same lipid-dependent parameter values, but the function $g_\diamond$ is multiplicatively scaled such that the net rate of the behaviour of interest matches the reference rate at steady state (mathematically, this is expressed as $G_\diamond\left(\infty\right) = \int_1^\infty g_\diamond\left(a\right) m\left(a,\infty\right) \, da = 1$ where $g_\diamond\left(a\right) \neq 1$). While it may appear excessive to consider both unscaled and scaled simulations for each scenario, we note that each approach has unique advantages and disadvantages. Unscaled simulations are relevant for understanding the impact of variability in lipid-dependent macrophage behaviour and/or the response of the model system to interventions that alter the lipid-dependence. However, as the net steady state rate of the behaviour of interest is not conserved from case to case, it is difficult to interpret the impact of the lipid-dependence itself. By correcting for this shortcoming, scaled simulations provide a more consistent and insightful means of comparison across simulated scenarios. However, the scaling value required in a given case is not known \emph{a priori} and must be determined by a process of estimation and refinement through repeated simulation. As such, the scaling process itself can be computationally demanding.

Our results and analysis indicate some interesting population level differences in the relative influence of lipid-dependent apoptosis, emigration and proliferation on plaque progression. For unscaled lipid-dependent proliferation simulations with net steady state proliferation rate $\rho G_\rho\left(\infty\right)$, we find that the steady state ODE solutions $M$, $P$, $A_M$, $A_P$ and $N$ can always be matched by an equivalent lipid-independent simulation with the constant proliferation rate $\rho G_\rho\left(\infty\right)$. Consequently, for all scaled lipid-dependent apoptosis simulations ($G_\rho\left(\infty\right) \approx 1$), we obtain essentially identical steady state ODE results across the board. These observations, it turns out, are unique to lipid-dependent proliferation and equivalent results do not hold for lipid-dependent apoptosis or lipid-dependent emigration simulations. Mathematically, this reflects the presence in the model equations of the terms $G_{\beta a}$ and $G_{\gamma a}$. These terms appear because the net rate of lipid removal from the live cell population by lipid-dependent apoptosis or lipid-dependent emigration depends on the distribution of lipid across the population. There is no equivalent $G_{\rho a}$ term in the model equations because the quantity of lipid added to the system by a proliferation event ($a_0$ in dimensional terms) is always fixed and does not depend on the internalised lipid content of the parent cell. Thus, compared to lipid-dependent proliferation, the model equations for lipid-dependent apoptosis or lipid-dependent emigration have added nonlinearity that increases the complexity of the underlying dynamics. This is emphasised by equation (\ref{dimless_AMbar_eqn}), where we see that lipid-dependent apoptosis and lipid-dependent emigration can dynamically alter $\bar{A}_M\left(t\right)$ in a way that is not observed at all in lipid-independent cases.

In our unscaled lipid-dependent apoptosis simulations, we observe the emergence of decaying oscillations in the ODE variables when the increase in $g_\beta\left(a\right)$ is sufficiently severe (i.e.\ large in both magnitude and steepness). Oscillatory solutions of this type appear to be unique to the lipid-dependent model as they have not been observed in earlier lipid-independent approaches \citep{Ford19a, Cham22}. Two factors that appear to be necessary to initiate these oscillations are: (1) macrophages ingest (primarily necrotic) lipid at such a rate that $A_M$ grows irrespective of a concurrent increase in the net apoptosis rate; and (2) this increase in $A_M$ stimulates an increase in macrophage recruitment. Note that this latter requirement may only be satisfied for certain values of the dimensionless parameter $\kappa$. We use $\kappa = 5$ in our simulations. However, for a smaller $\kappa$ value, the recruitment rate would likely be less sensitive to changes in $A_M$ and it is possible that oscillations may not emerge at all.

The observation that oscillatory solutions require a sufficiently rapid rate of lipid ingestion relative to apoptosis raises an interesting question about what might happen should $g_\beta$ not saturate and instead approach infinity at some finite $a$ (that is, if there exists a certain lipid load above which macrophages become unviable). It is not clear whether oscillations could occur in this case because $A_M$ may be unable to grow in the face of (potentially) very rapid lipid-induced apoptosis. Note that non-saturating lipid-dependent death rates may be generally interesting to investigate in this model because, alongside lipid ingestion rates that decrease with $a$, they offer another means to reduce (and, indeed, cap) the length of the $a$ domain.

A full understanding of the oscillatory dynamics that the lipid-dependent apoptosis model can permit is unlikely to be achieved without formal mathematical analysis. However, the nonlinearity of the function $g_\beta^s$ makes this analysis challenging. An approach that may allow progress is to remove the nonlinearity by approximating the saturating function $g_\beta^s$ with an equivalent step function that increases from 1 to $\delta_\beta$ at $a = a_\beta$. We will not pursue this possibility in the current work but it may form the basis of a future publication.       

For scaled lipid-dependent apoptosis, we find the surprising result that the steady state $M$, $P$ and $A_M$ values remain unchanged across all simulated cases. Our analysis shows that this happens because, at steady state in each simulation, the net rate of lipid loss from live cells due to apoptosis ($G_{\beta a}$) is exactly balanced by the rate of lipid gain via efferocytosis and necrotic lipid consumption ($\eta A_P + \theta N$). The balance between these terms fixes the steady state $A_M$ value in each simulation, and fixed steady state $M$ and $P$ follow because the recruitment rate, net apoptosis rate and all other relevant rates remain unchanged. Of course, while the relationship between $G_{\beta a}$ and $\eta A_P + \theta N$ causes steady state $M$, $P$ and $A_M$ to remain fixed, the very same relationship causes steady state $A_P$ and $N$ to vary. With increasing severity of change in scaled $g_\beta^s$, we find that the steady state values of $G_{\beta a}$, $A_P$ and $N$ all experience an increase. Overall, these results suggest that plaques whose cells have increased susceptibility to the cytotoxic effects of lipid accumulation can have larger necrotic cores at steady state even if the net macrophage apoptosis rate is held fixed. Based on the ODE for total system lipid $L\left(t\right)$, we further propose that this result reflects excess lipid accumulation in the plaque over time, primarily due to a sustained period of reduced lipid removal by emigration.

Our results for lipid-dependent emigration are arguably the most difficult to analyse and interpret because, even in the case of scaled $g_\gamma\left(a\right)$, none of the steady state ODE solutions remain fixed across our simulated cases. This reflects the influence of the term $G_{\gamma a}$ on the dynamics of $A_M$, which feeds through to the other ODE variables via the macrophage recruitment rate. Although not considered here, our results suggest that it may be worthwhile to perform a complementary set of lipid-dependent simulations where the functions $g_\gamma$ are scaled such that, at steady state, the quantity $G_{\gamma a}$ matches its base case value. In other words, scale the $g_\gamma$ to match the rate of lipid removal by macrophage emigration at steady state rather than the rate of macrophage emigration itself. While this approach is again unlikely to keep any of the individual steady state ODE solutions at their base case values, it may give additional insight to help unpick the intricacies of the simulated outcomes.

In our simulations with unscaled monotonic lipid-dependent emigration, we find that the steady state values of the ODE variables all increase with increasing rate of decline of $g_\gamma$ (decreasing $a_\gamma$). This is unsurprising because, as $a_\gamma$ decreases, the net emigration rate also decreases and the system therefore retains more cells and more lipids. In simulations with scaled $g_\gamma$, we observe the exact same trend in the steady state ODE values. This shows that the reduced net emigration rate in the unscaled cases is only partially responsible for the observed outcomes and, in fact, it is the form of the lipid-dependence that contributes much of the increase in the steady state ODE values relative to the base case. This observation reflects the fact that any cell that fails to emigrate for $a$ small becomes increasingly unlikely to emigrate at all. Cells can therefore persist in the system, acquiring lipid and eventually undergoing apoptosis. This leads to large ingested lipid quantities in both the live and apoptotic cell populations, which ultimately feed into enlarged necrotic cores. Our results show that the detrimental effects of monotonically decreasing lipid-dependent emigration can be substantial. Scaled simulations predict up to a 40\% increase in necrotic core size despite a 2- to 3-fold increase in the net rate of necrotic core consumption.    

As well as monotonic lipid-dependent emigration, we consider non-monotonic $g_\gamma$ where macrophages with moderate lipid loads have the highest emigration rates. These $g_\gamma$ impose reduced rates of emigration for macrophages with very small lipid loads on the basis that such cells (in the absence of proliferation) seem unlikely to exit the plaque \citep{Llod04}. Our results show that non-monotonic lipid-dependent emigration can be beneficial relative to the monotonic lipid-dependence. As cells are more likely to emigrate with a relatively large lipid load, this not only removes more lipid from the system but it also (in scaled cases at least) reduces the likelihood of cells acquiring large lipid loads before dying. Our simulations with parameter value $b_\gamma = 9$ are particularly interesting. In the unscaled case, we see a reduction in steady state $N$ relative to the base case. This is particularly remarkable because: (1) the steady state $G_\gamma$ is only half of the base case value; and (2) the emigration rate of cells with very large $a$ is up to 10 times smaller than in the base case. In the equivalent scaled case, however, we see a larger steady state $N$ relative to the base case. This is because extensive lipid removal from the system by emigration inhibits the immune response. Taken together, these results suggest that the presence of at least some heavily lipid-loaded (and highly inflammatory) macrophages in the plaque may be beneficial for stimulating the recruitment that is required for necrotic core consumption. Observations such as this may prove to be useful in the interpretation of results from experimental studies on plaque resolution and regression \citep{Rahm18}.

Our lipid-dependent emigration results provide novel insight into how the distribution of emigration events can influence the extent of lipid removal from the plaque and the overall plaque progression. These results, however, should be interpreted cautiously. We assume that the likelihood of macrophage emigration depends only on lipid load, whereas, in practice, this relationship would be time-dependent due to spatial factors such as the plaque size and the strength of the emigratory signals relative to other migratory cues (e.g.\ ``find me'' signals from apoptotic cells \citep{Koji17}). Alongside the current work, we are developing new spatio-temporal plaque formation models that include lipid-structured cell populations. We envisage that these models will allow us to better understand how macrophage emigration events are distributed with respect to $a$, over time and at steady state, in different parameter regimes and with different mechanisms of macrophage migration. Moreover, by including proliferation in these new models, we can improve our understanding of how the distribution of emigration events may be altered when macrophages reduce their lipid loads by dividing in the plaque.

In the current work, we perform two sets of simulations that include macrophage proliferation. We first investigate how lipid-independent proliferation interacts with lipid-dependence in the other kinetic terms, and we then investigate the impact of a monotonically decreasing lipid-dependent proliferation rate. When lipid-independent proliferation is combined with scaled lipid-dependent apoptosis or emigration, we observe, as expected, a general reduction in necrotic core size and average cell lipid loads. Overall, however, the qualitative trends in the steady state ODE solutions remain the same as in the proliferation-free cases. When $g_\beta$ is unscaled, we make the interesting observation that proliferation can improve the outcome of plaque formation by reducing the net apoptosis rate. Although the observed reduction at steady state is relatively small (around 8\%), note that the lipid-dependent apoptosis rates range from 2--6 times larger than the proliferation rate (and, as observed in proliferation-free cases, lipid-dependent apoptosis tends to drive cells up this range by increasing lipid consumption). For simulations with lipid-dependent proliferation, we identify two main findings. First, that the steady state ODE solutions depend only on the net steady state proliferation rate $\rho G_\rho$, and, second, that the steady state distribution of lipid across the live cell population is relatively insensitive to the form of $g_\rho$ (provided steady state $G_\rho$ is not too far from 1).   

Although we consider a monotonically decreasing form for $g_\rho$, there exists experimental evidence that the rate of lipid-dependent macrophage proliferation may peak at an intermediate lipid content \citep{Xu15}. This form of lipid-dependent proliferation can be easily included in the model by allowing $g_\rho$ to take the non-monotonic form defined in equation (\ref{dimless_g_nonmono}). An interesting implication of this non-monotonic lipid-dependence is that, unlike monotonically decreasing lipid-dependent proliferation, it can naturally explain the observation that the contribution of plaque macrophage proliferation increases over time (\citet{Lhot16}; i.e.\ as plaque-resident macrophages gradually accumulate lipid, the net population proliferation rate would grow). Of course, many other factors may contribute to this phenomenon, including the presence of a fibrous cap in mature plaques that may inhibit the rate of monocyte recruitment from the bloodstream. We have performed preliminary model investigations using non-monotonic lipid-dependent proliferation, but we omit the results here because we find that they are not substantially different to those with monotonic lipid-dependence. However, in simulations that consider more than one lipid-dependent behaviour at a time (e.g.\ lipid-dependent proliferation and lipid-dependent apoptosis), we envisage that the particular form of lipid-dependent proliferation may have a more significant impact on the outcome of plaque progression. For example, a lipid-dependent macrophage proliferation rate that peaks around the lipid load at which the lipid-dependent apoptosis rate begins to rise would potentially optimise the protective effect of proliferation against the otherwise detrimental effects of lipid cytotoxicity.

\section{Conclusions}
This paper extends the work of \citet{Ford19a} and \citet{Cham22} by developing a lipid-structured model of atherosclerotic plaque macrophages in which the rates of macrophage apoptosis, emigration and proliferation are modulated by the internalised lipid load. We model these lipid-dependent behaviours using dimensionless modulating functions whose features align qualitatively with a range of experimental observations. Our results indicate, particularly for apoptosis and emigration, that variations in macrophage behaviour across lipid loads can substantially alter plaque fate relative to cases without lipid-dependence. This work provides new insight into how macrophage lipid accumulation, and associated lipid-dependent effects, can shape the progression of atherosclerotic plaques.

\section*{Acknowledgements}
MGW and MRM acknowledge funding from an Australian Research Council Discovery Grant (DP200102071).

\bibliography{lipid_dependent_manuscript_arxiv}

\end{document}